\documentclass[longauth]{aa}

\usepackage{graphicx}
\usepackage{natbib}
\usepackage{scalerel}
\usepackage{comment}
\usepackage{placeins}
\usepackage{changepage}
\usepackage{lastpage}

\usepackage[table]{xcolor}
\usepackage[percent]{overpic}

\usepackage{txfonts}
\usepackage[pdfencoding=auto,psdextra]{hyperref}
\hypersetup{
    colorlinks=true,
    linkcolor=blue,
    filecolor=magenta,      
    urlcolor=blue,
    citecolor=blue
}
\makeatletter
\renewcommand*\aa@pageof{, page \thepage{} of \pageref*{LastPage}}
\makeatother

\usepackage[utf8]{inputenc}

\usepackage[switch, modulo]{lineno}

\renewcommand{\linenumbers}[0]{}

\usepackage{euclid}

\newcommand*{\EXT}{EXT\xspace}
\newcommand*{\Gaia}{\emph{Gaia}\xspace}
\newcommand*{\DRone}{DR1\xspace}
\newcommand{\orcid}[1]{} 

\begin{document}
\title{Euclid Data Release 1 (DR1)}
\subtitle{\Euclid photometric masks}   
\titlerunning{Photometric masks}
\authorrunning{Euclid collaboration: J. Odier et al.}
\author{Euclid Collaboration: J.~Odier\orcid{0000-0002-1650-2246}\thanks{\email{jerome.odier@lpsc.in2p3.fr}}\inst{\ref{aff1}}
\and J.~F.~Macias-Perez\orcid{0000-0002-5385-2763}\inst{\ref{aff1}}
\and G.~Alguero\inst{\ref{aff1}}
\and C.~Benoist\orcid{0009-0003-0880-6295}\inst{\ref{aff2}}
\and M.~Radovich\orcid{0000-0002-3585-866X}\inst{\ref{aff3}}
\and P.~Corcho-Caballero\orcid{0000-0001-6327-7080}\inst{\ref{aff4}}
\and H.~Aussel\orcid{0000-0002-1371-5705}\inst{\ref{aff5}}
\and P.~Casenove\orcid{0009-0006-6736-1670}\inst{\ref{aff6}}
\and M.~Ruffenach\orcid{0000-0001-8772-7330}\inst{\ref{aff6}}
\and S.~Farrens\orcid{0000-0002-9594-9387}\inst{\ref{aff5}}
\and B.~Kubik\orcid{0009-0006-5823-4880}\inst{\ref{aff7}}
\and M.~K\"ummel\orcid{0000-0003-2791-2117}\inst{\ref{aff8}}
\and S.~Andreon\orcid{0000-0002-2041-8784}\inst{\ref{aff9}}
\and C.~Baccigalupi\orcid{0000-0002-8211-1630}\inst{\ref{aff10},\ref{aff11},\ref{aff12},\ref{aff13}}
\and M.~Baldi\orcid{0000-0003-4145-1943}\inst{\ref{aff14},\ref{aff15},\ref{aff16}}
\and A.~Balestra\orcid{0000-0002-6967-261X}\inst{\ref{aff3}}
\and S.~Bardelli\orcid{0000-0002-8900-0298}\inst{\ref{aff15}}
\and P.~Battaglia\orcid{0000-0002-7337-5909}\inst{\ref{aff15}}
\and A.~Biviano\orcid{0000-0002-0857-0732}\inst{\ref{aff11},\ref{aff10}}
\and M.~Bolzonella\orcid{0000-0003-3278-4607}\inst{\ref{aff15}}
\and E.~Branchini\orcid{0000-0002-0808-6908}\inst{\ref{aff17},\ref{aff18},\ref{aff9}}
\and M.~Brescia\orcid{0000-0001-9506-5680}\inst{\ref{aff19},\ref{aff20}}
\and J.~Brinchmann\orcid{0000-0003-4359-8797}\inst{\ref{aff21},\ref{aff22},\ref{aff23}}
\and S.~Camera\orcid{0000-0003-3399-3574}\inst{\ref{aff24},\ref{aff25},\ref{aff26}}
\and G.~Ca\~nas-Herrera\orcid{0000-0003-2796-2149}\inst{\ref{aff27}}
\and V.~Capobianco\orcid{0000-0002-3309-7692}\inst{\ref{aff26}}
\and C.~Carbone\orcid{0000-0003-0125-3563}\inst{\ref{aff28}}
\and F.~J.~Castander\orcid{0000-0001-7316-4573}\inst{\ref{aff29},\ref{aff30}}
\and M.~Castellano\orcid{0000-0001-9875-8263}\inst{\ref{aff31}}
\and G.~Castignani\orcid{0000-0001-6831-0687}\inst{\ref{aff15}}
\and S.~Cavuoti\orcid{0000-0002-3787-4196}\inst{\ref{aff20},\ref{aff32}}
\and A.~Cimatti\inst{\ref{aff33}}
\and G.~Congedo\orcid{0000-0003-2508-0046}\inst{\ref{aff34}}
\and L.~Conversi\orcid{0000-0002-6710-8476}\inst{\ref{aff35},\ref{aff36}}
\and Y.~Copin\orcid{0000-0002-5317-7518}\inst{\ref{aff7}}
\and A.~Costille\inst{\ref{aff37}}
\and M.~Cropper\orcid{0000-0003-4571-9468}\inst{\ref{aff38}}
\and A.~Da~Silva\orcid{0000-0002-6385-1609}\inst{\ref{aff39},\ref{aff40}}
\and R.~da~Silva\orcid{0000-0003-4788-677X}\inst{\ref{aff31},\ref{aff41}}
\and G.~De~Lucia\orcid{0000-0002-6220-9104}\inst{\ref{aff11}}
\and X.~Dupac\inst{\ref{aff36}}
\and S.~Escoffier\orcid{0000-0002-2847-7498}\inst{\ref{aff42}}
\and M.~Farina\orcid{0000-0002-3089-7846}\inst{\ref{aff43}}
\and R.~Farinelli\inst{\ref{aff15}}
\and S.~Ferriol\inst{\ref{aff7}}
\and S.~Fotopoulou\orcid{0000-0002-9686-254X}\inst{\ref{aff44}}
\and N.~Fourmanoit\orcid{0009-0005-6816-6925}\inst{\ref{aff42}}
\and M.~Frailis\orcid{0000-0002-7400-2135}\inst{\ref{aff11}}
\and L.~Gabarra\orcid{0000-0002-8486-8856}\inst{\ref{aff45}}
\and S.~Galeotta\orcid{0000-0002-3748-5115}\inst{\ref{aff11}}
\and K.~George\orcid{0000-0002-1734-8455}\inst{\ref{aff46}}
\and C.~Giocoli\orcid{0000-0002-9590-7961}\inst{\ref{aff15},\ref{aff16}}
\and F.~Grupp\inst{\ref{aff47},\ref{aff8}}
\and L.~Guzzo\orcid{0000-0001-8264-5192}\inst{\ref{aff48},\ref{aff9},\ref{aff49}}
\and W.~G.~Hartley\orcid{0000-0002-4563-5101}\inst{\ref{aff50}}
\and S.~V.~H.~Haugan\orcid{0000-0001-9648-7260}\inst{\ref{aff51}}
\and S.~Hemmati\orcid{0000-0003-2226-5395}\inst{\ref{aff52}}
\and H.~Hoekstra\orcid{0000-0002-0641-3231}\inst{\ref{aff27}}
\and F.~Hormuth\inst{\ref{aff53}}
\and K.~Jahnke\orcid{0000-0003-3804-2137}\inst{\ref{aff54}}
\and M.~Jhabvala\inst{\ref{aff55}}
\and S.~Kermiche\orcid{0000-0002-0302-5735}\inst{\ref{aff42}}
\and A.~Kiessling\orcid{0000-0002-2590-1273}\inst{\ref{aff56}}
\and M.~Kunz\orcid{0000-0002-3052-7394}\inst{\ref{aff57}}
\and H.~Kurki-Suonio\orcid{0000-0002-4618-3063}\inst{\ref{aff58},\ref{aff59}}
\and P.~Liebing\orcid{0009-0005-8043-7150}\inst{\ref{aff38}}
\and S.~Ligori\orcid{0000-0003-4172-4606}\inst{\ref{aff26}}
\and P.~B.~Lilje\orcid{0000-0003-4324-7794}\inst{\ref{aff51}}
\and I.~Lloro\orcid{0000-0001-5966-1434}\inst{\ref{aff60}}
\and M.~Magliocchetti\orcid{0000-0001-9158-4838}\inst{\ref{aff43}}
\and O.~Marggraf\orcid{0000-0001-7242-3852}\inst{\ref{aff61}}
\and M.~Martinelli\orcid{0000-0002-6943-7732}\inst{\ref{aff31},\ref{aff62}}
\and N.~Martinet\orcid{0000-0003-2786-7790}\inst{\ref{aff37}}
\and R.~J.~Massey\orcid{0000-0002-6085-3780}\inst{\ref{aff63}}
\and N.~Mauri\orcid{0000-0001-8196-1548}\inst{\ref{aff33},\ref{aff16}}
\and H.~J.~McCracken\orcid{0000-0002-9489-7765}\inst{\ref{aff64}}
\and E.~Medinaceli\orcid{0000-0002-4040-7783}\inst{\ref{aff15}}
\and M.~Meneghetti\orcid{0000-0003-1225-7084}\inst{\ref{aff15},\ref{aff16}}
\and E.~Merlin\orcid{0000-0001-6870-8900}\inst{\ref{aff3}}
\and G.~Meylan\orcid{0000-0001-6503-0209}\inst{\ref{aff65}}
\and P.~Monaco\orcid{0000-0003-2083-7564}\inst{\ref{aff66},\ref{aff11},\ref{aff12},\ref{aff10}}
\and M.~Moresco\orcid{0000-0002-7616-7136}\inst{\ref{aff67},\ref{aff15}}
\and L.~Moscardini\orcid{0000-0002-3473-6716}\inst{\ref{aff67},\ref{aff15},\ref{aff16}}
\and R.~Nakajima\orcid{0009-0009-1213-7040}\inst{\ref{aff61}}
\and C.~Neissner\orcid{0000-0001-8524-4968}\inst{\ref{aff68},\ref{aff69}}
\and R.~C.~Nichol\orcid{0000-0003-0939-6518}\inst{\ref{aff70}}
\and S.-M.~Niemi\orcid{0009-0005-0247-0086}\inst{\ref{aff71}}
\and C.~Padilla\orcid{0000-0001-7951-0166}\inst{\ref{aff68}}
\and F.~Pasian\orcid{0000-0002-4869-3227}\inst{\ref{aff11}}
\and J.~A.~Peacock\orcid{0000-0002-1168-8299}\inst{\ref{aff34}}
\and W.~J.~Percival\orcid{0000-0002-0644-5727}\inst{\ref{aff72},\ref{aff73},\ref{aff74}}
\and V.~Pettorino\orcid{0000-0002-4203-9320}\inst{\ref{aff71}}
\and S.~Pires\orcid{0000-0002-0249-2104}\inst{\ref{aff5}}
\and G.~Polenta\orcid{0000-0003-4067-9196}\inst{\ref{aff41}}
\and M.~Poncet\inst{\ref{aff6}}
\and L.~A.~Popa\inst{\ref{aff75}}
\and L.~Pozzetti\orcid{0000-0001-7085-0412}\inst{\ref{aff15}}
\and A.~Refregier\inst{\ref{aff76}}
\and A.~Renzi\orcid{0000-0001-9856-1970}\inst{\ref{aff77},\ref{aff78},\ref{aff15}}
\and J.~Rhodes\orcid{0000-0002-4485-8549}\inst{\ref{aff56}}
\and G.~Riccio\inst{\ref{aff20}}
\and E.~Romelli\orcid{0000-0003-3069-9222}\inst{\ref{aff11}}
\and M.~Roncarelli\orcid{0000-0001-9587-7822}\inst{\ref{aff15}}
\and B.~Rusholme\orcid{0000-0001-7648-4142}\inst{\ref{aff52}}
\and R.~Saglia\orcid{0000-0003-0378-7032}\inst{\ref{aff8},\ref{aff47}}
\and Z.~Sakr\orcid{0000-0002-4823-3757}\inst{\ref{aff79},\ref{aff80},\ref{aff81}}
\and M.~Schirmer\orcid{0000-0003-2568-9994}\inst{\ref{aff54}}
\and P.~Schneider\orcid{0000-0001-8561-2679}\inst{\ref{aff61}}
\and A.~Secroun\orcid{0000-0003-0505-3710}\inst{\ref{aff42}}
\and E.~Sihvola\orcid{0000-0003-1804-7715}\inst{\ref{aff58},\ref{aff59}}
\and P.~Simon\inst{\ref{aff61}}
\and C.~Sirignano\orcid{0000-0002-0995-7146}\inst{\ref{aff77},\ref{aff78}}
\and G.~Sirri\orcid{0000-0003-2626-2853}\inst{\ref{aff16}}
\and J.~Skottfelt\orcid{0000-0003-1310-8283}\inst{\ref{aff82}}
\and L.~Stanco\orcid{0000-0002-9706-5104}\inst{\ref{aff78}}
\and P.~Tallada-Cresp\'{i}\orcid{0000-0002-1336-8328}\inst{\ref{aff83},\ref{aff69}}
\and H.~I.~Teplitz\orcid{0000-0002-7064-5424}\inst{\ref{aff84}}
\and I.~Tereno\orcid{0000-0002-4537-6218}\inst{\ref{aff39},\ref{aff40}}
\and N.~Tessore\orcid{0000-0002-9696-7931}\inst{\ref{aff38}}
\and S.~Toft\orcid{0000-0003-3631-7176}\inst{\ref{aff85},\ref{aff86}}
\and R.~Toledo-Moreo\orcid{0000-0002-2997-4859}\inst{\ref{aff87},\ref{aff88}}
\and I.~Tutusaus\orcid{0000-0002-3199-0399}\inst{\ref{aff29},\ref{aff30},\ref{aff80}}
\and J.~Valiviita\orcid{0000-0001-6225-3693}\inst{\ref{aff58},\ref{aff59}}
\and T.~Vassallo\orcid{0000-0001-6512-6358}\inst{\ref{aff11},\ref{aff46}}
\and A.~Veropalumbo\orcid{0000-0003-2387-1194}\inst{\ref{aff9},\ref{aff18},\ref{aff17}}
\and Y.~Wang\orcid{0000-0002-4749-2984}\inst{\ref{aff52}}
\and J.~Weller\orcid{0000-0002-8282-2010}\inst{\ref{aff8},\ref{aff47}}
\and A.~Zacchei\orcid{0000-0003-0396-1192}\inst{\ref{aff11},\ref{aff10}}
\and G.~Zamorani\orcid{0000-0002-2318-301X}\inst{\ref{aff15}}
\and F.~M.~Zerbi\orcid{0000-0002-9996-973X}\inst{\ref{aff9}}
\and E.~Zucca\orcid{0000-0002-5845-8132}\inst{\ref{aff15}}
\and V.~Allevato\orcid{0000-0001-7232-5152}\inst{\ref{aff20}}
\and M.~Ballardini\orcid{0000-0003-4481-3559}\inst{\ref{aff89},\ref{aff90},\ref{aff15}}
\and S.~Borgani\orcid{0000-0001-6151-6439}\inst{\ref{aff66},\ref{aff10},\ref{aff11},\ref{aff12},\ref{aff91}}
\and E.~Bozzo\orcid{0000-0002-8201-1525}\inst{\ref{aff50}}
\and C.~Burigana\orcid{0000-0002-3005-5796}\inst{\ref{aff92},\ref{aff93}}
\and M.~Calabrese\orcid{0000-0002-2637-2422}\inst{\ref{aff94},\ref{aff28}}
\and F.~Caro\orcid{0009-0003-1053-0507}\inst{\ref{aff31}}
\and T.~Castro\orcid{0000-0002-6292-3228}\inst{\ref{aff95},\ref{aff11},\ref{aff10}}
\and S.~Conseil\orcid{0000-0002-3657-4191}\inst{\ref{aff96},\ref{aff7}}
\and D.~Di~Ferdinando\inst{\ref{aff16}}
\and J.~A.~Escartin~Vigo\inst{\ref{aff47}}
\and G.~Fabbian\orcid{0000-0002-3255-4695}\inst{\ref{aff97}}
\and T.~Gasparetto\orcid{0000-0002-7913-4866}\inst{\ref{aff31}}
\and A.~Gregorio\orcid{0000-0003-4028-8785}\inst{\ref{aff66},\ref{aff11},\ref{aff12}}
\and C.~M.~Gutierrez\orcid{0000-0001-7854-783X}\inst{\ref{aff98},\ref{aff99}}
\and Y.~Kang\orcid{0009-0000-8588-7250}\inst{\ref{aff100}}
\and A.~Loureiro\orcid{0000-0002-4371-0876}\inst{\ref{aff101},\ref{aff102}}
\and R.~Maoli\orcid{0000-0002-6065-3025}\inst{\ref{aff103},\ref{aff31}}
\and R.~B.~Metcalf\orcid{0000-0003-3167-2574}\inst{\ref{aff67},\ref{aff15}}
\and M.~P\"ontinen\orcid{0000-0001-5442-2530}\inst{\ref{aff58}}
\and E.~Sarpa\orcid{0000-0002-1256-655X}\inst{\ref{aff11}}
\and V.~Scottez\orcid{0009-0008-3864-940X}\inst{\ref{aff104},\ref{aff105}}
\and M.~Sereno\orcid{0000-0003-0302-0325}\inst{\ref{aff15},\ref{aff16}}
\and K.~Tanidis\orcid{0000-0001-9843-5130}\inst{\ref{aff106}}
\and M.~Tenti\orcid{0000-0002-4254-5901}\inst{\ref{aff16}}
\and M.~Wiesmann\orcid{0009-0000-8199-5860}\inst{\ref{aff51}}
\and Abdurro'uf\orcid{0000-0002-5258-8761}\inst{\ref{aff107}}
\and J.~A.~Acevedo~Barroso\orcid{0000-0002-9654-1711}\inst{\ref{aff56}}
\and Y.~Akrami\orcid{0000-0002-2407-7956}\inst{\ref{aff79},\ref{aff108}}
\and G.~Angora\orcid{0000-0002-0316-6562}\inst{\ref{aff20},\ref{aff89}}
\and S.~Anselmi\orcid{0000-0002-3579-9583}\inst{\ref{aff78},\ref{aff77},\ref{aff109}}
\and M.~Archidiacono\orcid{0000-0003-4952-9012}\inst{\ref{aff48},\ref{aff49}}
\and G.~Aric\`o\orcid{0000-0002-2802-2928}\inst{\ref{aff16}}
\and F.~Atrio-Barandela\orcid{0000-0002-2130-2513}\inst{\ref{aff110}}
\and M.~Aubert\orcid{0009-0002-7667-8814}\inst{\ref{aff7}}
\and M.~Baes\orcid{0000-0002-3930-2757}\inst{\ref{aff111}}
\and N.~Bartolo\inst{\ref{aff77},\ref{aff78},\ref{aff3}}
\and L.~Bazzanini\orcid{0000-0003-0727-0137}\inst{\ref{aff92},\ref{aff89}}
\and J.~R.~Bermejo-Climent\inst{\ref{aff112},\ref{aff113}}
\and D.~Bertacca\orcid{0000-0002-2490-7139}\inst{\ref{aff77},\ref{aff3},\ref{aff78}}
\and M.~Bethermin\orcid{0000-0002-3915-2015}\inst{\ref{aff114}}
\and F.~Bouch\`e\orcid{0000-0002-4663-1786}\inst{\ref{aff115},\ref{aff32}}
\and A.~Calabro\orcid{0000-0003-2536-1614}\inst{\ref{aff31}}
\and C.~S.~Carvalho\inst{\ref{aff116}}
\and N.~Chartab\orcid{0000-0003-3691-937X}\inst{\ref{aff52}}
\and F.~Cogato\orcid{0000-0003-4632-6113}\inst{\ref{aff15}}
\and T.~E.~Collett\orcid{0000-0001-5564-3140}\inst{\ref{aff117}}
\and M.~Correnti\orcid{0000-0001-6464-3257}\inst{\ref{aff31},\ref{aff41}}
\and B.~Csizi\orcid{0000-0003-3227-6581}\inst{\ref{aff100}}
\and O.~Cucciati\orcid{0000-0002-9336-7551}\inst{\ref{aff15}}
\and T.~de~Boer\orcid{0000-0001-5486-2747}\inst{\ref{aff118}}
\and G.~Desprez\orcid{0000-0001-8325-1742}\inst{\ref{aff4}}
\and A.~D\'iaz-S\'anchez\orcid{0000-0003-0748-4768}\inst{\ref{aff119}}
\and S.~Di~Domizio\orcid{0000-0003-2863-5895}\inst{\ref{aff17},\ref{aff18}}
\and V.~Duret\orcid{0009-0009-0383-4960}\inst{\ref{aff120}}
\and A.~Enia\orcid{0000-0002-0200-2857}\inst{\ref{aff121},\ref{aff15}}
\and A.~Farina\orcid{0009-0000-3420-929X}\inst{\ref{aff9},\ref{aff18}}
\and A.~Fontana\orcid{0000-0003-3820-2823}\inst{\ref{aff31}}
\and A.~Franco\orcid{0000-0002-4761-366X}\inst{\ref{aff122},\ref{aff123},\ref{aff124}}
\and N.~Frusciante\orcid{0000-0002-7375-1230}\inst{\ref{aff125},\ref{aff126}}
\and A.~Fumagalli\orcid{0009-0004-0300-2535}\inst{\ref{aff11}}
\and Y.~Fu\orcid{0000-0002-0759-0504}\inst{\ref{aff27},\ref{aff4}}
\and K.~Ganga\orcid{0000-0001-8159-8208}\inst{\ref{aff127}}
\and R.~Gavazzi\orcid{0000-0002-5540-6935}\inst{\ref{aff37},\ref{aff64}}
\and E.~Gaztanaga\orcid{0000-0001-9632-0815}\inst{\ref{aff29},\ref{aff30},\ref{aff117}}
\and Z.~Ghaffari\orcid{0000-0002-6467-8078}\inst{\ref{aff11},\ref{aff10}}
\and F.~Gianotti\orcid{0000-0003-4666-119X}\inst{\ref{aff15}}
\and G.~Gozaliasl\orcid{0000-0002-0236-919X}\inst{\ref{aff128},\ref{aff58}}
\and M.~Guidi\orcid{0000-0001-9408-1101}\inst{\ref{aff14},\ref{aff15}}
\and H.~Hildebrandt\orcid{0000-0002-9814-3338}\inst{\ref{aff129}}
\and J.~Hjorth\orcid{0000-0002-4571-2306}\inst{\ref{aff130}}
\and D.~Karagiannis\orcid{0000-0002-4927-0816}\inst{\ref{aff131},\ref{aff89},\ref{aff132}}
\and J.~Kim\orcid{0000-0003-2776-2761}\inst{\ref{aff45}}
\and C.~C.~Kirkpatrick\inst{\ref{aff133}}
\and A.~Kov\'acs\orcid{0000-0002-5825-579X}\inst{\ref{aff113},\ref{aff112}}
\and I.~Kova{\v{c}}i{\'{c}}\orcid{0000-0001-6751-3263}\inst{\ref{aff111}}
\and K.~Koyama\orcid{0000-0001-6727-6915}\inst{\ref{aff117}}
\and S.~Kruk\orcid{0000-0001-8010-8879}\inst{\ref{aff36}}
\and M.~C.~Lam\orcid{0000-0002-9347-2298}\inst{\ref{aff34}}
\and M.~Lembo\orcid{0000-0002-5271-5070}\inst{\ref{aff64}}
\and M.~D.~Lepinzan\orcid{0000-0003-1287-9801}\inst{\ref{aff11}}
\and L.~Linke\orcid{0000-0002-2622-8113}\inst{\ref{aff100}}
\and F.~Mannucci\orcid{0000-0002-4803-2381}\inst{\ref{aff134}}
\and C.~J.~A.~P.~Martins\orcid{0000-0002-4886-9261}\inst{\ref{aff135},\ref{aff22}}
\and M.~Migliaccio\inst{\ref{aff136},\ref{aff137}}
\and M.~Miluzio\inst{\ref{aff36},\ref{aff138}}
\and G.~Morgante\inst{\ref{aff15}}
\and K.~Naidoo\orcid{0000-0002-9182-1802}\inst{\ref{aff117},\ref{aff54}}
\and A.~Navarro-Alsina\orcid{0000-0002-3173-2592}\inst{\ref{aff61}}
\and S.~Nesseris\orcid{0000-0002-0567-0324}\inst{\ref{aff79}}
\and L.~Nicastro\orcid{0000-0001-8534-6788}\inst{\ref{aff15}}
\and F.~Oppizzi\orcid{0000-0003-3904-8370}\inst{\ref{aff18},\ref{aff78},\ref{aff77}}
\and F.~Pace\orcid{0000-0001-8039-0480}\inst{\ref{aff24},\ref{aff25},\ref{aff26},\ref{aff40}}
\and M.~J.~Page\orcid{0000-0002-6689-6271}\inst{\ref{aff38}}
\and D.~Paoletti\orcid{0000-0003-4761-6147}\inst{\ref{aff15},\ref{aff93}}
\and F.~Passalacqua\orcid{0000-0002-8606-4093}\inst{\ref{aff78}}
\and K.~Paterson\orcid{0000-0001-8340-3486}\inst{\ref{aff54}}
\and C.~Pattison\orcid{0000-0003-3272-2617}\inst{\ref{aff117}}
\and A.~Pisani\orcid{0000-0002-6146-4437}\inst{\ref{aff42}}
\and L.~Pizzuti\orcid{0000-0001-5654-7580}\inst{\ref{aff139}}
\and J.~J.~Popp\orcid{0000-0002-3724-1727}\inst{\ref{aff140}}
\and D.~Potter\orcid{0000-0002-0757-5195}\inst{\ref{aff141}}
\and M.~Poulain\orcid{0000-0002-7664-4510}\inst{\ref{aff142}}
\and G.~W.~Pratt\inst{\ref{aff5}}
\and S.~Quai\orcid{0000-0002-0449-8163}\inst{\ref{aff67},\ref{aff15}}
\and F.~Ricci\orcid{0000-0001-5742-5980}\inst{\ref{aff143},\ref{aff31}}
\and G.~Rodighiero\orcid{0000-0002-9415-2296}\inst{\ref{aff77},\ref{aff3}}
\and W.~Roster\orcid{0000-0002-9149-6528}\inst{\ref{aff47}}
\and M.~Sahl\'en\orcid{0000-0003-0973-4804}\inst{\ref{aff144}}
\and D.~B.~Sanders\orcid{0000-0002-1233-9998}\inst{\ref{aff118}}
\and A.~Schneider\orcid{0000-0001-7055-8104}\inst{\ref{aff141}}
\and S.~Serjeant\orcid{0000-0002-0517-7943}\inst{\ref{aff140}}
\and L.~C.~Smith\orcid{0000-0002-3259-2771}\inst{\ref{aff145}}
\and E.~Soubrie\orcid{0000-0001-9295-1863}\inst{\ref{aff97}}
\and I.~Szapudi\orcid{0000-0003-2274-0301}\inst{\ref{aff118}}
\and G.~Testera\orcid{0000-0003-2970-766X}\inst{\ref{aff18}}
\and R.~Teyssier\orcid{0000-0001-7689-0933}\inst{\ref{aff146}}
\and S.~Tosi\orcid{0000-0002-7275-9193}\inst{\ref{aff17},\ref{aff9},\ref{aff18}}
\and C.~Uhlemann\orcid{0000-0001-7831-1579}\inst{\ref{aff147}}
\and C.~Valieri\inst{\ref{aff16}}
\and A.~Viitanen\orcid{0000-0001-9383-786X}\inst{\ref{aff50},\ref{aff133},\ref{aff31}}
\and S.~Vinciguerra\orcid{0009-0005-4018-3184}\inst{\ref{aff37}}
\and M.~von~Wietersheim-Kramsta\orcid{0000-0003-4986-5091}\inst{\ref{aff63},\ref{aff148}}
\and D.~Scott\orcid{0000-0002-6878-9840}\inst{\ref{aff149}}}
										   
\institute{Univ. Grenoble Alpes, CNRS, Grenoble INP, LPSC-IN2P3, 53, Avenue des Martyrs, 38000 Grenoble, France\label{aff1}
\and
Universit\'e C\^{o}te d'Azur, Observatoire de la C\^{o}te d'Azur, CNRS, Laboratoire Lagrange, Bd de l'Observatoire, CS 34229, 06304 Nice cedex 4, France\label{aff2}
\and
INAF-Osservatorio Astronomico di Padova, Via dell'Osservatorio 5, 35122 Padova, Italy\label{aff3}
\and
Kapteyn Astronomical Institute, University of Groningen, PO Box 800, 9700 AV Groningen, The Netherlands\label{aff4}
\and
Universit\'e Paris-Saclay, Universit\'e Paris Cit\'e, CEA, CNRS, AIM, 91191, Gif-sur-Yvette, France\label{aff5}
\and
Centre National d'Etudes Spatiales -- Centre spatial de Toulouse, 18 avenue Edouard Belin, 31401 Toulouse Cedex 9, France\label{aff6}
\and
Universit\'e Lyon 1, CNRS/IN2P3, IP2I Lyon, Villeurbanne, France\label{aff7}
\and
Universit\"ats-Sternwarte M\"unchen, Fakult\"at f\"ur Physik, Ludwig-Maximilians-Universit\"at M\"unchen, Scheinerstr.~1, 81679 M\"unchen, Germany\label{aff8}
\and
INAF-Osservatorio Astronomico di Brera, Via Brera 28, 20122 Milano, Italy\label{aff9}
\and
IFPU, Institute for Fundamental Physics of the Universe, via Beirut 2, 34151 Trieste, Italy\label{aff10}
\and
INAF-Osservatorio Astronomico di Trieste, Via G. B. Tiepolo 11, 34143 Trieste, Italy\label{aff11}
\and
INFN, Sezione di Trieste, Via Valerio 2, 34127 Trieste TS, Italy\label{aff12}
\and
SISSA, International School for Advanced Studies, Via Bonomea 265, 34136 Trieste TS, Italy\label{aff13}
\and
Dipartimento di Fisica e Astronomia, Universit\`a di Bologna, Via Gobetti 93/2, 40129 Bologna, Italy\label{aff14}
\and
INAF-Osservatorio di Astrofisica e Scienza dello Spazio di Bologna, Via Piero Gobetti 93/3, 40129 Bologna, Italy\label{aff15}
\and
INFN-Sezione di Bologna, Viale Berti Pichat 6/2, 40127 Bologna, Italy\label{aff16}
\and
Dipartimento di Fisica, Universit\`a di Genova, Via Dodecaneso 33, 16146, Genova, Italy\label{aff17}
\and
INFN-Sezione di Genova, Via Dodecaneso 33, 16146, Genova, Italy\label{aff18}
\and
Department of Physics "E. Pancini", University Federico II, Via Cinthia 6, 80126, Napoli, Italy\label{aff19}
\and
INAF-Osservatorio Astronomico di Capodimonte, Via Moiariello 16, 80131 Napoli, Italy\label{aff20}
\and
European Southern Observatory, Karl-Schwarzschild-Str.~2, 85748 Garching, Germany\label{aff21}
\and
Instituto de Astrof\'isica e Ci\^encias do Espa\c{c}o, Universidade do Porto, CAUP, Rua das Estrelas, PT4150-762 Porto, Portugal\label{aff22}
\and
Faculdade de Ci\^encias da Universidade do Porto, Rua do Campo de Alegre, 4150-007 Porto, Portugal\label{aff23}
\and
Dipartimento di Fisica, Universit\`a degli Studi di Torino, Via P. Giuria 1, 10125 Torino, Italy\label{aff24}
\and
INFN-Sezione di Torino, Via P. Giuria 1, 10125 Torino, Italy\label{aff25}
\and
INAF-Osservatorio Astrofisico di Torino, Via Osservatorio 20, 10025 Pino Torinese (TO), Italy\label{aff26}
\and
Leiden Observatory, Leiden University, Einsteinweg 55, 2333 CC Leiden, The Netherlands\label{aff27}
\and
INAF-IASF Milano, Via Alfonso Corti 12, 20133 Milano, Italy\label{aff28}
\and
Institute of Space Sciences (ICE, CSIC), Campus UAB, Carrer de Can Magrans, s/n, 08193 Barcelona, Spain\label{aff29}
\and
Institut d'Estudis Espacials de Catalunya (IEEC),  Edifici RDIT, Campus UPC, 08860 Castelldefels, Barcelona, Spain\label{aff30}
\and
INAF-Osservatorio Astronomico di Roma, Via Frascati 33, 00078 Monteporzio Catone, Italy\label{aff31}
\and
INFN -- Sezione di Napoli, Via Cinthia 6, 80126, Napoli, Italy\label{aff32}
\and
Dipartimento di Fisica e Astronomia "Augusto Righi" - Alma Mater Studiorum Universit\`a di Bologna, Viale Berti Pichat 6/2, 40127 Bologna, Italy\label{aff33}
\and
Institute for Astronomy, University of Edinburgh, Royal Observatory, Blackford Hill, Edinburgh EH9 3HJ, UK\label{aff34}
\and
European Space Agency/ESRIN, Largo Galileo Galilei 1, 00044 Frascati, Roma, Italy\label{aff35}
\and
ESAC/ESA, Camino Bajo del Castillo, s/n., Urb. Villafranca del Castillo, 28692 Villanueva de la Ca\~nada, Madrid, Spain\label{aff36}
\and
Aix-Marseille Universit\'e, CNRS, CNES, LAM, Marseille, France\label{aff37}
\and
Mullard Space Science Laboratory, University College London, Holmbury St Mary, Dorking, Surrey RH5 6NT, UK\label{aff38}
\and
Departamento de F\'isica, Faculdade de Ci\^encias, Universidade de Lisboa, Edif\'icio C8, Campo Grande, PT1749-016 Lisboa, Portugal\label{aff39}
\and
Instituto de Astrof\'isica e Ci\^encias do Espa\c{c}o, Faculdade de Ci\^encias, Universidade de Lisboa, Campo Grande, 1749-016 Lisboa, Portugal\label{aff40}
\and
Space Science Data Center, Italian Space Agency, via del Politecnico snc, 00133 Roma, Italy\label{aff41}
\and
Aix-Marseille Universit\'e, CNRS/IN2P3, CPPM, Marseille, France\label{aff42}
\and
INAF-Istituto di Astrofisica e Planetologia Spaziali, via del Fosso del Cavaliere, 100, 00100 Roma, Italy\label{aff43}
\and
School of Physics, HH Wills Physics Laboratory, University of Bristol, Tyndall Avenue, Bristol, BS8 1TL, UK\label{aff44}
\and
Department of Physics, University of Oxford, Keble Road, Oxford OX1 3RH, UK\label{aff45}
\and
University Observatory, LMU Faculty of Physics, Scheinerstr.~1, 81679 Munich, Germany\label{aff46}
\and
Max Planck Institute for Extraterrestrial Physics, Giessenbachstr. 1, 85748 Garching, Germany\label{aff47}
\and
Dipartimento di Fisica "Aldo Pontremoli", Universit\`a degli Studi di Milano, Via Celoria 16, 20133 Milano, Italy\label{aff48}
\and
INFN-Sezione di Milano, Via Celoria 16, 20133 Milano, Italy\label{aff49}
\and
Department of Astronomy, University of Geneva, ch. d'Ecogia 16, 1290 Versoix, Switzerland\label{aff50}
\and
Institute of Theoretical Astrophysics, University of Oslo, P.O. Box 1029 Blindern, 0315 Oslo, Norway\label{aff51}
\and
Caltech/IPAC, 1200 E. California Blvd., Pasadena, CA 91125, USA\label{aff52}
\and
Felix Hormuth Engineering, Goethestr. 17, 69181 Leimen, Germany\label{aff53}
\and
Max-Planck-Institut f\"ur Astronomie, K\"onigstuhl 17, 69117 Heidelberg, Germany\label{aff54}
\and
NASA Goddard Space Flight Center, Greenbelt, MD 20771, USA\label{aff55}
\and
Jet Propulsion Laboratory, California Institute of Technology, 4800 Oak Grove Drive, Pasadena, CA, 91109, USA\label{aff56}
\and
Universit\'e de Gen\`eve, D\'epartement de Physique Th\'eorique and Centre for Astroparticle Physics, 24 quai Ernest-Ansermet, CH-1211 Gen\`eve 4, Switzerland\label{aff57}
\and
Department of Physics, P.O. Box 64, University of Helsinki, 00014 Helsinki, Finland\label{aff58}
\and
Helsinki Institute of Physics, Gustaf H{\"a}llstr{\"o}min katu 2, University of Helsinki, 00014 Helsinki, Finland\label{aff59}
\and
SKAO, Jodrell Bank, Lower Withington, Macclesfield SK11 9FT, UK\label{aff60}
\and
Universit\"at Bonn, Argelander-Institut f\"ur Astronomie, Auf dem H\"ugel 71, 53121 Bonn, Germany\label{aff61}
\and
INFN-Sezione di Roma, Piazzale Aldo Moro, 2 - c/o Dipartimento di Fisica, Edificio G. Marconi, 00185 Roma, Italy\label{aff62}
\and
Department of Physics, Institute for Computational Cosmology, Durham University, South Road, Durham, DH1 3LE, UK\label{aff63}
\and
Institut d'Astrophysique de Paris, UMR 7095, CNRS, and Sorbonne Universit\'e, 98 bis boulevard Arago, 75014 Paris, France\label{aff64}
\and
Institute of Physics, Laboratory of Astrophysics, Ecole Polytechnique F\'ed\'erale de Lausanne (EPFL), Observatoire de Sauverny, 1290 Versoix, Switzerland\label{aff65}
\and
Dipartimento di Fisica - Sezione di Astronomia, Universit\`a di Trieste, Via Tiepolo 11, 34131 Trieste, Italy\label{aff66}
\and
Dipartimento di Fisica e Astronomia "Augusto Righi" - Alma Mater Studiorum Universit\`a di Bologna, via Piero Gobetti 93/2, 40129 Bologna, Italy\label{aff67}
\and
Institut de F\'{i}sica d'Altes Energies (IFAE), The Barcelona Institute of Science and Technology, Campus UAB, 08193 Bellaterra (Barcelona), Spain\label{aff68}
\and
Port d'Informaci\'{o} Cient\'{i}fica, Campus UAB, C. Albareda s/n, 08193 Bellaterra (Barcelona), Spain\label{aff69}
\and
School of Mathematics and Physics, University of Surrey, Guildford, Surrey, GU2 7XH, UK\label{aff70}
\and
European Space Agency/ESTEC, Keplerlaan 1, 2201 AZ Noordwijk, The Netherlands\label{aff71}
\and
Waterloo Centre for Astrophysics, University of Waterloo, Waterloo, Ontario N2L 3G1, Canada\label{aff72}
\and
Department of Physics and Astronomy, University of Waterloo, Waterloo, Ontario N2L 3G1, Canada\label{aff73}
\and
Perimeter Institute for Theoretical Physics, Waterloo, Ontario N2L 2Y5, Canada\label{aff74}
\and
Institute of Space Science, Str. Atomistilor, nr. 409 M\u{a}gurele, Ilfov, 077125, Romania\label{aff75}
\and
Institute for Particle Physics and Astrophysics, Dept. of Physics, ETH Zurich, Wolfgang-Pauli-Strasse 27, 8093 Zurich, Switzerland\label{aff76}
\and
Dipartimento di Fisica e Astronomia "G. Galilei", Universit\`a di Padova, Via Marzolo 8, 35131 Padova, Italy\label{aff77}
\and
INFN-Padova, Via Marzolo 8, 35131 Padova, Italy\label{aff78}
\and
Instituto de F\'isica Te\'orica UAM-CSIC, Campus de Cantoblanco, 28049 Madrid, Spain\label{aff79}
\and
Institut de Recherche en Astrophysique et Plan\'etologie (IRAP), Universit\'e de Toulouse, CNRS, UPS, CNES, 14 Av. Edouard Belin, 31400 Toulouse, France\label{aff80}
\and
Universit\'e St Joseph; Faculty of Sciences, Beirut, Lebanon\label{aff81}
\and
Centre for Electronic Imaging, Open University, Walton Hall, Milton Keynes, MK7~6AA, UK\label{aff82}
\and
Centro de Investigaciones Energ\'eticas, Medioambientales y Tecnol\'ogicas (CIEMAT), Avenida Complutense 40, 28040 Madrid, Spain\label{aff83}
\and
Infrared Processing and Analysis Center, California Institute of Technology, Pasadena, CA 91125, USA\label{aff84}
\and
Cosmic Dawn Center (DAWN)\label{aff85}
\and
Niels Bohr Institute, University of Copenhagen, Jagtvej 128, 2200 Copenhagen, Denmark\label{aff86}
\and
Universidad Polit\'ecnica de Cartagena, Departamento de Electr\'onica y Tecnolog\'ia de Computadoras,  Plaza del Hospital 1, 30202 Cartagena, Spain\label{aff87}
\and
European University of Technology EUt+, European Union\label{aff88}
\and
Dipartimento di Fisica e Scienze della Terra, Universit\`a degli Studi di Ferrara, Via Giuseppe Saragat 1, 44122 Ferrara, Italy\label{aff89}
\and
Istituto Nazionale di Fisica Nucleare, Sezione di Ferrara, Via Giuseppe Saragat 1, 44122 Ferrara, Italy\label{aff90}
\and
ICSC - Centro Nazionale di Ricerca in High Performance Computing, Big Data e Quantum Computing, Via Magnanelli 2, Bologna, Italy\label{aff91}
\and
INAF, Istituto di Radioastronomia, Via Piero Gobetti 101, 40129 Bologna, Italy\label{aff92}
\and
INFN-Bologna, Via Irnerio 46, 40126 Bologna, Italy\label{aff93}
\and
Astronomical Observatory of the Autonomous Region of the Aosta Valley (OAVdA), Loc. Lignan 39, I-11020, Nus (Aosta Valley), Italy\label{aff94}
\and
Department of Mathematical Physics, Institute of Physics, University of S\~ao Paulo, R. do Mat\~ao 1371, 05508-090, S\~ao Paulo, SP, Brazil\label{aff95}
\and
Centre de Recherche Astrophysique de Lyon, UMR5574, CNRS, Universit\'e Claude Bernard Lyon 1, ENS de Lyon, 69230, Saint-Genis-Laval, France\label{aff96}
\and
Universit\'e Paris-Saclay, CNRS, Institut d'astrophysique spatiale, 91405, Orsay, France\label{aff97}
\and
Instituto de Astrof\'{\i}sica de Canarias, E-38205 La Laguna, Tenerife, Spain\label{aff98}
\and
Universidad de La Laguna, Dpto. Astrof\'\i sica, E-38206 La Laguna, Tenerife, Spain\label{aff99}
\and
Universit\"at Innsbruck, Institut f\"ur Astro- und Teilchenphysik, Technikerstr. 25/8, 6020 Innsbruck, Austria\label{aff100}
\and
Oskar Klein Centre for Cosmoparticle Physics, Department of Physics, Stockholm University, Stockholm, SE-106 91, Sweden\label{aff101}
\and
Astrophysics Group, Blackett Laboratory, Imperial College London, London SW7 2AZ, UK\label{aff102}
\and
Dipartimento di Fisica, Sapienza Universit\`a di Roma, Piazzale Aldo Moro 2, 00185 Roma, Italy\label{aff103}
\and
Institut d'Astrophysique de Paris, 98bis Boulevard Arago, 75014, Paris, France\label{aff104}
\and
ICL, Junia, Universit\'e Catholique de Lille, LITL, 59000 Lille, France\label{aff105}
\and
Center for Astrophysics and Cosmology, University of Nova Gorica, Nova Gorica, Slovenia\label{aff106}
\and
Department of Astronomy, Indiana University, 727 East Third Street, Bloomington, IN 47405, USA\label{aff107}
\and
CERCA/ISO, Department of Physics, Case Western Reserve University, 10900 Euclid Avenue, Cleveland, OH 44106, USA\label{aff108}
\and
Laboratoire Univers et Th\'eorie, Observatoire de Paris, Universit\'e PSL, Universit\'e Paris Cit\'e, CNRS, 92190 Meudon, France\label{aff109}
\and
Departamento de F{\'\i}sica Fundamental. Universidad de Salamanca. Plaza de la Merced s/n. 37008 Salamanca, Spain\label{aff110}
\and
Universiteit Gent, Department of Physics and Astronomy, Proeftuinstraat 86 N3, 9000 Ghent, Belgium
\label{aff111}
\and
Konkoly Observatory, HUN-REN CSFK, MTA Centre of Excellence, Budapest, Konkoly Thege Mikl\'os {\'u}t 15-17. H-1121, Hungary\label{aff112}
\and
MTA-CSFK Lend\"ulet Large-Scale Structure Research Group, Konkoly-Thege Mikl\'os \'ut 15-17, H-1121 Budapest, Hungary\label{aff113}
\and
Universit\'e de Strasbourg, CNRS, Observatoire astronomique de Strasbourg, UMR 7550, 67000 Strasbourg, France\label{aff114}
\and
Scuola Superiore Meridionale, Via Mezzocannone 4, 80138, Napoli, Italy\label{aff115}
\and
Instituto de Astrof\'isica e Ci\^encias do Espa\c{c}o, Faculdade de Ci\^encias, Universidade de Lisboa, Tapada da Ajuda, 1349-018 Lisboa, Portugal\label{aff116}
\and
Institute of Cosmology and Gravitation, University of Portsmouth, Portsmouth PO1 3FX, UK\label{aff117}
\and
Institute for Astronomy, University of Hawaii, 2680 Woodlawn Drive, Honolulu, HI 96822, USA\label{aff118}
\and
Departamento F\'isica Aplicada, Universidad Polit\'ecnica de Cartagena, Campus Muralla del Mar, 30202 Cartagena, Murcia, Spain\label{aff119}
\and
Univ. Grenoble Alpes, CNRS, IPAG, 38000 Grenoble, France\label{aff120}
\and
CEA Saclay, DFR/IRFU, Service d'Astrophysique, Bat. 709, 91191 Gif-sur-Yvette, France\label{aff121}
\and
Department of Mathematics and Physics E. De Giorgi, University of Salento, Via per Arnesano, CP-I93, 73100, Lecce, Italy\label{aff122}
\and
INFN, Sezione di Lecce, Via per Arnesano, CP-193, 73100, Lecce, Italy\label{aff123}
\and
INAF-Sezione di Lecce, c/o Dipartimento Matematica e Fisica, Via per Arnesano, 73100, Lecce, Italy\label{aff124}
\and
Dipartimento di Fisica "E. Pancini", Universita degli Studi di Napoli Federico II, Via Cinthia 6, 80126, Napoli, Italy\label{aff125}
\and
INFN Sezione di Napoli, Universita degli Studi di Napoli "Federico II", Compl. Univ. di Monte S. Angelo, Edificio G, Via Cinthia, I-80126, Napoli, Italy\label{aff126}
\and
Universit\'e Paris Cit\'e, CNRS, Astroparticule et Cosmologie, 75013 Paris, France\label{aff127}
\and
Department of Computer Science, Aalto University, PO Box 15400, Espoo, FI-00 076, Finland\label{aff128}
\and
Ruhr University Bochum, Faculty of Physics and Astronomy, Astronomical Institute (AIRUB), German Centre for Cosmological Lensing (GCCL), 44780 Bochum, Germany\label{aff129}
\and
DARK, Niels Bohr Institute, University of Copenhagen, Jagtvej 155, 2200 Copenhagen, Denmark\label{aff130}
\and
Van Swinderen Institute, University of Groningen, Nijenborgh 3, 9747 AG Groningen, The Netherlands\label{aff131}
\and
Department of Physics and Astronomy, University of the Western Cape, Bellville, Cape Town, 7535, South Africa\label{aff132}
\and
Department of Physics and Helsinki Institute of Physics, Gustaf H\"allstr\"omin katu 2, University of Helsinki, 00014 Helsinki, Finland\label{aff133}
\and
INAF-Osservatorio Astrofisico di Arcetri, Largo E. Fermi 5, 50125, Firenze, Italy\label{aff134}
\and
Centro de Astrof\'{\i}sica da Universidade do Porto, Rua das Estrelas, 4150-762 Porto, Portugal\label{aff135}
\and
Dipartimento di Fisica, Universit\`a di Roma Tor Vergata, Via della Ricerca Scientifica 1, Roma, Italy\label{aff136}
\and
INFN, Sezione di Roma 2, Via della Ricerca Scientifica 1, Roma, Italy\label{aff137}
\and
HE Space for European Space Agency (ESA), Camino bajo del Castillo, s/n, Urbanizacion Villafranca del Castillo, Villanueva de la Ca\~nada, 28692 Madrid, Spain\label{aff138}
\and
Dipartimento di Fisica ``G. Occhialini", Universit\`a degli Studi di Milano Bicocca, Piazza della Scienza 3, 20126 Milano, Italy\label{aff139}
\and
School of Physical Sciences, The Open University, Milton Keynes, MK7 6AA, UK\label{aff140}
\and
Department of Astrophysics, University of Zurich, Winterthurerstrasse 190, 8057 Zurich, Switzerland\label{aff141}
\and
Space physics and astronomy research unit, University of Oulu, Pentti Kaiteran katu 1, FI-90014 Oulu, Finland\label{aff142}
\and
Department of Mathematics and Physics, Roma Tre University, Via della Vasca Navale 84, 00146 Rome, Italy\label{aff143}
\and
Theoretical astrophysics, Department of Physics and Astronomy, Uppsala University, Box 524, 751 20 Uppsala, Sweden\label{aff144}
\and
Institute of Astronomy, University of Cambridge, Madingley Road, Cambridge CB3 0HA, UK\label{aff145}
\and
Department of Astrophysical Sciences, Peyton Hall, Princeton University, Princeton, NJ 08544, USA\label{aff146}
\and
Fakult\"at f\"ur Physik, Universit\"at Bielefeld, Postfach 100131, 33501 Bielefeld, Germany\label{aff147}
\and
Department of Physics, Centre for Extragalactic Astronomy, Durham University, South Road, Durham, DH1 3LE, UK\label{aff148}
\and
Department of Physics and Astronomy, University of British Columbia, Vancouver, BC V6T 1Z1, Canada\label{aff149}}    
 
\date{\today}
\abstract{This paper describes the Euclid \DRone photometric masks, which trace spatial variations in the survey properties (e.g. satellite position angles, integration time, sky tiling, and survey footprint); in the instrumental performance of \Euclid\ and the external data (\EXT) it uses (e.g. survey coverage, depth, PSF width, quality flags, systematic effects and astrometric uncertainties); and in possible contamination from sky emission (e.g. galactic extinction and zodiacal light emission). Among this, the effective coverage mask is particularly important as it defines the effective survey area and can be used to exclude spurious objects when selecting \Euclid\ photometric samples for scientific analyses. The \Euclid\ photometric masks are built for each observed sky region (tile) and are provided as full-sky \texttt{HEALPix} maps (nested ordering and explicit indexing) at $N_{\mathrm{side}}\,{=}\,$16\,384, equating to a resolution of \ang{;;12.88}, for both the Euclid Wide Survey (EWS) and Euclid Deep Survey (EDS). We estimate the total area of the \DRone EWS (EDS) to be 2008.8~ (74.7)\,deg$^2$  and find that 82~(81)\% of this area is preserved by the effective coverage mask. We observe that \Euclid's instrumental properties are consistent with expectations, showing  only minor relative variations across the sky. For \DRone EWS the standard deviation of the depth (computed for a 2$^{\prime \prime}$ aperture and at 5 $\sigma$) is below 0.2 AB mag, with median values of 25.69, 24.14, 24.3, and 24.29 for the \Euclid\ \IE, \YE, \JE, and \HE bands. Finally, we expect these photometric masks to be key to many science applications and in particular for cosmological analyses.}
\maketitle
\section{\label{sc:Intro}Introduction}
The \Euclid\ satellite mission \citep{EuclidSkyOverview}, launched in July 2023, aims to constrain the main cosmological parameters through two primary cosmological probes: gravitational lensing and galaxy clustering \citep{Amendola18,Blanchard-EP7}. To this end, it will collect data over the 14\,000 $\text{deg}^2$ Euclid Wide Survey (EWS) and the approximately $63\,\text{deg}^2$ Euclid Deep Survey \citep[EDS;][]{Scaramella-EP1}\footnote{\url{https://www.cosmos.esa.int/web/euclid/euclid-survey}}.

\Euclid\ has two instruments \citep{Scaramella-EP1}: the Visible instrument \citep[VIS,][]{EuclidSkyVIS} and the Near Infrared Spectrometer and Photometer \citep[NISP,][]{EuclidSkyNISP}. VIS was designed to provide exquisite imaging in the optical domain for galaxy detection and shear measurements. It consists of a single visible band, \IE. NISP provides photometric imaging in three bands, \YE, \JE, and \HE, for detecting distant galaxies, as well as spectroscopic redshifts for galaxy clustering. The \Euclid\ data are combined with external data \citep[referred to as `EXT', see][for more details]{DR1-TP006} from ground-based photometric surveys such as the Dark Energy Survey \citep[DES,][]{Abbott2021}, the Canada-France Imaging Survey \citep[CFIS,][]{Ibata2017}, the Panoramic Survey Telescope and Rapid Response System on Maui \citep[Pan-STARRS,][]{Magnier2020}, The Waterloo-Hawaii-IfA G-band Survey (WHIGS), and the Wide Imaging with Subaru HSC of the Euclid Sky \citep[WHISHES,][]{2018PASJ...70S...4A} to determine photometric redshifts for all VIS- and NISP-detected objects \citep{DR1-TP011}.

The processing of \Euclid\ data \citep[see][for a detailed description]{DR1-TP033} is organised within the \Euclid\ Science Ground Segment (SGS) into pipelines and processing functions (PFs). Each PF is developed and maintained by a corresponding Organisational Unit (OU) within the SGS. This paper describes the OU-LE3 (Organizational Unit Level 3) VMPZ-ID (Visibility Mask PhotoZ-Internal Data) PF (Processing Function) and associated pipelines contributing to the core products of Euclid Data Release 1 \citep[\DRone][]{DR1cite}, also known as \DRone Foundation. An overview of the DR1 Foundation products is given in \citet{DR1-TP001}. The DR1 sky survey \citep{DR1-TP027} comprises about $2000\,{\rm deg}^2$ of the EWS and the three EDS subfields: Euclid Deep Field Fornax (EDF-F), South (EDF-S), and North (EDF-N). Some Euclid Auxiliary Fields and calibration fields are also released.

The main VMPZ-ID products are \texttt{HEALPix}\footnote{Hierarchical Equal Area isoLatitude Pixelation of a sphere, see\\
    \url{https://healpix.sourceforge.io/}} \citep{2005ApJ...622..759G} sky maps, which are called photometric masks hereafter, tracking spatial variations in the properties of the \Euclid\ survey, the instrumental performance of VIS and NISP, the quality of EXT data, and foreground sky emission. These variations could propagate into the final \Euclid\ catalogues and bias the estimation of cosmological parameters, as reported for other photometric surveys, such as KiDS and DES \citep[see][and references therein]{2026arXiv260114484W,2025A&A...694A.259Y}. These masks will be also useful in selecting \Euclid\ galaxy samples for statistical studies in non-cosmological analyses.

The main inputs to the VMPZ-ID PF are 
merged sky images and associated RMS noise and flags \citep[see][for further description]{DR1-TP009}. Star catalogues and star polygons including the central diffraction lobe and diffraction spikes, which are constructed from the \Gaia survey, are also used \citep{2016A&A...595A...1G}.

This document is structured as follows. Section~\ref{sec:pipeline} presents the VMPZ-ID pipeline and the products delivered in \DRone. Section~\ref{sec:detailedmasks} details the VMPZ-ID photometric masks, Sect.~\ref{sec:wide} discusses the main EWS photometric masks, and Sect.~\ref{sec:deep} describes the masks for the EDFs. Conclusions are presented in Sect.~\ref{sec:conclu}.

\section{VMPZ-ID pipeline and products}
\label{sec:pipeline}
The DR1 VMPZ-ID pipeline is based on the Q1 pipeline described in \citet{Q1-TP001} and has since been improved and extended to include additional sky masks. Information associated to the VIS, NIR, and \EXT\ MER \citep[MERge processing,][]{DR1-TP009} images at exposure (e.g. integration time, star polygons, satellite position angles, PSF width) and tile (e.g. image footprint, mask flags and noise) levels is combined into a series of masks in the \texttt{HEALPix} format using nested ordering, explicit indexing and $N_{\mathrm{side}}\,{=}\,16\,384$, which corresponds to a resolution of approximately \ang{;;12.88}. This resolution was selected as a good compromise between usability, allowing for loading an all-sky mask in the memory of a \Euclid\ Science Data Centre node, and preserving as much as possible of the fine spatial information given by the pixel scale masks of the original MER images \citep[][]{DR1-TP009}. VMPZ-ID also produces, at the same resolution, the so-called {\tt DpdHealpixInfoMapVMPZ} product, which enables satellite and instrumental properties to be tracked and projected onto the sky. This allows straightforward cross-checks between spatial information (for example, the noise at a given point in the survey) and temporal information (for example, the value of a temperature sensor on the \Euclid\ focal plane), and facilitates searches for possible effects on \Euclid\ performance. Finally, random galaxy catalogues are also produced to mimic spatial variations in the combined coverage of all bands used for photometry.

\begin{table*}[htbp!]
    \caption{Summary of the main characteristics of the photometric masks produced by VMPZ-ID.}

    \newlength{\tabcs}
    \setlength{\tabcs}{0mm}

    \centering

    \makebox[\textwidth][c]{%
        \scalebox{0.975}{%
            \begin{tabular}{@{\hskip 0mm}l@{\hskip -3mm}c@{\hskip 2mm}c@{\hskip -7mm}c@{\hskip 0mm}c @{\hskip 0mm}c@{\hskip -2mm}c@{\hskip 0mm}c}
                \hline\hline
                \noalign{\vskip 1pt}

                Mask  & Mask  & Mask  & Information & Numerical & Units & Per-band \\
                name & family & product type & map type & type & & mask \\

                \hline
                \noalign{\vskip 1pt}

                Tile index & Survey & {\small \texttt{DpdHealpixInfoMapVMPZ}} & Tile & int64 & \dots & no\\
                Solar aspect angle & Survey/Satellite & {\small \texttt{DpdHealpixInfoMapVMPZ}} & SAA & float32 & \textrm{deg} & no \\
                Alpha angle & Survey/Satellite & {\small \texttt{DpdHealpixInfoMapVMPZ}} & AA & float32 & \textrm{deg} & no \\
                Beta angle & Survey/Satellite & {\small \texttt{DpdHealpixInfoMapVMPZ}} & BA & float32 & \textrm{deg} & no \\

                \hline
                \noalign{\vskip 1pt}

                Integration time & Survey & {\small \texttt{DpdHealpixInfoMapVMPZ}} & Exposures & float32 & \textrm{sec} & \Euclid only \\
                Footprint & Survey/Instrument & {\small \texttt{DpdHealpixFootprintMaskVMPZ}} & \dots & float32 & \dots & yes \\
                Coverage & Survey/Instrument & {\small \texttt{DpdHealpixCoverageMaskVMPZ}} & \dots & float32 & \dots & yes \\
                Effective coverage & Survey/Instrument & {\tiny \texttt{DpdHealpixEffectiveCoverageMaskVMPZ}} & \dots & float32 & \dots & combined \\

                \hline
                \noalign{\vskip 1pt}

                Bit mask & Instrument & {\small \texttt{DpdHealpixBitMaskVMPZ}} & \dots & uint32 & \dots & yes \\
                Depth & Instrument & {\small \texttt{DpdHealpixDepthMapVMPZ}} & \dots & float32 & \textrm{AB mag} & yes \\
                RMS noise & Instrument & {\small \texttt{DpdHealpixInfoMapVMPZ}} & Noise & float32 & \textrm{ADU} & yes \\
                Point-spread function & Instrument & {\small \texttt{DpdHealpixInfoMapVMPZ}} & PSF & float32 & \textrm{arcsec} & yes \\
                Persistence & Instrument & {\small \texttt{DpdHealpixInfoMapVMPZ}} & Persistence & float32 & \dots & NIR only \\
                NIR PRMS & Instrument & {\small \texttt{DpdHealpixInfoMapVMPZ}} & NIR\_PRMS & float32 & \textrm{mas} & NIR only \\
                Pan-STARRS artefacts & Instrument & {\small \texttt{DpdHealpixInfoMapVMPZ}} & PanStarrsArtefacts & float32 & \dots & Pan-STARRS only \\

                \hline
                \noalign{\vskip 1pt}

                Zodiacal light & Sky/Instrument & {\small \texttt{DpdHealpixInfoMapVMPZ}} & ZodiacalLight & float32 & $\textrm{erg}\,\textrm{cm}^{-2}\textrm{\r{A}}^{-1}$ & \Euclid only \\
                & & & & & $\textrm{arcsec}^{-2}$ & \\
                Galactic extinction & Sky/Instrument & {\small \texttt{DpdHealpixInfoMapVMPZ}} & GalacticExtinction & float32 & mag & \Euclid only \\

                \hline
                \noalign{\vskip 1pt}
                Star number density & Sky emission & {\small \texttt{DpdHealpixInfoMapVMPZ}} & StarNumberDensity & float32 & $\textrm{counts}/\textrm{pixel}$ & no \\
                Star mean flux & Sky emission & {\small \texttt{DpdHealpixInfoMapVMPZ}} &  StarBrightnessMean & float32 & $ \textrm{e}^{-}/\textrm{s} $ & \Gaia $G$ \\
                Star RMS flux & Sky emission & {\small \texttt{DpdHealpixInfoMapVMPZ}} &  StarBrightnessRms & float32 & $\textrm{e}^{-}/\textrm{s}$ & \Gaia $G$ \\
                Star std flux & Sky emission & {\small \texttt{DpdHealpixInfoMapVMPZ}} & StarBrightnessStd & float32 & $\textrm{e}^{-}/\textrm{s}$ & \Gaia $G$ \\

                \hline
            \end{tabular}%
        }%
    }
     \footnotesize
 \tablefoot{For each mask, the table gives the mask family (indicating whether it is related to survey properties, satellite status, instrument performance, or sky emission), the information map type as defined in the text, information map type (only for the \texttt{DpdHealpixInfoMapVMPZ} masks) and the numerical type. It also indicates whether the masks are produced per band, including \Euclid\ and \EXT\ bands. \Gaia $G$ band is defined in \url{https://www.cosmos.esa.int/web/gaia/edr3-passbands/}}
    \label{tab:photomasks}
\end{table*}

\subsection{Pipeline}
VMPZ-ID is the first OU-LE3 PF to be executed after OU-LE2 processing \citep{DR1-TP033}. The photometric masks are constructed from MER data products and \Gaia cutouts \citep{DR1-TP009} at the tile \citep[see tessellation scheme in][]{Scaramella-EP1} level using two main pipelines named \texttt{VMPZ\_Star\_Pipeline} and \texttt{VMPZ\_Mosaic\_Pipeline}. These two pipelines are triggered by the operations team, as described in \citet{DR1-TP033}, as soon as the MER \citep[see][for details]{DR1-TP009} data products are available for a given tile. The \texttt{VMPZ\_Star\_Pipeline} mainly produces masks related to properties of the stellar distribution on the sky. The \texttt{VMPZ\_Mosaic\_Pipeline} handles survey, satellite, instrumental and sky properties. 

\subsection{Main data products}

The main products of the \texttt{ VMPZ\_Star\_Pipeline} and \texttt{VMPZ\_Mosaic\_Pipeline} pipelines consist of the following six main types of photometric masks and a random catalogue.
\begin{itemize}
    \item {\texttt{DpdHealpixFootprintMaskVMPZ}:} integer values, equal to 1 if the sky region has been observed and 0 otherwise.\\ For \DRone, instances have been produced for the \Euclid\ bands, \IE, \JE, \HE, and \YE, and for all available external bands \citep{DR1-TP006}. \\

    \item {\texttt{DpdHealpixCoverageVMPZ}:} float values between 0 and 1, representing the coverage, i.e., the fraction of \texttt{HEALPix}
    pixels covered by the available valid data for each of the \Euclid\ and \EXT\ bands, as for \texttt{DpdHealpixFootprintMaskVMPZ}. \\

    \item {\texttt{DpdHealpixEffectiveCoverageVMPZ}:} float values between 0 and 1, representing the effective coverage, i.e., the fraction of \texttt{HEALPix} pixels covered by valid data combining the \Euclid\ and \EXT\ bands. \\

    \item {\texttt{DpdHealpixBitMaskVMPZ}:} unsigned integer values whose bits are associated with the flags of the \texttt{MerBksMosaic} \citep[see][for details]{DR1-TP009}. \\

    \item {\texttt{DpdHealpixDepthMapVMPZ}:} float values representing the RMS-derived depth for each \Euclid\ and \EXT\ band. \\

    \item {\texttt{DpdHealpixInfoMapVMPZ}:} float values representing different quantities of interest. To discriminate between the masks associated to different physical quantities we define the so-called \texttt{InfoMap} type as described in Sect.~\ref{sec:detailedmasks} and Table.\ref{tab:photomasks}.  \\

    \item {\texttt{DpdHealpixRandomCatalogueVMPZ}:} generic container for storing simulated galaxy catalogues.
\end{itemize}

\section{Detailed mask description}
\label{sec:detailedmasks}
The VMPZ-ID pipelines have produced 104 photometric masks for \DRone. Table~\ref{tab:photomasks} summarises the main properties of these masks. A more detailed description of each mask is given below. We warn the reader that these masks present in some cases redundant or correlated information.

\subsection{Survey-definition masks and random catalogues}
The main masks accounting for the \Euclid\ survey coverage and the spatial distribution of the \EXT\ data are presented here.
\begin{itemize}

    \item {Tile index}: for each \texttt{HEALPix} pixel, it encodes the number of the MER tile on which it lies. This is stored as a \texttt{DpdHealpixInfoMapVMPZ} product with information map type \texttt{Tile}. Tile numbers are given as signed 64-bit integers. \\

    \item {Footprint}: defines the observed or valid sky area. This is stored as a \texttt{DpdHealpixFootprintMaskVMPZ} product with a weight of 1 for valid \texttt{HEALPix} pixels. It is available for the \Euclid\ and \EXT\ bands. The mask is constructed from the \texttt{MerBksMosaic} maps, with observed sky regions set to 1 and the others to 0. Additional flagging at the MER processing level, encoded in the \texttt{MerBksMosaic} RMS maps, is also taken into account. MER tile pixels with zero RMS or above a given fixed threshold, $\sigma_{\mathrm{thresh}}\,{=}\,10\,000$ ADU, are considered unobserved and treated accordingly when computing the footprint. Finally, all MER tile pixels lying within a \texttt{HEALPix} pixel are combined using a logical \texttt{OR} operator. \\

    \item {Coverage}: determines the relative coverage, float number between 0 and 1, for each \texttt{HealPix} pixel and for each \Euclid\ and \EXT\ band.
    The coverage per band (\Euclid\ and \EXT) is calculated from the \texttt{MerBksMosaic} maps. MER tile pixels for which the RMS value is zero or exceeds $\sigma_{\mathrm{thresh}}$ are considered unobserved. Furthermore, for the \Euclid\ bands, polygonal masks are used to exclude sky regions affected by bright-star emission. For each \texttt{HEALPix} pixel, the coverage is given by the fraction of valid MER tile pixels. \\

    \item {Effective coverage}: defines the fraction of sky simultaneously observed in all photometric bands. An initial effective coverage mask is constructed at the MER mosaic resolution ($19\,200\times19\,200$ pixels), before projection onto the \texttt{HEALPix} grid. As defined the initial effective coverage mask is binary (either 0 or 1) while the \texttt{HEALPix} one can have float values between 0 and 1. In this initial mask, for each available VIS, NIR, or \EXT\ band, MER tile pixels for which the RMS value is zero or above $\sigma_{\mathrm{thresh}}$ are considered unobserved. In some cases, multiple \EXT\ data sets in equivalent photometric bands (e.g. HSC $R$, HSC $R2$, HSC $I$, HSC $I2$) are used to cover the same tile. To handle this, effective \EXT bands are defined; for each of them, multiple input data sets are available, and their validity masks are first combined using a logical \texttt{OR} operator. The masks from the different effective bands are then combined using a logical \texttt{AND} operator, thereby retaining only regions observed in all effective bands.

    Regions affected by bright stars are subsequently excluded. For the \IE, \YE, \JE, and \HE\ bands, when stellar polygons describing stellar lobes and diffraction spikes are available, they are directly rasterised onto the initial MER effective coverage grid. For the \EXT\ data, an additional mask is constructed using \Gaia star catalogues. Stars with magnitude in the $G$ red band $G_{\rm RP}<14$ are masked\footnote{The \Gaia $G$ red band magnitude, $G_{\rm RP}$, is defined in \url{https://www.cosmos.esa.int/web/gaia/edr3-passbands/}} with circular regions whose radii, expressed in arcmin, follow the empirical relation (see left panel of Fig.~\ref{fig:star_radius_and bleeding})
    \begin{equation}
        \begin{aligned}
            r_\ast(G_{\rm RP})={}&-0.01972 \, G_{\rm RP}^3+0.65849 \, G_{\rm RP}^2 \\
            &-7.31734 \, G_{\rm RP}+27.75053.
        \end{aligned}
    \end{equation}
    
    The resulting radii are enlarged by 20\% when constructing the circular masks to minimise the number of spurious detections while preserving galaxy counts in the final \Euclid\ catalogues. Bleeding trails \citep[see][for a discussion]{DESAI201667} are additionally masked for stars with $G_{\rm RP}<10.3$; their width is described by the empirical polynomial (see Fig.~\ref{fig:star_radius_and bleeding}~b)
    \begin{equation}
        \begin{aligned}
            r_{\rm bleed}(G_{\rm RP})={}&+0.06158 \, G_{\rm RP}^4-1.98886 \, G_{\rm RP}^3\\
            &+23.65131 \, G_{\rm RP}^2-123.78714 \, G_{\rm RP}\\
            &+245.65947,
        \end{aligned}
    \end{equation}
    with the polynomial value saturated at $G_{\rm RP}=4$ for the brightest stars. The resulting rectangular masks have a magnitude-dependent total angular width of $2\,r_{\rm bleed}$ and a fixed extent of $\ang{;0.488;}$ in the perpendicular direction to the bleed.

    A pre-selected set of extended objects is also masked. This set comprises globular clusters and planetary nebulae from the New General Catalogue\footnote{Bleeding trails on the OpenNGC (NGC/IC objects) catalogue. See \url{https://www.legacysurvey.org/dr9/external/} and \url{https://github.com/mattiaverga/OpenNGC/}\,.} (NGC) and nearby galaxies from the Siena Galaxy Atlas\footnote{Based on the 2020 version of the Siena Galaxy Atlas, SGA-2020. See \url{https://www.legacysurvey.org/dr9/external/} and \url{https://www.legacysurvey.org/sga/sga2020/}\,.} (SGA). NGC objects are masked using their radii, while SGA galaxies with an approximate Vega magnitude of ${\tt MAG\_LEDA}<16.5$ are masked using elliptical regions Bleeding on their major and minor axes and position angles, with their dimensions enlarged by 30\%. The resulting initial effective coverage mask is projected and averaged onto an $N_{\rm side}\,{=}\,16\,384$ \texttt{HEALPix} grid to produce the final effective coverage mask. This mask is saved as a \texttt{DpdHealpixEffectiveCoverageMaskVMPZ} product.
    
    Additionally, the initial effective coverage mask is used to construct a flag called \texttt{EFF\_COV} for each object in the \texttt{DpdMerFinalCatalog}. This flag is 1 if the object position\footnote{Notice that we do not take into account the object extension} lies on a pixel with non-zero initial effective coverage, and 0 otherwise. This flag is stored in the second extension of the effective coverage FITS\footnote{Flexible Image Transport System: \url{https://heasarc.gsfc.nasa.gov/docs/heasarc/fits.html}} file. \\

    \begin{figure*}[htbp!]
        \centering
        \includegraphics[width=0.38\hsize]{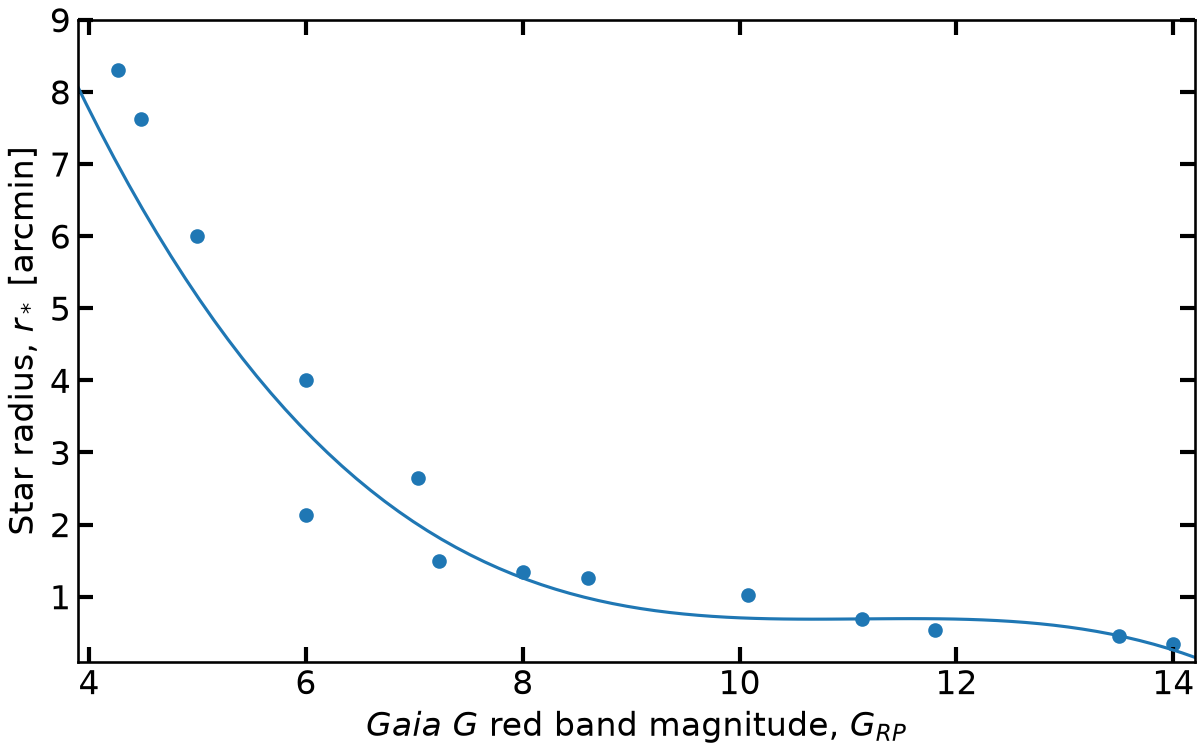}
        \hspace{1cm}
        \includegraphics[width=0.38\hsize]{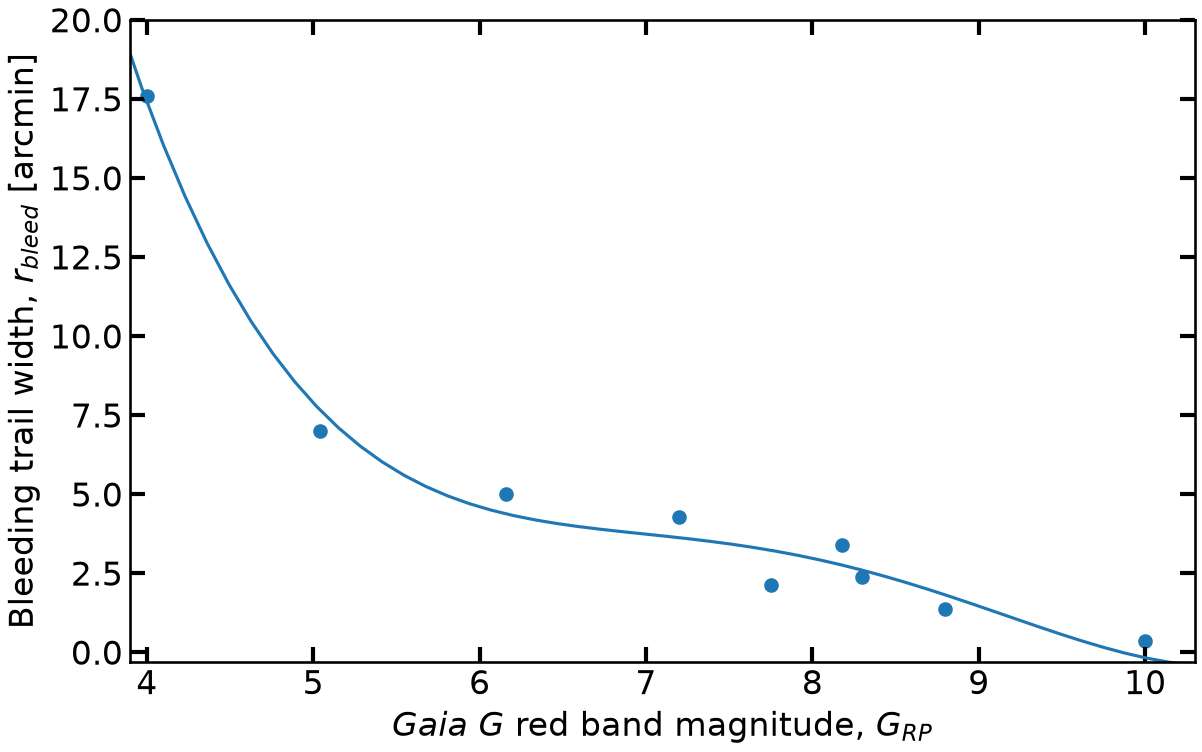}
        \caption{\textit{Left}: empirical radius around \Gaia stars as a function of \Gaia red band magnitude. \textit{Right}: empirical bleeding-trail width around \Gaia stars as a function of \Gaia red band magnitude.}
        \label{fig:star_radius_and bleeding}
    \end{figure*}

    \item {Random catalogues}: random galaxy catalogues are generated from a uniform distribution weighted at each position by the effective coverage value. The positions of the simulated galaxies (RA and Dec) are stored as tables in \texttt{DpdHealpixRandomCatalogueVMPZ} products for each tile. Figure~\ref{fig:mask_102015162_cl} shows an example of a random catalogue.

    \begin{figure*}[htbp!]
        \centering
        \includegraphics[angle=0,width=0.30\hsize]{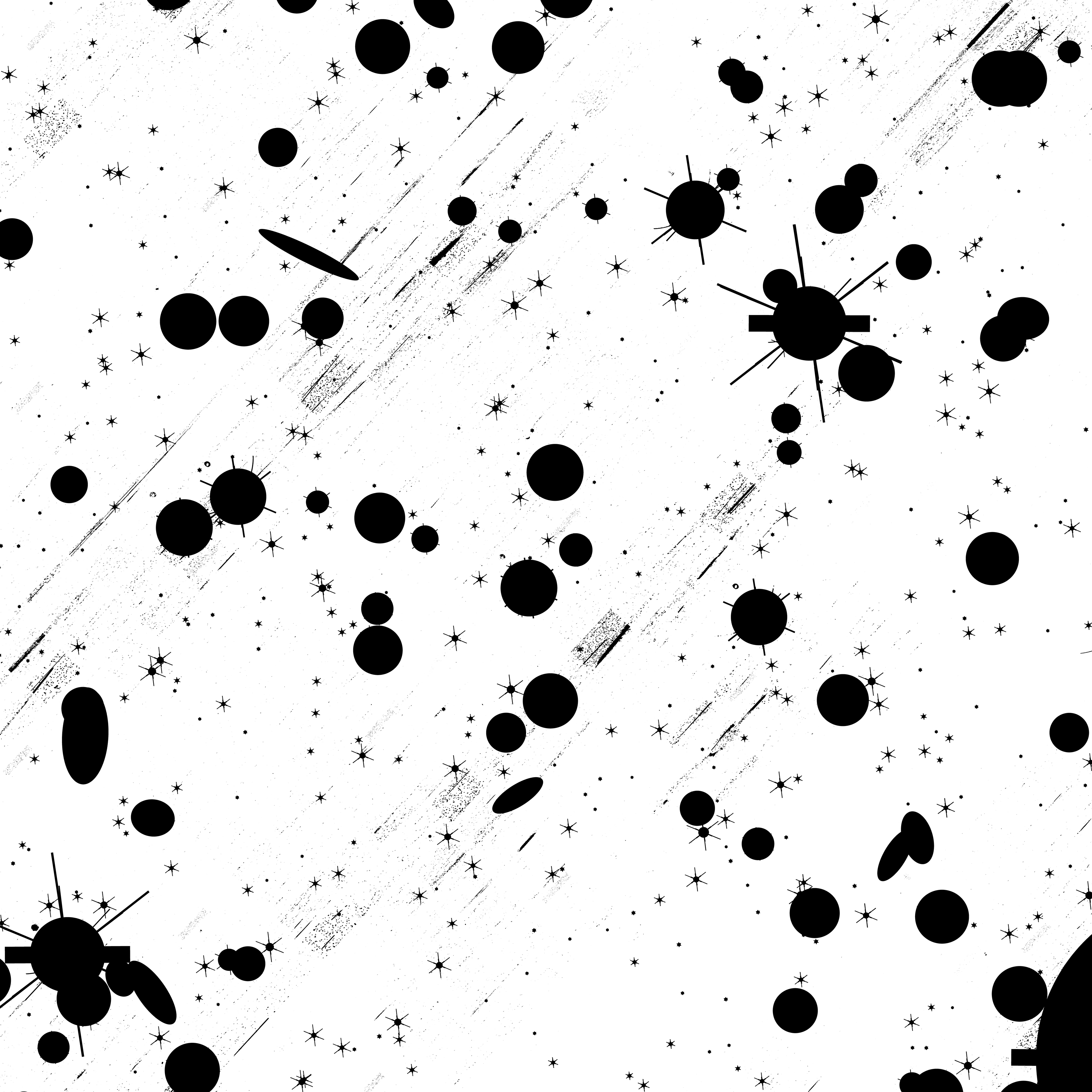}
        \raisebox{-5.5mm}{%
            \includegraphics[angle=0,width=0.30\hsize]{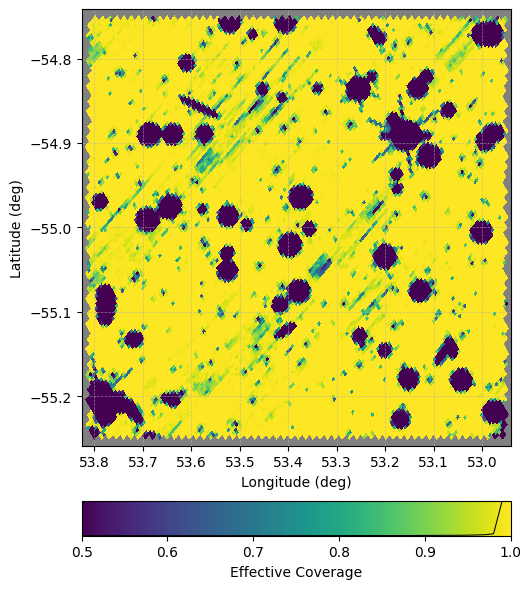}
            \includegraphics[angle=0,width=0.30\hsize]{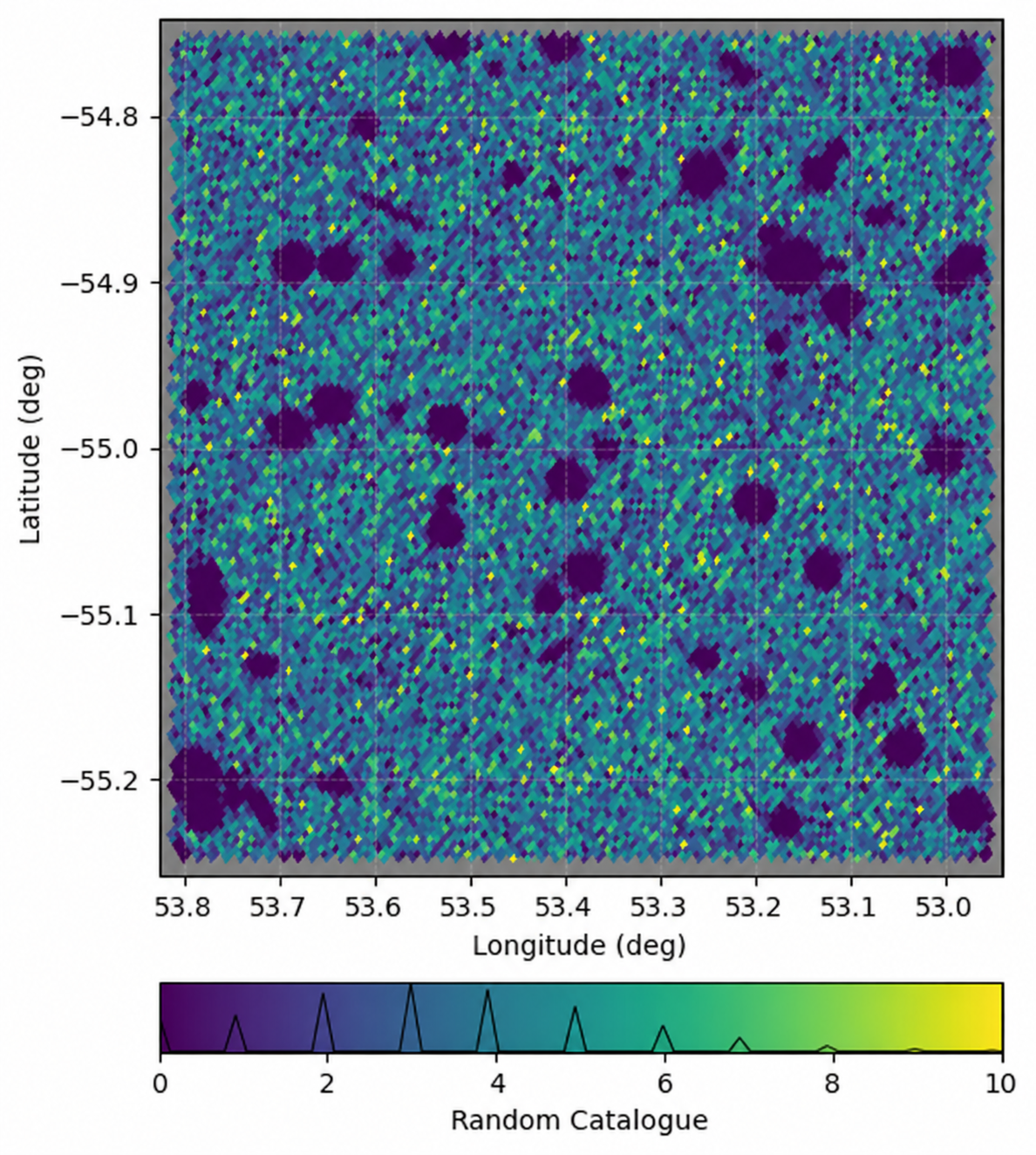}
        }
        \caption{\textit{Left}: initial effective coverage binary mask at MER resolution for tile number $102015162$. The circular masks around bright stars are clearly visible, as well as the additional masks for diffraction spikes and bleeding trails. \textit{Middle}: final combined \texttt{HEALPix} effective coverage mask for the same tile. Because of the significantly lower resolution of the effective coverage mask, the original masking is smeared out, reducing the survey coverage. \textit{Right}: example of random catalogue derived from the effective coverage mask for a fixed mean galaxy density of $ 80 $ $ \text{galaxies}\,\text{arcmin}^{-2} $.}
        \label{fig:mask_102015162_cl}
    \end{figure*}

\end{itemize}

\subsection{Survey-related masks}
\label{sub:survey_masks}
Masks related to the satellite's position and integration time during operations are presented here.
\begin{itemize}
    \item {Solar aspect angle}: defines the \Euclid\ satellite solar aspect angle in degrees as described in \citet[][]{Scaramella-EP1}. This is stored in a \texttt{DpdHealpixInfoMapVMPZ} information map of type \texttt{SAA}. It is computed from the MER layering files, which contain both the integration time and the satellite angles for each exposure. The final mask is obtained by computing the exposure-time-weighted mean value for each \texttt{HEALPix} pixel. \\

    \item {Alpha/beta angles}: defines the \Euclid\ satellite Alpha and Beta angles in degrees \citep[][]{Scaramella-EP1}. Both are computed and stored like \texttt{SAA}, using information map types \texttt{AA} and \texttt{BA}, respectively. \\

    \item {Exposures}: for each of the \Euclid\ bands, this represents the total integration time in seconds per \texttt{HEALPix} pixel and is calculated from the MER layering files described above. This is stored as a \texttt{DpdHealpixInfoMapVMPZ} product.
\end{itemize}

\subsection{Instrumental-property masks}
For each of the \Euclid\ and \EXT\ bands, masks of the main instrumental properties are obtained from the \texttt{MerBksMosaic} maps. For each \texttt{HEALPix} pixel, the weighted average of the instrumental properties is computed. As for the previously described masks, zero weight is assigned to MER tile pixels with RMS equal to zero or above a given threshold representing bad or corrupted data.

\begin{itemize}
    \item {RMS noise}: represents the average root mean square (RMS) noise of the \Euclid\ and \EXT\ images. This is computed from the \texttt{MerBksMosaic} RMS noise maps in ADU units, which are averaged and projected onto \texttt{HEALPix} maps. The resulting maps are stored as \texttt{DpdHealpixInfoMapVMPZ} products of information map type \texttt{Noise}. For unobserved regions, the RMS noise value is set to zero. \\

    \item {Depth}: represents the average depth at each sky position and is stored as \texttt{DpdHealpixDepthMapVMPZ} products. This is computed from the \Euclid\ and \EXT\ \texttt{MerBksMosaic} RMS maps according to the following definition:
    \begin{equation}m=-2.5\log_{10}\left(\mathrm{RMS}\;d_{\mathrm{thresh}}\sqrt{\frac{A_{\mathrm{aper}}}{A_{\mathrm{pixel}}}}\,\right)+\textrm{ZP},
    \end{equation}
    \noindent where $\mathrm{RMS}$ and $\mathrm{ZP}$ are the RMS noise value and photometric zero point, respectively, as defined in the \texttt{MerBksMosaic} maps. The detection threshold, $d_{\mathrm{thresh}}$, is set to 5, corresponding to a 5$\,\sigma$ detection, while $A_{\mathrm{aper}}$ and $A_{\mathrm{pixel}}$ represent the detection aperture and pixel areas, respectively. A 2$\arcsec$-diameter aperture is used to compute the depth. \\

    \item {PSF FWHM}: defines the average full width at half maximum (FWHM), in arcsec, at each sky position for the \Euclid\ and \EXT\ bands. This is stored in a DpdHealpixInfoMapVMPZ product of information map type \texttt{PSF}. For the \texttt{MerBksMosaic} maps, FWHM estimates are available only at the positions of bright stars in the tile field. A two-pass 2D interpolation (cubic spline then nearest-neighbour on the remaining not sampled areas) is therefore performed across the tile area, starting from a sufficiently coarse grid to avoid artefacts in the final derived mask.\\

    \item {Bit mask:} is an effective bit mask stored as a \texttt{DpdHealpixBitMaskVMPZ} product. For each \texttt{HEALPix} pixel in the final mask, it is computed by applying a bitwise \texttt{OR} operation to the bit-mask values of all \texttt{MerBksMosaic} pixels falling within it. The meaning of the different bits is described in the \texttt{MER} data model presented in \citet{DR1-TP009}. \\

    \item {Persistence:} reflects the overall contamination of the NIR photometric data by persistence~\citep[see][for a detailed description of persistence effects]{DR1-TP038} at each sky position. This is stored as a \texttt{DpdHealpixInfoMapVMPZ} product with information map type \texttt{Persistence}. The \texttt{HEALPix} persistence mask is derived from bit 13 of the \texttt{MerBksMosaic} bit mask, which flags regions severely affected by persistence. It ranges from 0 to 1 and gives the fraction of each \texttt{HEALPix} pixel affected by persistence, but does not provide information on the level of persistence.\\

    \item {NIR PRMS:} represents the maximum RMS offset in milliarcseconds (mas) of the NIR stack astrometric residuals. The maximum RMS offset is computed for each stacked catalogue generated for a specific observation ID at a resolution of about 30$^{\prime \prime}$, following the procedure described in \citet[][]{DR1-TP004}. The maps for all observation IDs corresponding to the EWS are then projected and averaged onto a common \texttt{HEALPix} grid. The resulting mask is stored as a \texttt{DpdHealpixInfoMapVMPZ} product with information map type \texttt{NIR\_PRMS}.\\

    \item {Pan-STARRS artefacts:} defines regions in the Pan-STARRS $i$-band \EXT\ images \citep[see][]{DR1-TP006} that are affected by systematic effects. This is stored as a \texttt{DpdHealpixInfoMapVMPZ} product with information map type \texttt{PanStarrsArtefacts}. The mask is built from the MER final catalogues \citep[see][]{DR1-TP009} by selecting sources with anomalous Pan-STARRS/VIS flux ratios, defined by default as $F_{\rm PS}/F_{\rm VIS}\,{>}\,3.5$ or $F_{\rm PS}/F_{\rm VIS}\,{<}\,1/3.5$, and requiring a flux-ratio signal-to-noise ${>}\,3$.
    Their spatial distribution is smoothed with an Epanechnikov kernel density estimator\footnote{The Epanechnikov kernel density estimator is defined as
$\hat{f}(x) = \frac{1}{nh}\sum_{i=1}^{n} K\left(\frac{x-x_i}{h}\right)$,
where $K(u) = \frac{3}{4}(1-u^2)$ for $|u| \leq 1$ and $0$ otherwise,
$n$ is the sample size, and $h$ is the smoothing bandwidth.} (KDE) with a $30''$ bandwidth, evaluated on a regular $10''$ RA--Dec grid.
    An adaptive density threshold derived from the KDE value distribution defines candidate contaminated regions, after which connected regions smaller than 30 grid pixels are discarded. The resulting binary mask is then projected onto a \texttt{HEALPix} grid, recording whether each \texttt{HEALPix} pixel is masked. This results into a binary mask.
\end{itemize}

\subsection{Sky emission masks}
Following the models presented in~\citet[][]{Scaramella-EP1}, per-tile masks of zodiacal light emission and Galactic extinction are constructed for the \Euclid\ bands.

\begin{itemize}
    \item {Galactic extinction:} defines, for each sky position, the total extinction in magnitudes for each of the \Euclid\ bands, as defined in equation 5 of~\citet[][]{Scaramella-EP1}. This is stored as a \texttt{DpdHealpixInfoMapVMPZ} product with information map type \texttt{GalacticExtinction}.

    \item {Zodiacal light:} gives, for each sky position, the zodiacal light flux for each \Euclid\ band normalised by the effective wavelength of the band, $F/\lambda_{*}$, as defined in table~4 of~\citet[][]{Scaramella-EP1}. This is stored as a \texttt{DpdHealpixInfoMapVMPZ} product with information map type \texttt{ZodiacalLight}.
\end{itemize}

\subsection{Star masks}
The masks monitoring stellar density and flux across the sky are presented here. They are built by the {\it VMPZ\_Star\_Pipeline} using the \texttt{DpdExtGaiaCutout} products, which contain a reduced \Gaia star catalogue for each tile. Only stars with a \Gaia G-band flux above zero and a valid parallax angle are considered.

\begin{itemize}
    \item {Star number density:} represents the number density of stars in each \texttt{HEALPix} pixel. This is stored as a \texttt{DpdHealpixInfoMapVMPZ} product with information map type \texttt{StarNumberDensity}.\\

    \item {Mean star brightness:} provides the mean flux of all stars lying within a given \texttt{HEALPix} pixel. This is stored as a \texttt{DpdHealpixInfoMapVMPZ} product with information map type \texttt{StarBrightnessMean}.\\

    \item {Star brightness RMS:} provides the RMS flux of all stars lying within a given \texttt{HEALPix} pixel. This is stored as a \texttt{DpdHealpixInfoMapVMPZ} product with information map type \texttt{StarBrightnessRms}.\\

    \item {Star brightness standard deviation:} provides the standard deviation of the flux of all stars lying within a given \texttt{HEALPix} pixel. this is stored as a \texttt{DpdHealpixInfoMapVMPZ} product with information map type \texttt{StarBrightnessStd}.

\end{itemize}
Examples for most of the masks are given in Appendix~\ref{sec:app_zoomin} and are discussed in Sect.~\ref{sec:wsmask_highres}.
\section{EWS masks}
\label{sec:wide}
\subsection{Full survey masks}
\label{sec:ewsfullmasks}

\begin{figure*}[p!]
    \centering

    \vspace{0.25cm}

    \begin{overpic}[width=0.49\hsize]{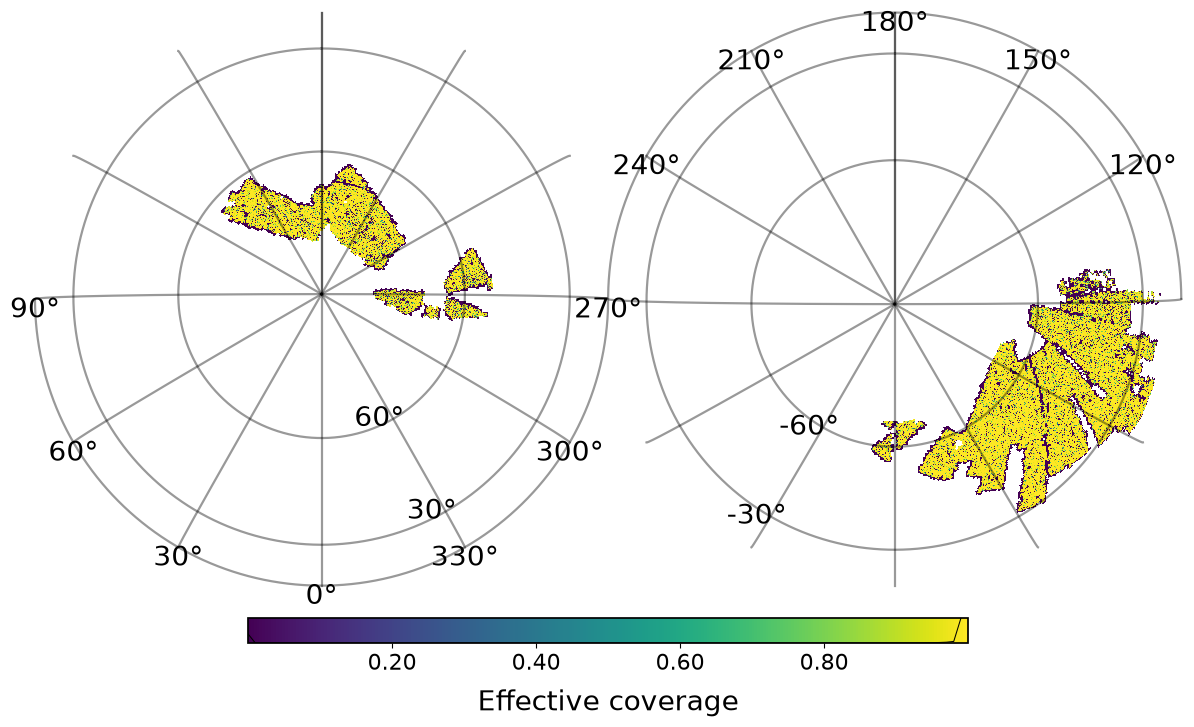}
        \put(2,57){\large(a)}
    \end{overpic}
    \hfill
    \begin{overpic}[width=0.49\hsize]{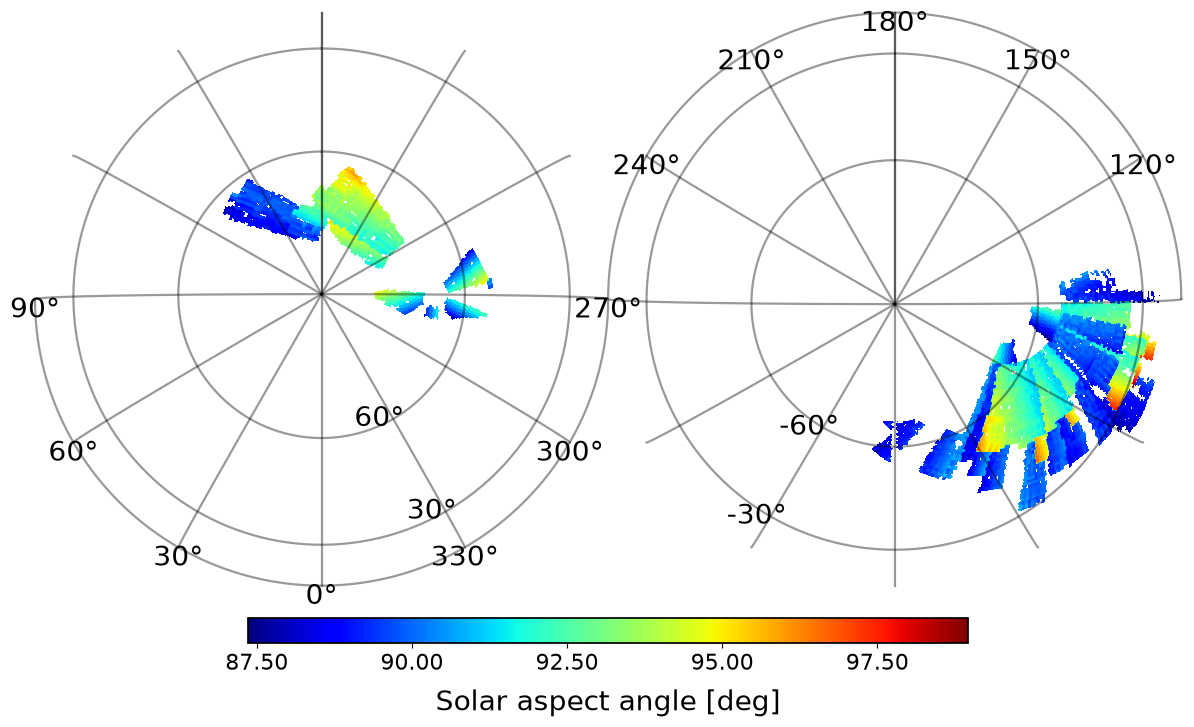}
        \put(2,57){\large(b)}
    \end{overpic}

    \vspace{0.66cm}

    \begin{overpic}[width=0.49\hsize]{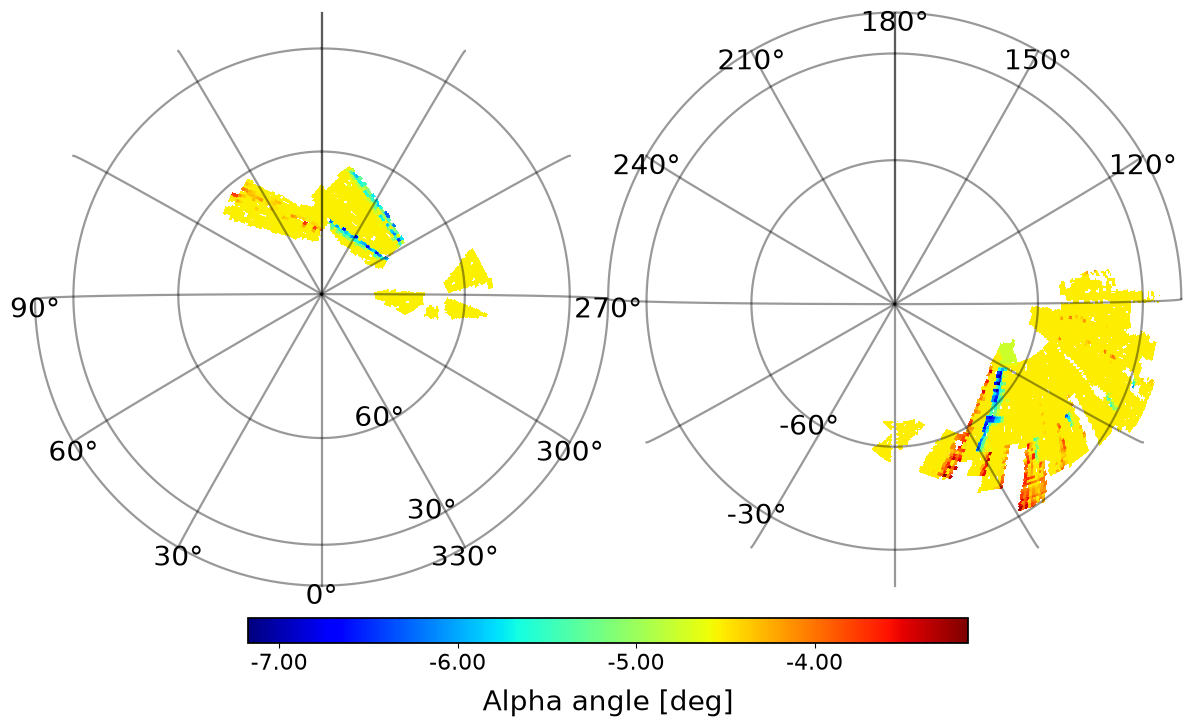}
        \put(2,57){\large(c)}
    \end{overpic}
    \hfill
    \begin{overpic}[width=0.49\hsize]{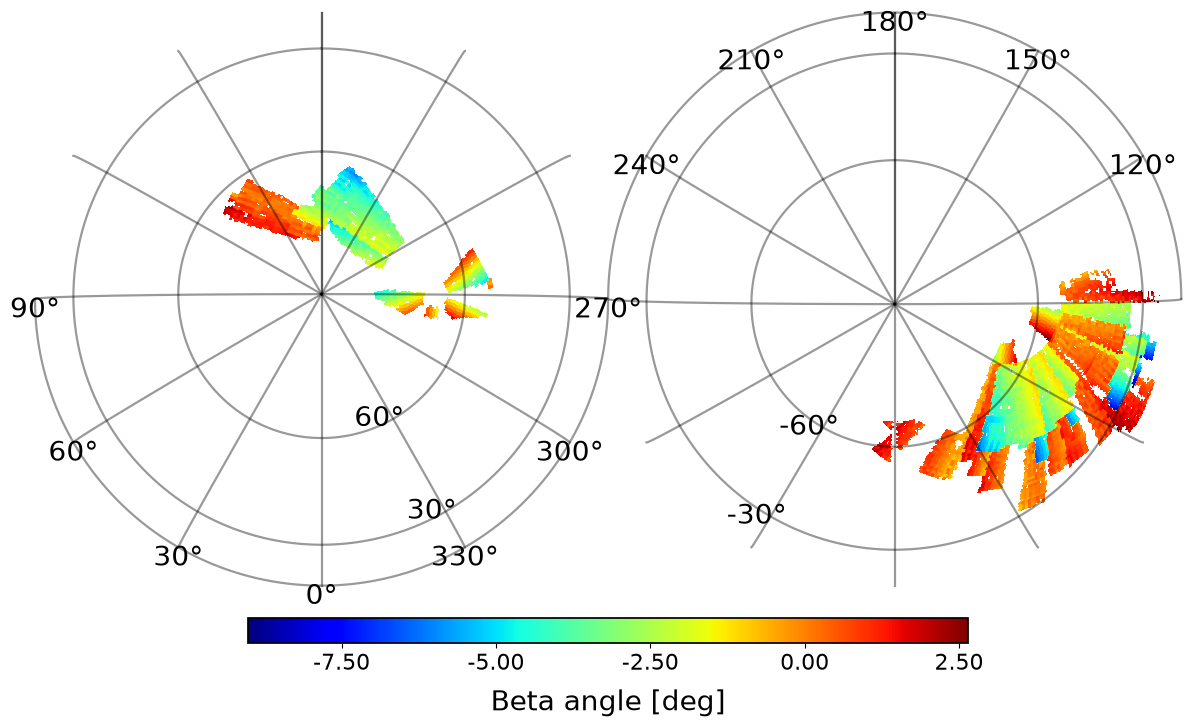}
        \put(2,57){\large(d)}
    \end{overpic}

    \vspace{0.66cm}

    \begin{overpic}[width=0.49\hsize]{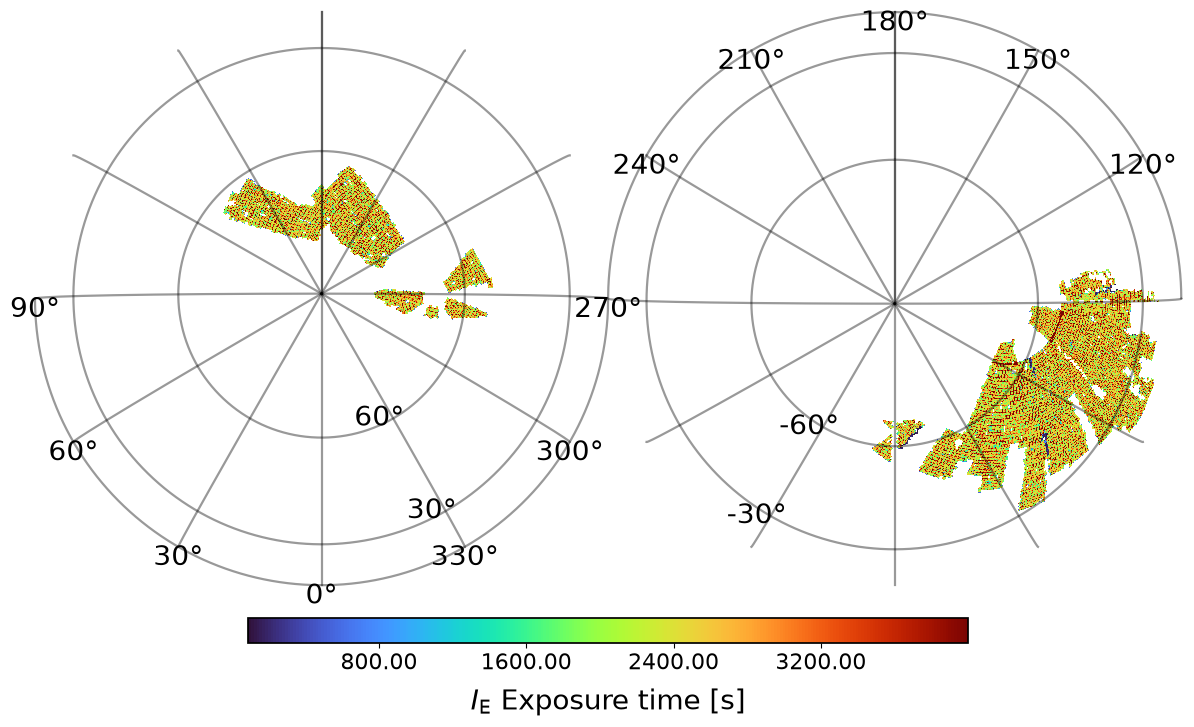}
        \put(2,57){\large(e)}
    \end{overpic}
    \hfill
    \begin{overpic}[width=0.49\hsize]{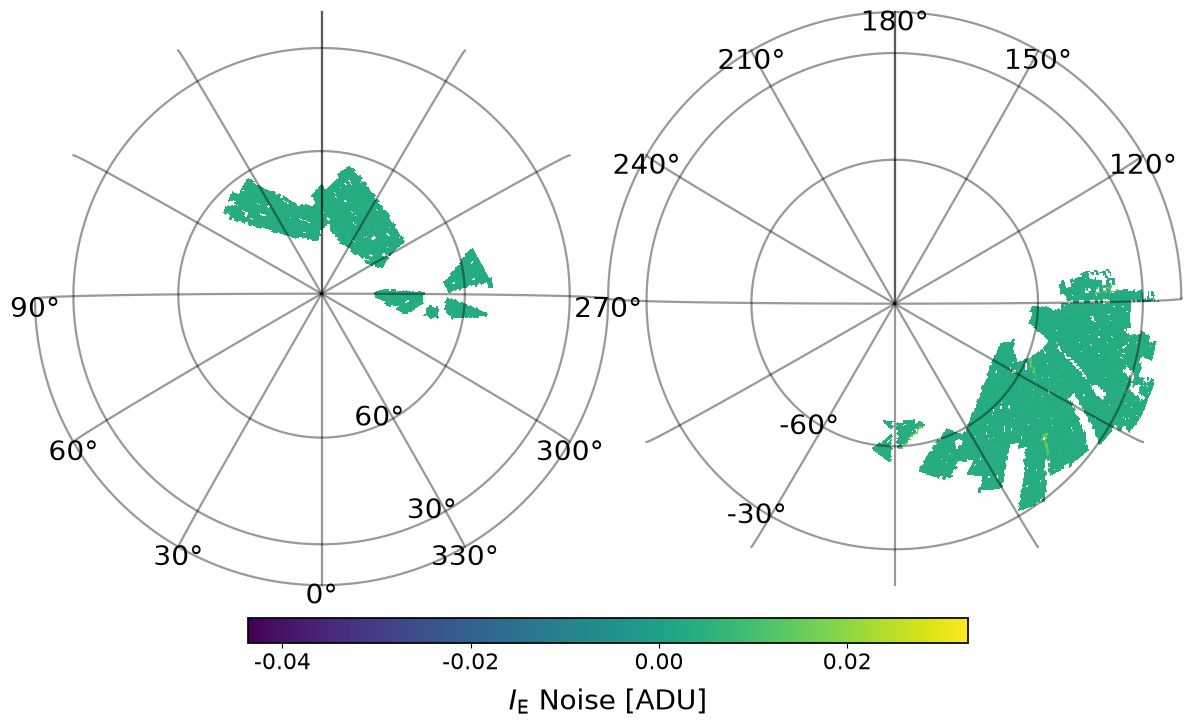}
        \put(2,57){\large(f)}
    \end{overpic}

    \vspace{0.66cm}

    \begin{overpic}[width=0.49\hsize]{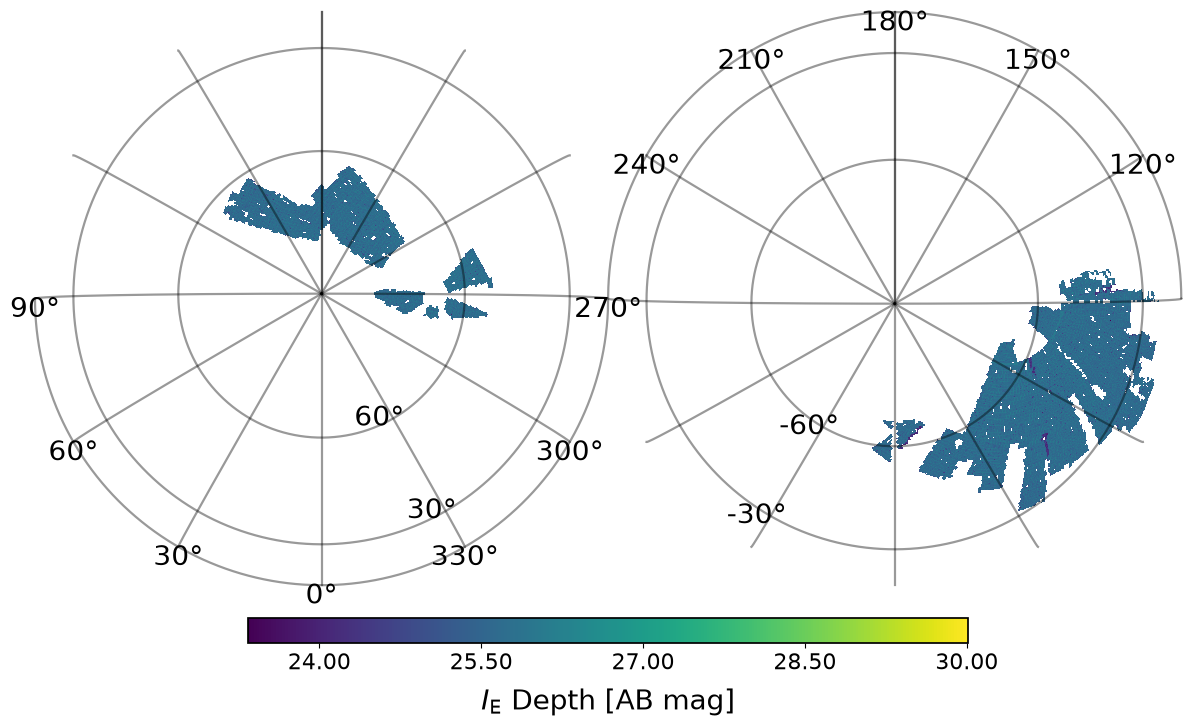}
        \put(2,57){\large(g)}
    \end{overpic}
    \hfill
    \begin{overpic}[width=0.49\hsize]{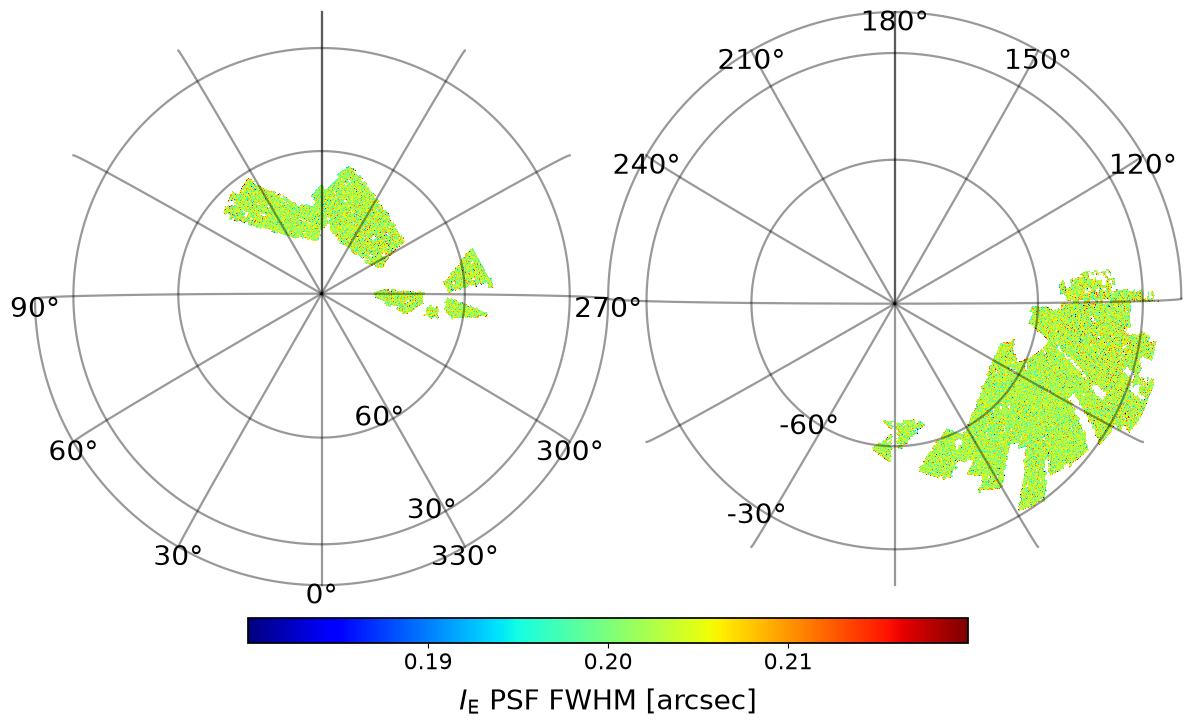}
        \put(2,57){\large(h)}
    \end{overpic}

    \vspace{0.25cm}

    \caption{DR1 EWS, full-sky orthographic projections of a selection of the VMPZ-ID masks including: (a) effective coverage; (b) solar aspect angle; (c) alpha angle; (d) beta angle; (e) VIS integration time; (f) VIS RMS noise; (g) VIS depth; and (h) VIS PSF FWHM.}
    \label{fig:wide_fullsky_dr1}
\end{figure*}

\begin{table*}[htbp!]
    \caption{\DRone EWS mean, median, standard deviation, and percentiles of the distributions of integration time, RMS noise, depth, and PSF FWHM for the four \Euclid\ bands. \label{tab:instrumentprop}}
    \centering
    \begin{small}
        \begin{tabular*}{\textwidth}{@{\extracolsep{\fill}}clrrrrrrrr}
            \hline\hline
            \noalign{\vskip 2pt}
            Band & Quantity & Mean & Median & Std & $Q_{16}$ & $Q_{25}$ & $Q_{75}$ & $Q_{84}$ & IQR\tablefootmark{a}\\
            \noalign{\vskip 1pt}
            \hline
            \noalign{\vskip 2pt}
            & Exposures [s] & 2735 & 2421 & 922 & 1861 & 2332 & 3071 & 3632 & 740 \\
            $ \IE $ & Noise [ADU] & 0.003882 & 0.003716 & 0.001423 & 0.003238 & 0.003421 & 0.004085 & 0.004239 & 0.0006632 \\
            & Depth [AB mag] & 25.68 & 25.69 & 0.224 & 25.56 & 25.59 & 25.78 & 25.84 & 0.186 \\
            & PSF FWHM [arcsec] & 0.2029 & 0.2028 & 0.004051 & 0.1991 & 0.2004 & 0.2053 & 0.2066 & 0.004964 \\
            \hline
            \noalign{\vskip 2pt}
            & Exposures [s] & 338 & 349 & 117 & 262 & 262 & 349 & 436 & 87 \\
            $ \YE $  & Noise [ADU] & 2.089 & 2.077 & 0.4039 & 1.794 & 1.867 & 2.253 & 2.333 & 0.3855 \\
            & Depth [AB mag] & 24.15 & 24.14 & 0.1699 & 24.01 & 24.05 & 24.25 & 24.3 & 0.2038 \\
            & PSF FWHM [arcsec] & 0.4879 & 0.4883 & 0.01016 & 0.4791 & 0.4825 & 0.4939 & 0.4968 & 0.01142 \\
            \hline
            \noalign{\vskip 2pt}
            & Exposures [s] & 338 & 349 & 118 & 262 & 262 & 349 & 436 & 87 \\
            $ \JE $ & Noise [ADU] & 2.393 & 2.379 & 0.4584 & 2.057 & 2.139 & 2.574 & 2.67 & 0.4349 \\
            & Depth [AB mag] & 24.3 & 24.29 & 0.1717 & 24.16 & 24.2 & 24.41 & 24.45 & 0.2009 \\
            & PSF FWHM [arcsec] & 0.5138 & 0.5143 & 0.01071 & 0.5045 & 0.5081 & 0.5201 & 0.5232 & 0.01206 \\
            \hline
            \noalign{\vskip 2pt}
            & Exposures [s] & 339 & 349 & 117 & 262 & 262 & 349 & 436 & 87 \\
            $ \HE $ & Noise [ADU] & 2.194 & 2.183 & 0.4215 & 1.887 & 1.963 & 2.364 & 2.45 & 0.4013 \\
            & Depth [AB mag] & 24.29 & 24.28 & 0.1704 & 24.16 & 24.2 & 24.4 & 24.44 & 0.202 \\
            & PSF FWHM [arcsec] & 0.5589 & 0.5593 & 0.009184 & 0.5512 & 0.5542 & 0.5642 & 0.5666 & 0.009917 \\
            \hline
        \end{tabular*}
    \end{small}
    \setlength{\VSpaceBeforeTabFoot}{0pt}
    \par\vspace{-0.75mm}
    \tablefoot{
        \tablefoottext{a}{Interquartile range $\equiv Q_{75} - Q_{25}$.}
    }
\end{table*}
\begin{figure}[htbp!]
    \centering
    \includegraphics[angle=0,width=0.9\hsize]{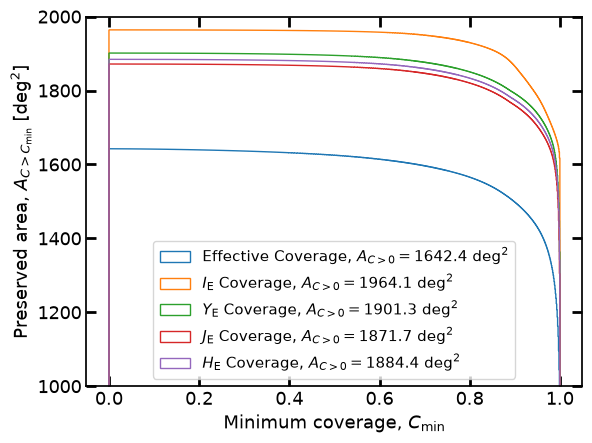}
    \caption{For \DRone EWS, the integrated coverage weighted area $A_{C > C_{\mathrm{min}}}$ for \texttt{HEALPix} pixels with coverage above a given minimum value, $C_{\mathrm{min}}$. This is shown for the effective coverage mask and the individual coverage masks for the four \Euclid\ bands. We recall that the effective coverage also includes the \EXT\ data.}
    \label{fig:cumul_Coverage}
\end{figure}

\begin{figure*}[htbp!]
    \centering
    \includegraphics[angle=0,width=0.33\hsize]{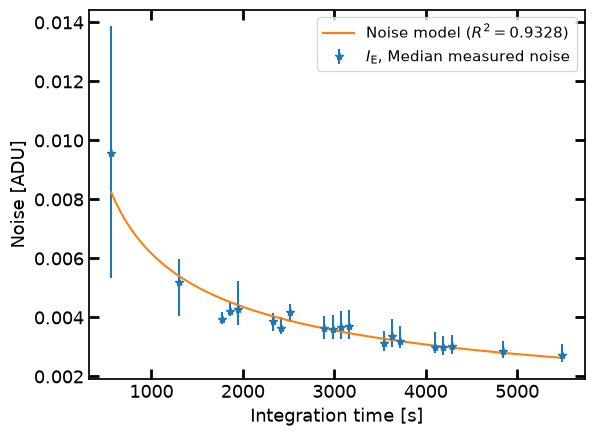}
    \includegraphics[angle=0,width=0.33\hsize]{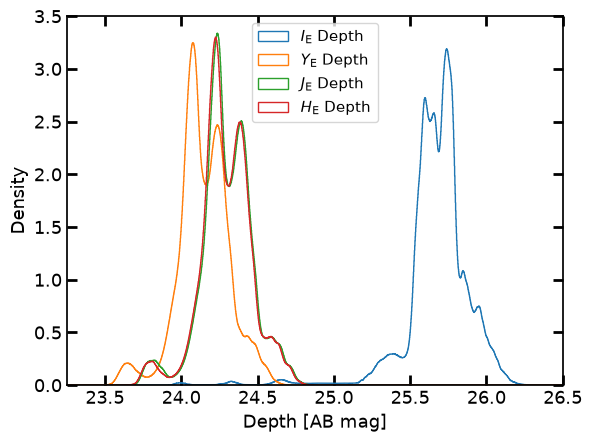}
    \includegraphics[angle=0,width=0.321\hsize]{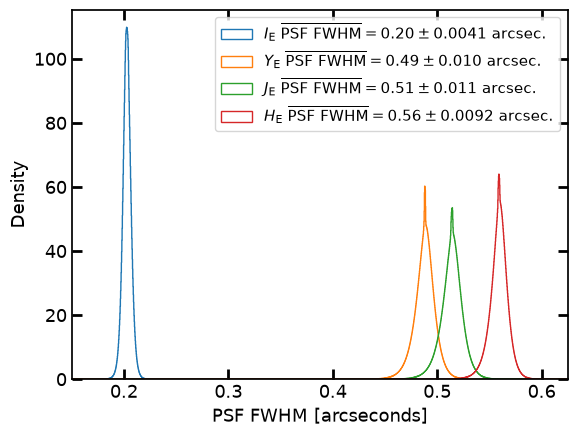}
    \caption{For \DRone EWS, the panels show, from left to right: the VIS RMS noise as a function of integration time, fitted with $\sigma(t) = \sqrt{A\,t_{\rm ref}/t + B}$, where $A$ is the exposure-dependent variance at the reference integration time $t_{\rm ref}$ and $B$ is the exposure-independent term, with $R^2$ denoting the coefficient of determination of the fit; and the 1D distributions of depth and PSF FWHM for the four \Euclid\ bands.}
    \label{fig:wide_distribution}
\end{figure*}

A subset of the \DRone EWS masks is shown in full-sky orthographic projection in Fig.~\ref{fig:wide_fullsky_dr1}, including the North (left hemisphere) and South (right hemisphere) regions. The effective coverage, presented in Fig.~\ref{fig:wide_fullsky_dr1}~(a), is crucial for scientific analyses to account for variations in sky coverage. Overall, the effective coverage is close to 1 (full coverage), except in regions around stars and near the edges of the coverage (see Fig.~\ref{fig:mask_102015162_cl}), which appear as blue dots in the figure. This is also clearly illustrated in Fig.~\ref{fig:cumul_Coverage}, which displays the integrated coverage weighted area as a function of minimum selected coverage. The expected total area of the \DRone EWS is 2008.8\,deg$^2$ (computed from the sum of the area of the regions for which we have one VIS exposure), of which 1642.4\,deg$^2$ is preserved by the effective coverage mask. This effective area is obtained by summing the WEIGHT values in the effective coverage mask, multiplied by the \texttt{HEALPix} pixel area. Figure~\ref{fig:wide_fullsky_dr1} also shows that the effective area for pixels with effective coverage above 0.8 corresponds to about 99\% of the total effective area presented above. The orientation angles of the \Euclid\ satellite are shown in Fig.~\ref{fig:wide_fullsky_dr1}~(b), (c), and (d). As expected, they are all within survey specifications. Moreover, the beta angle and the solar aspect angle are anti-correlated.

Figure~\ref{fig:wide_fullsky_dr1}~(e) shows that the VIS (\IE band) integration time is highly uniform across the survey, with only weak large-scale variations. This uniformity is expected from the survey design, such that most of the survey area is covered by 3- or 4-single-dither exposure \citep{DR1-TP027}. Table~\ref{tab:instrumentprop} shows that the typical integration time values in the EWS are about 2735 and 338\, for the VIS and NISP instruments, respectively. As a consequence, the VIS \IE\ RMS noise and depth masks also show small spatial variations across the survey. Similar conclusions can be drawn for the NISP bands. The left panel of Fig.~\ref{fig:wide_distribution} shows that the noise decreases as expected with increasing integration time. The 1D distribution of depth for the four \Euclid\ bands is shown in the middle panel of Fig.~\ref{fig:wide_distribution}. The distributions have narrow, well-defined peaks, with median depth values of 25.69, 24.14, 24.3, and 24.29 AB mag for the \IE, \YE, \JE and \HE bands, respectively. The standard deviation is below 0.2 AB mag for all bands. A more detailed statistical description of the EWS RMS noise and depth distributions is also given in Table~\ref{tab:instrumentprop}.

The EWS VIS PSF FWHM mask, which represents the VIS PSF FWHM, is shown in Fig.~\ref{fig:wide_fullsky_dr1}~(h). A very uniform spatial distribution is again observed. Similar results are obtained for the NISP bands. A statistical description of the 1D distribution of the PSF FWHM for the VIS and NISP bands is also given in Table~\ref{tab:instrumentprop}. The median FWHM values are \ang{;;0.2}, \ang{;;0.48}, \ang{;;0.51}, and \ang{;;0.56} for the \IE, \YE, \JE, and \HE bands, respectively. The standard deviation of the spatial variation is below \ang{;;0.01}.

\subsection{High-resolution view}
\label{sec:wsmask_highres}
Additional examples of the EWS masks are presented in Appendix~\ref{sec:app_zoomin} for observation Patch ID 4. Figure~\ref{fig:patch4_surveymasks} shows the main survey masks, including the tile index, alpha angle, beta angle, solar aspect angle, and integration time for the four \Euclid\ bands. The latter shows clear spatial patterns corresponding mainly to longer integration time in the overlap regions between adjacent tiles. These patterns are also reflected in the RMS noise and depth maps, as shown in Fig.~\ref{fig:patch4_instrumentmasks}. This figure also presents the PSF FWHM, NIR PRMS, and persistence masks. The measured FWHM is larger in tile-overlap regions, where multiple exposures are combined. The NIR PRMS mask also shows these spatial variations with smaller NIR pointing RMS residuals in the overlapping regions. The persistence mask additionally exhibits tile-correlated structures, although they are less distinct and more difficult to interpret.

Finally, Fig.~\ref{fig:patch4_skyemissionmasks} presents the main sky emission masks, including Galactic extinction, zodiacal light, star number density, mean star brightness, star brightness RMS, and star brightness standard deviation. Regions with larger stellar number density and mean brightness correspond to regions flagged on the effective coverage mask presented in Fig.~\ref{fig:patch4_instrumentmasks}. This figure also shows the expected correlation between persistence and bright-star positions. 

\section{EDF masks}
\label{sec:deep}

\begin{figure*}[p!]
    \centering

    \begin{overpic}[width=0.31\hsize]{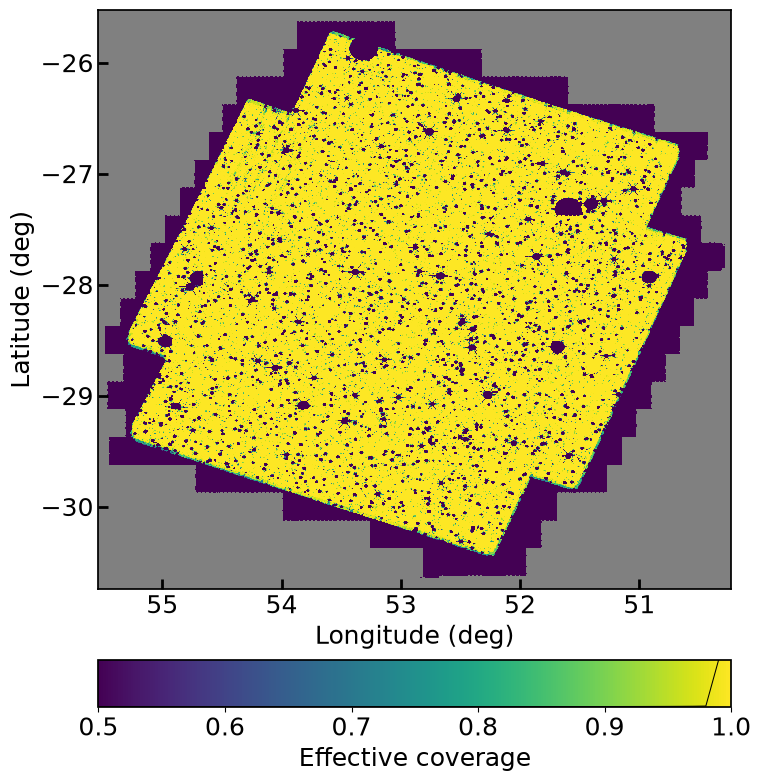}
        \put(45,102){\large EDF-F} \put(-7,30){\rotatebox{90}{\textcolor{black}{Effective coverage}}}
    \end{overpic}
    \begin{overpic}[width=0.31\hsize]{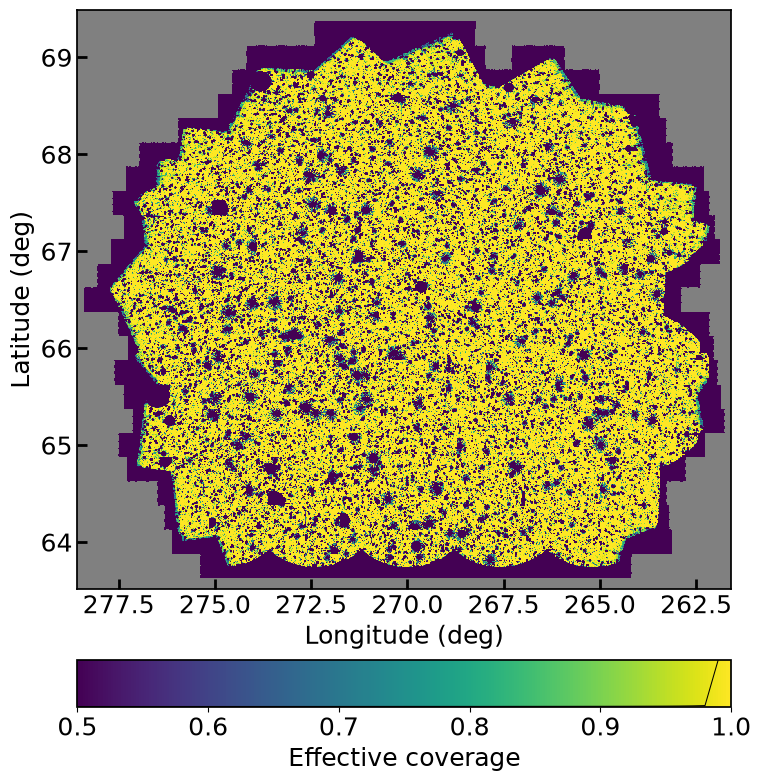}
        \put(45,102){\large EDF-N}
    \end{overpic}
    \begin{overpic}[width=0.31\hsize]{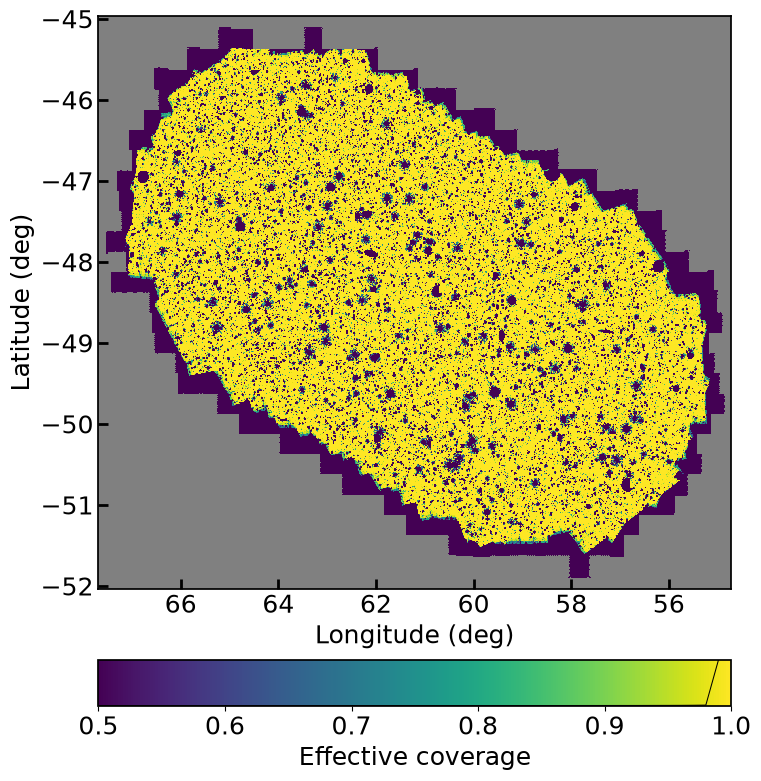}
        \put(45,102){\large EDF-S}
    \end{overpic}

    \vspace{0.1cm}

    \begin{overpic}[width=0.31\hsize]{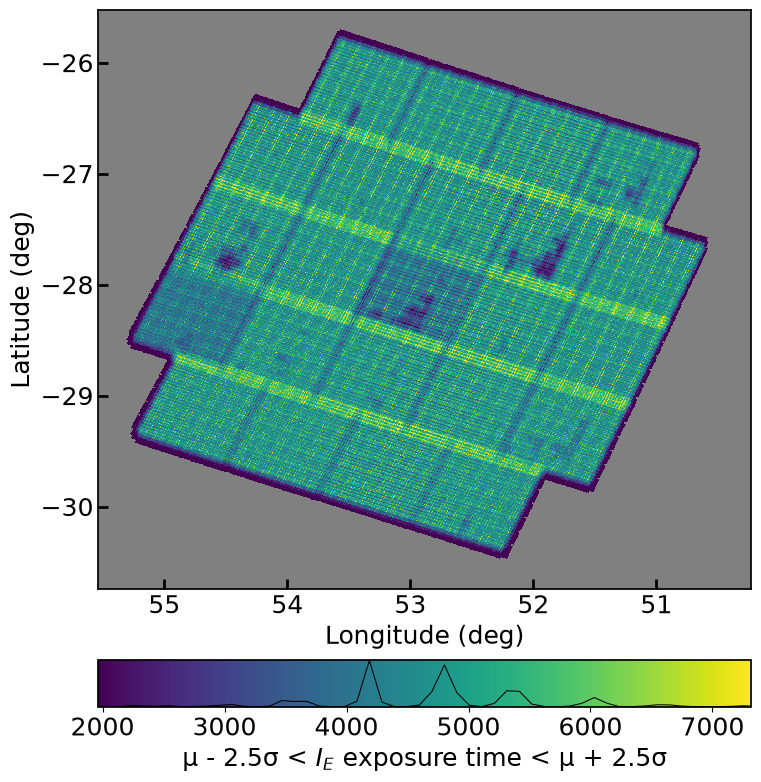}
        \put(-7,30){\rotatebox{90}{\textcolor{black}{\IE integration time}}}
    \end{overpic}
    \includegraphics[angle=0,width=0.31\hsize]{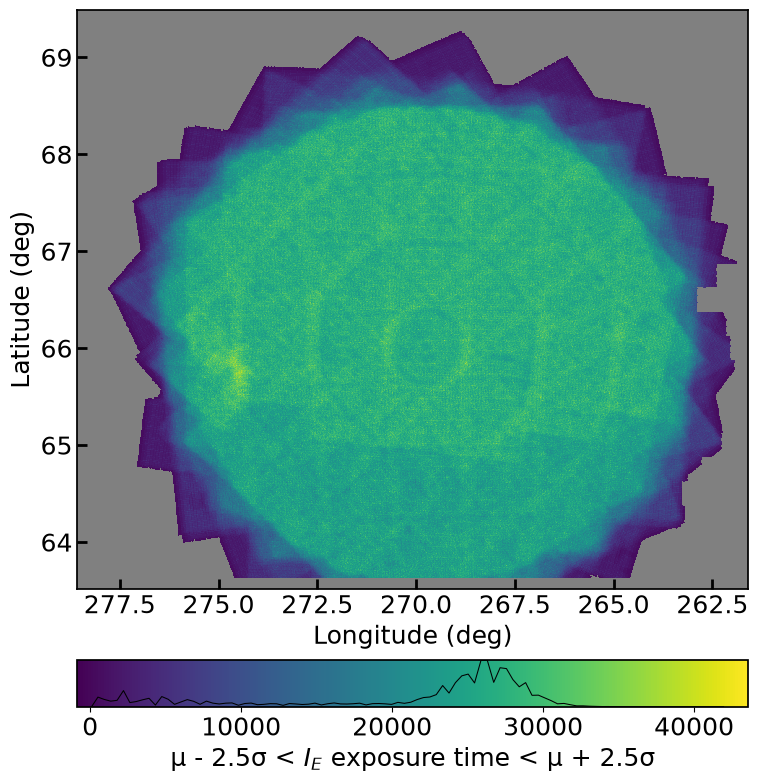}
    \includegraphics[angle=0,width=0.31\hsize]{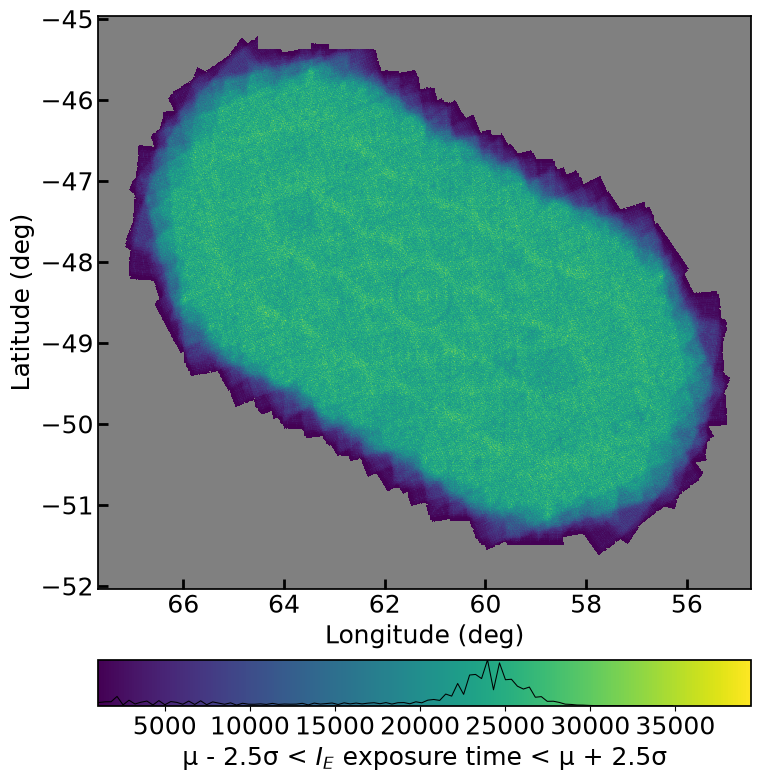}

    \vspace{0.1cm}

    \begin{overpic}[width=0.31\hsize]{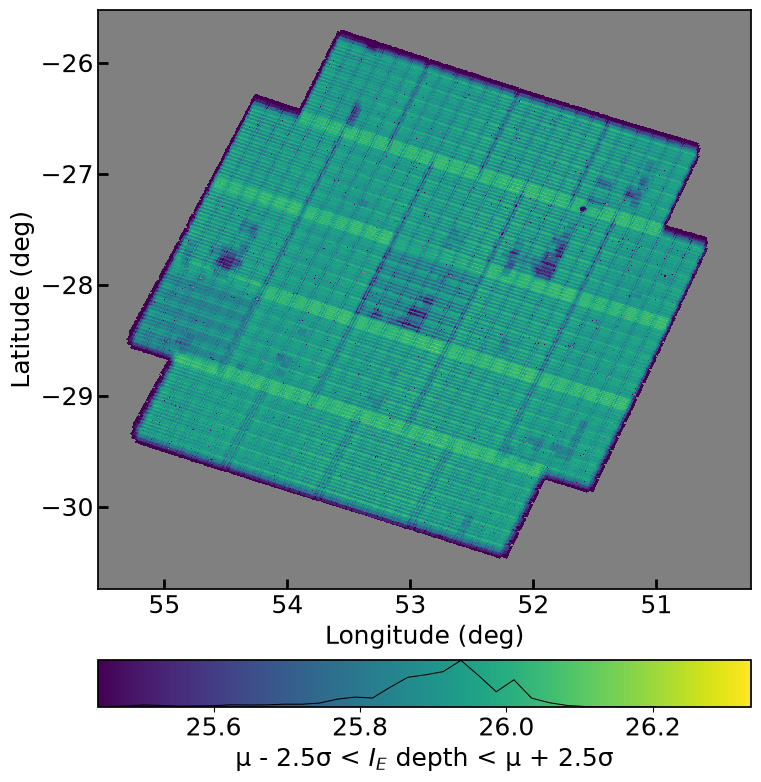}
        \put(-7,40){\rotatebox{90}{\textcolor{black}{\IE depth}}}
    \end{overpic}    
    \includegraphics[angle=0,width=0.31\hsize]{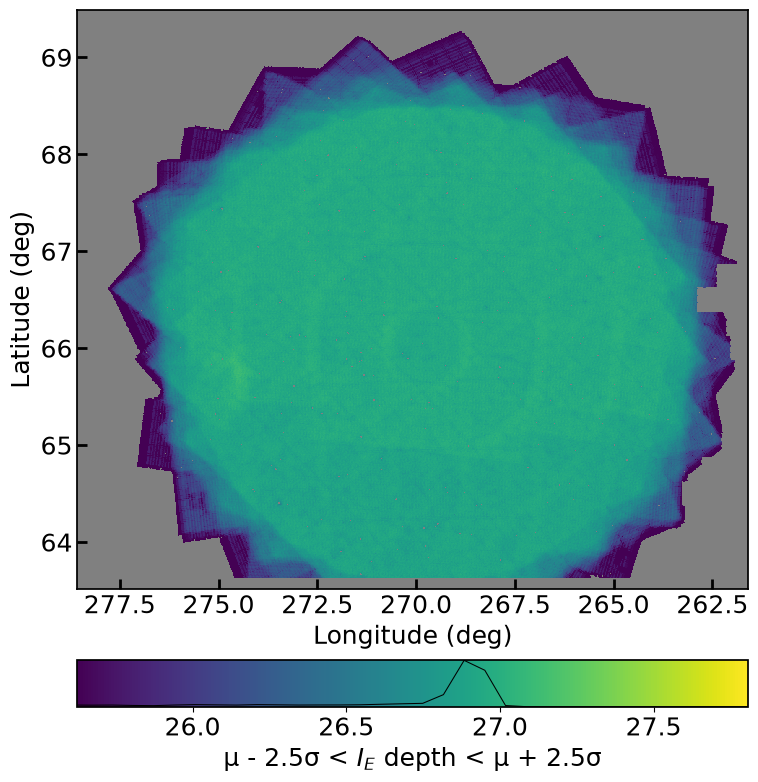}
    \includegraphics[angle=0,width=0.31\hsize]{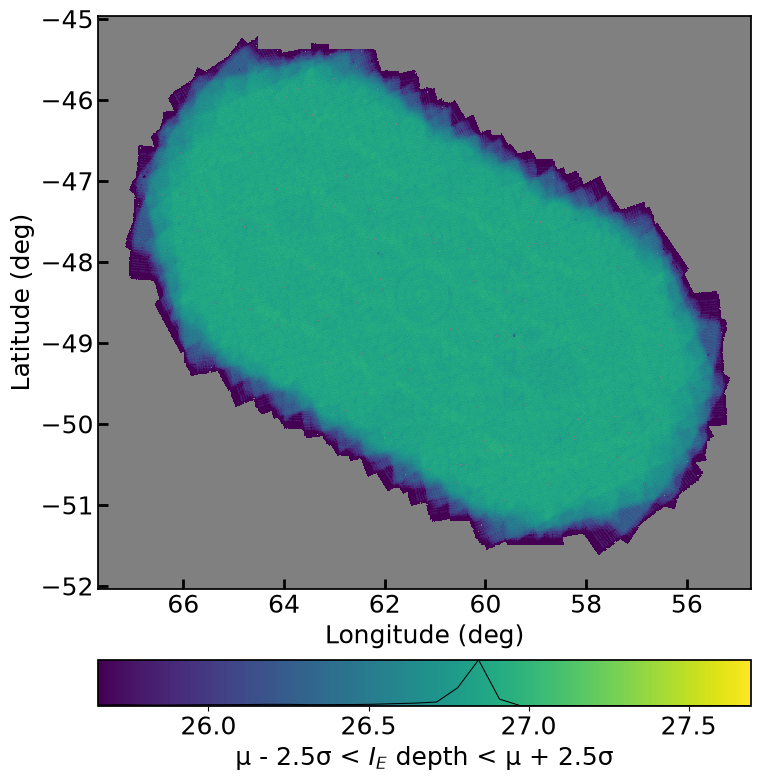}

    \vspace{0.1cm}

    \begin{overpic}[width=0.31\hsize]{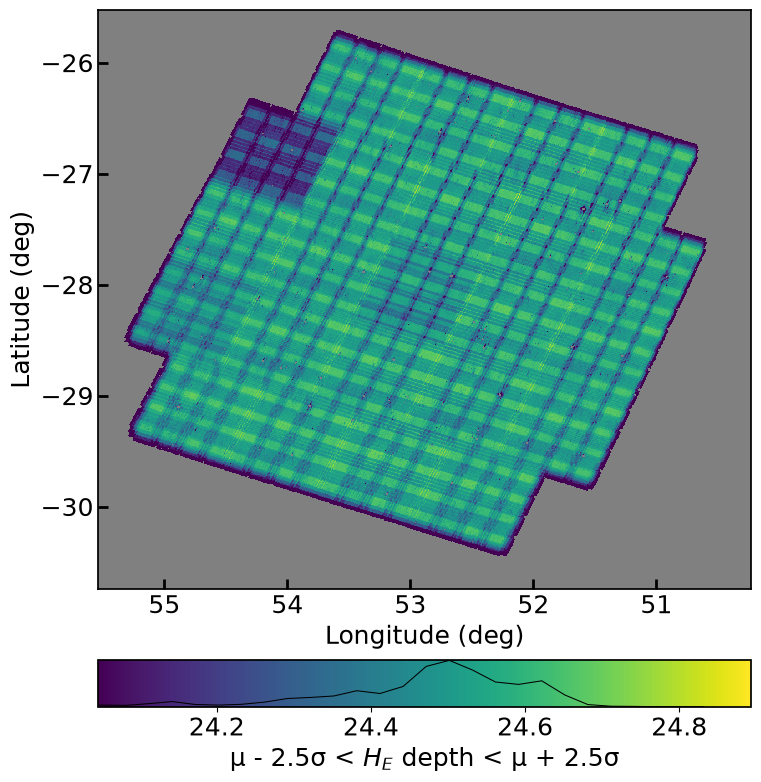}
        \put(-7,40){\rotatebox{90}{\textcolor{black}{\HE depth}}}
    \end{overpic}   
    \includegraphics[angle=0,width=0.31\hsize]{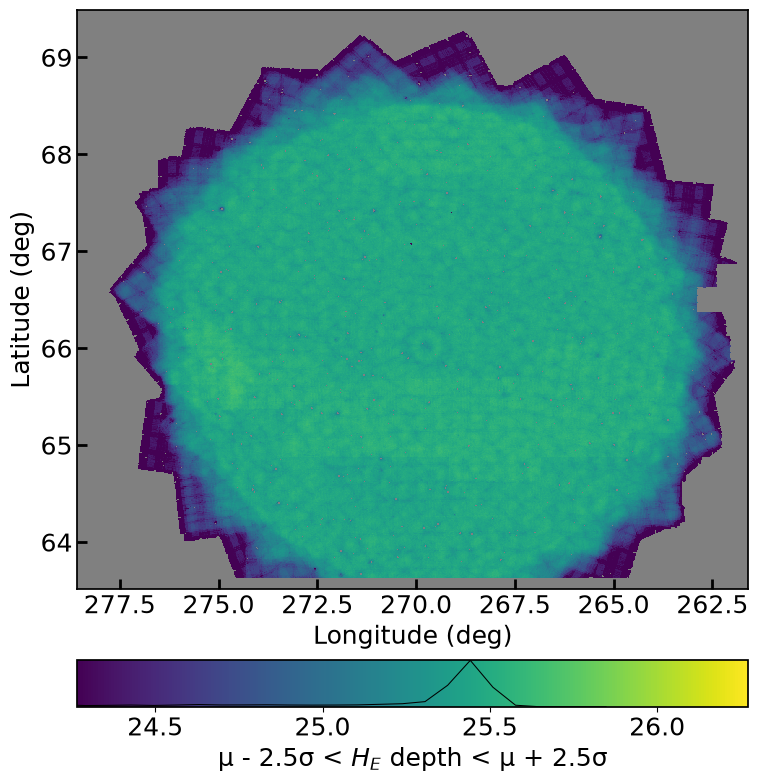}
    \includegraphics[angle=0,width=0.31\hsize]{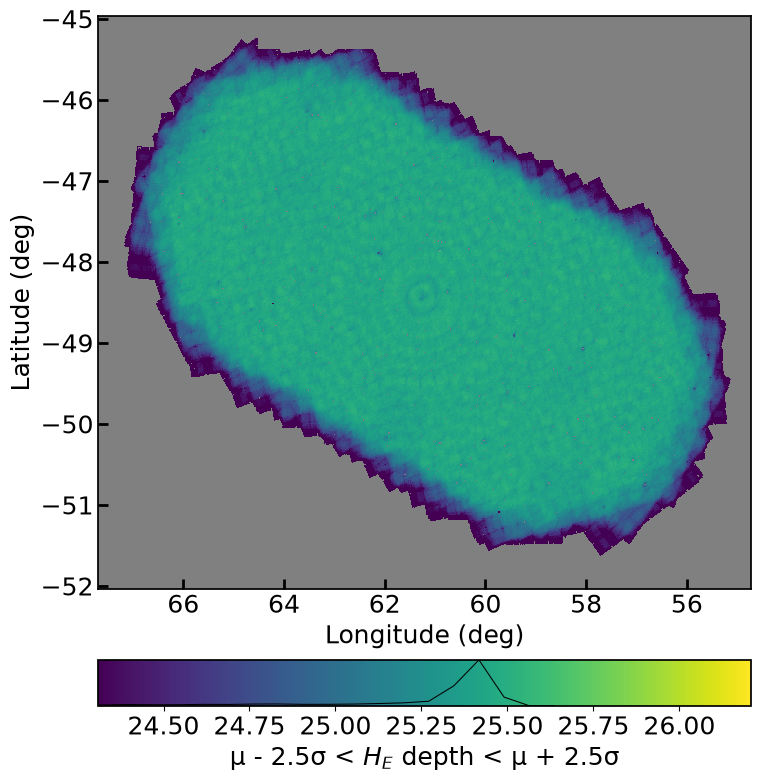}

    \caption{Examples of VMPZ-ID photometric masks for the EDF-F (left), EDF-N (middle), and EDF-S (right). From top to bottom, the panels show the effective coverage, integration time and depth for \IE, and the depth for \HE.
        \label{fig:deep_masks}}
\end{figure*}

\begin{figure*}[htbp!]
    \centering

    \begin{overpic}[width=0.334\hsize]{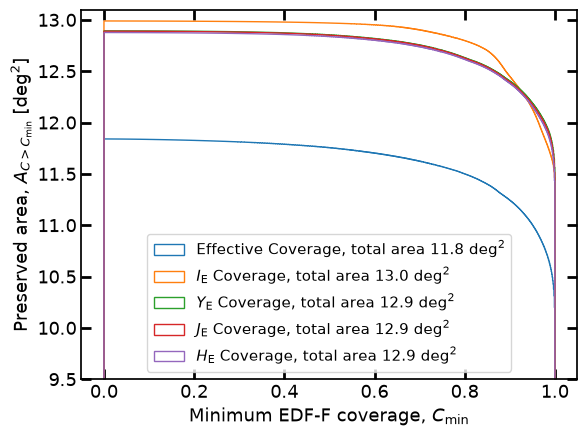}
        \put(40,62){EDF-F}
    \end{overpic}
    \begin{overpic}[width=0.325\hsize]{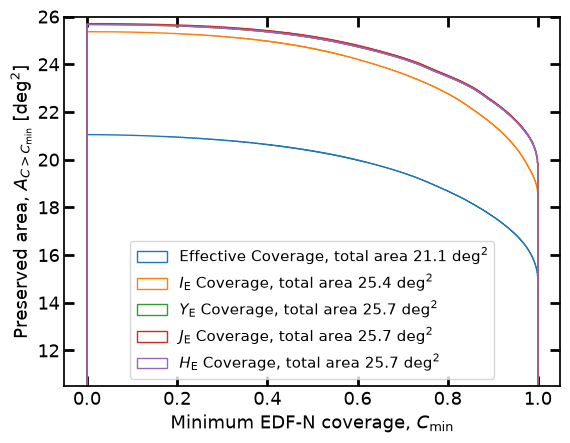}
        \put(40,62){EDF-N}
    \end{overpic}
    \begin{overpic}[width=0.325\hsize]{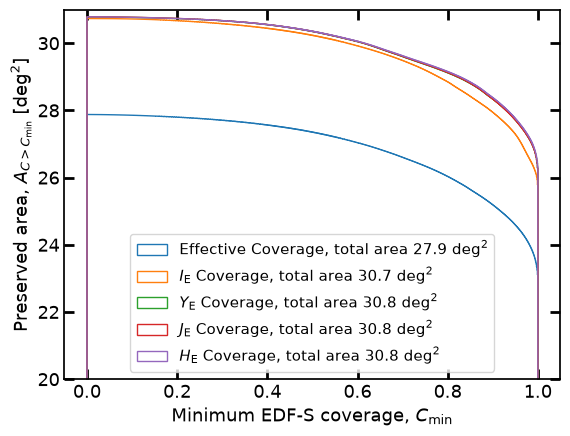}
        \put(40,62){EDF-S}
    \end{overpic}

    \begin{overpic}[width=0.33\hsize]{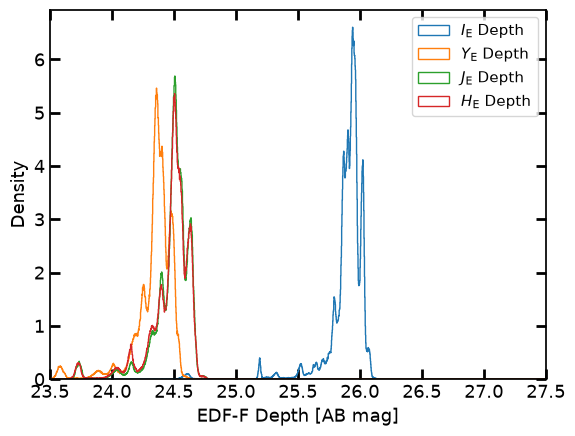}
        \put(40,66){EDF-F}
    \end{overpic}
    \begin{overpic}[width=0.33\hsize]{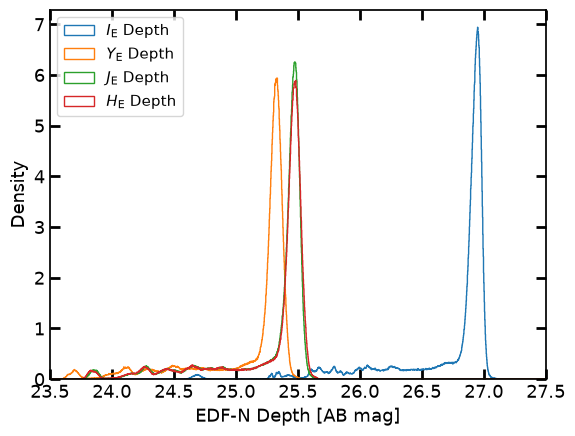}
        \put(40,66){EDF-N}
    \end{overpic}
    \begin{overpic}[width=0.33\hsize]{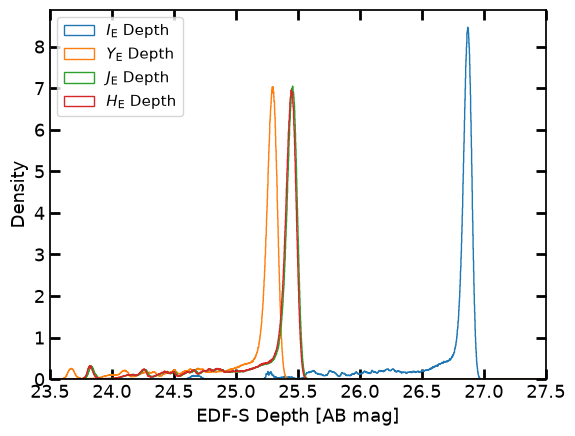}
        \put(40,66){EDF-S}
    \end{overpic}

    \caption{\textit{Top}: integrated coverage weighted area $A_{C > C_{\mathrm{min}}}$ for \texttt{HEALPix} pixels with coverage above a given minimum value, $C_{\mathrm{min}}$. \textit{Bottom}: depth distributions for the four \Euclid\ bands. From left to right results for the EDF-F, EDF-N, and EDF-S regions are presented.\label{fig:deep_Coverage}}
\end{figure*}

Figure~\ref{fig:deep_masks} presents examples of the main VMPZ-ID photometric masks in the EDF-F (left panel), EDF-N, (middle panel), and EDF-S (right panel). The effective coverage masks clearly show fully masked regions (dark blue in the figure) due to star contamination, as discussed in previous sections. The integrated coverage weighted area is shown in the top panels of Fig.~\ref{fig:deep_Coverage} for the three EDFs. Following the discussion in Sect.~\ref{sec:ewsfullmasks}, the total (effective) observed area is 13.2 (11.8), 28.5 (21.1), and 33.0 (27.9)\,deg$^2$ for the EDF-F, EDF-N, and EDF-S, respectively. Thus, the effective coverage mask preserves 89.3, 74.0, and 84.5\% of this area. Furthermore, 75, 72, and 92\% of the preserved regions have effective coverage greater than 0.8.

Figure~\ref{fig:deep_masks} also shows that the \IE integration time has small spatial variations for EDF-N and EDF-S, while more significant structure is observed for EDF-F. This can be explained by the significantly larger number of exposures and multiple exposure angles in EDF-N and EDF-S compared with EDF-F. A similar behaviour is observed for the \IE and \HE depth masks. This is also visible in the bottom panel of Fig.~\ref{fig:deep_Coverage}, where the 1D depth distributions are shown for the four \Euclid\ bands and the three EDFs. A detailed statistical description of these distributions is given in Tables~\ref{tab:instrumentprop_EDFF}, \ref{tab:instrumentprop_EDFN}, and \ref{tab:instrumentprop_EDFS}. The median \IE depth (integration time) values in AB mag (seconds) are 25.93 (\num{4842}), 26.91 (\num{25330}), and 26.85 (\num{23560}) for EDF-F, EDF-N, and EDF-S, respectively. These values are consistent with the corresponding numbers of visits, namely 2, 11, and 10 \citep{DR1-TP027}.

\section{Summary and conclusions}
\label{sec:conclu}
In this paper, a series of photometric masks has been presented in the form of \texttt{HEALPix} maps at a resolution of $N_{\mathrm{side}}\,{=}\,16\,384$ for each tile of the \DRone EWS and EDS. These masks were built using the OU-LE3 VMPZ-ID PF and pipelines within the \Euclid\ SGS. They represent spatial variations in the observational properties across the \DRone survey (integration time and satellite position angles), the instrumental properties for the \Euclid\ and \EXT\ bands (coverage, instrumental noise and depth, PSF width, systematic flags, NIR persistence, and astrometric uncertainties), foreground sky emission (Galactic extinction and zodiacal light), and the stellar distribution (star number density and star flux properties).

A mask representing the overall effective coverage of \DRone and combining information from the \Euclid\ and \EXT\ bands is also provided. This mask also accounts for bright-star contamination, including diffraction spikes and bleeding effects in the \Euclid\ and \EXT\ images. Overall, the effective coverage mask reduces the available area by 18.1\% in the \DRone EWS and by 11 to 26\% in the EDFs. Nevertheless, the effective coverage mask should be used to avoid spurious objects and artefacts in the data when selecting Euclid \DRone photometric galaxy samples for scientific analyses.

The instrumental performance in the VIS and NISP photometric bands is very stable across the survey. Spatial variations in instrumental RMS roise are strongly correlated with integration time, as expected. For the \DRone EWS, the standard deviation of depth (for a 2$^{\prime \prime}$ aperture) is below 0.2 AB mag, with median values of 25.69, 24.14, 24.3, and 24.29 for the \IE, \YE, \JE, and \HE bands, respectively. Furthermore, the spatial standard deviations of the PSF FWHM are of the order of \ang{;;0.01}, with median values of \ang{;;0.2}, \ang{;;0.48}, \ang{;;0.51}, and \ang{;;0.56} for the \IE, \YE, \JE, and \HE bands, respectively. At small angular scales, however, some structured patterns in integration time are observed.

For the \DRone EDS, multiple visits at different observation angles tend to smooth out spatial variations in the instrumental properties, particularly in EDF-N and EDF-S. Median depth values of 26.91, 25.30, 25.44, and 25.45 are achieved for the \IE, \YE, \JE, and \HE bands, respectively. The standard deviation of depth is below 0.16, 0.44, and 0.40 AB mag for EDF-F, EDF-N, and EDF-S, respectively.

The VMPZ-ID photometric masks will be particularly relevant for forthcoming \Euclid\ photometric astrophysical and cosmological analyses. They will be key for monitoring and correcting potential systematic spatial variations in galaxy number density, which could otherwise bias important derived quantities, such as cosmological parameter estimates.


\begin{acknowledgements}
    \AckEC
    \AckDRone
\end{acknowledgements}

\bibliography{Euclid,vmpzid}

\begin{thebibliography}{25}
\expandafter\ifx\csname natexlab\endcsname\relax\def\natexlab#1{#1}\fi

\bibitem[{{Abbott} {et~al.}(2021){Abbott}, {Adam{\'o}w}, {Aguena}, {Allam}, {Amon}, {Annis}, {Avila}, {Bacon}, {Banerji}, {Bechtol}, {Becker}, {Bernstein}, {Bertin}, {Bhargava}, {Bridle}, {Brooks}, {Burke}, {Carnero Rosell}, {Carrasco Kind}, {Carretero}, {Castander}, {Cawthon}, {Chang}, {Choi}, {Conselice}, {Costanzi}, {Crocce}, {da Costa}, {Davis}, {De Vicente}, {DeRose}, {Desai}, {Diehl}, {Dietrich}, {Drlica-Wagner}, {Eckert}, {Elvin-Poole}, {Everett}, {Evrard}, {Ferrero}, {Fert{\'e}}, {Flaugher}, {Fosalba}, {Friedel}, {Frieman}, {Garc{\'\i}a-Bellido}, {Gaztanaga}, {Gelman}, {Gerdes}, {Giannantonio}, {Gill}, {Gruen}, {Gruendl}, {Gschwend}, {Gutierrez}, {Hartley}, {Hinton}, {Hollowood}, {Honscheid}, {Huterer}, {James}, {Jeltema}, {Johnson}, {Kent}, {Kron}, {Kuehn}, {Kuropatkin}, {Lahav}, {Li}, {Lidman}, {Lin}, {MacCrann}, {Maia}, {Manning}, {Maloney}, {March}, {Marshall}, {Martini}, {Melchior}, {Menanteau}, {Miquel}, {Morgan}, {Myles}, {Neilsen}, {Ogando}, {Palmese}, {Paz-Chinch{\'o}n}, {Petravick},
  {Pieres}, {Plazas}, {Pond}, {Rodriguez-Monroy}, {Romer}, {Roodman}, {Rykoff}, {Sako}, {Sanchez}, {Santiago}, {Scarpine}, {Serrano}, {Sevilla-Noarbe}, {Smith}, {Smith}, {Soares-Santos}, {Suchyta}, {Swanson}, {Tarle}, {Thomas}, {To}, {Tremblay}, {Troxel}, {Tucker}, {Turner}, {Varga}, {Walker}, {Wechsler}, {Weller}, {Wester}, {Wilkinson}, {Yanny}, {Zhang}, {Nikutta}, {Fitzpatrick}, {Jacques}, {Scott}, {Olsen}, {Huang}, {Herrera}, {Juneau}, {Nidever}, {Weaver}, {Adean}, {Correia}, {de Freitas}, {Freitas}, {Singulani}, {Vila-Verde}, \& {Linea Science Server}}]{Abbott2021}
{Abbott}, T.~M.~C., {Adam{\'o}w}, M., {Aguena}, M., {et~al.} 2021, \apjs, 255, 20

\bibitem[{{Aihara} {et~al.}(2018){Aihara}, {Arimoto}, {Armstrong}, {Arnouts}, {Bahcall}, {Bickerton}, {Bosch}, {Bundy}, {Capak}, {Chan}, {Chiba}, {Coupon}, {Egami}, {Enoki}, {Finet}, {Fujimori}, {Fujimoto}, {Furusawa}, {Furusawa}, {Goto}, {Goulding}, {Greco}, {Greene}, {Gunn}, {Hamana}, {Harikane}, {Hashimoto}, {Hattori}, {Hayashi}, {Hayashi}, {He{\l}miniak}, {Higuchi}, {Hikage}, {Ho}, {Hsieh}, {Huang}, {Huang}, {Ikeda}, {Imanishi}, {Inoue}, {Iwasawa}, {Iwata}, {Jaelani}, {Jian}, {Kamata}, {Karoji}, {Kashikawa}, {Katayama}, {Kawanomoto}, {Kayo}, {Koda}, {Koike}, {Kojima}, {Komiyama}, {Konno}, {Koshida}, {Koyama}, {Kusakabe}, {Leauthaud}, {Lee}, {Lin}, {Lin}, {Lupton}, {Mandelbaum}, {Matsuoka}, {Medezinski}, {Mineo}, {Miyama}, {Miyatake}, {Miyazaki}, {Momose}, {More}, {More}, {Moritani}, {Moriya}, {Morokuma}, {Mukae}, {Murata}, {Murayama}, {Nagao}, {Nakata}, {Niida}, {Niikura}, {Nishizawa}, {Obuchi}, {Oguri}, {Oishi}, {Okabe}, {Okamoto}, {Okura}, {Ono}, {Onodera}, {Onoue}, {Osato}, {Ouchi}, {Price}, {Pyo},
  {Sako}, {Sawicki}, {Shibuya}, {Shimasaku}, {Shimono}, {Shirasaki}, {Silverman}, {Simet}, {Speagle}, {Spergel}, {Strauss}, {Sugahara}, {Sugiyama}, {Suto}, {Suyu}, {Suzuki}, {Tait}, {Takada}, {Takata}, {Tamura}, {Tanaka}, {Tanaka}, {Tanaka}, {Tanaka}, {Terai}, {Terashima}, {Toba}, {Tominaga}, {Toshikawa}, {Turner}, {Uchida}, {Uchiyama}, {Umetsu}, {Uraguchi}, {Urata}, {Usuda}, {Utsumi}, {Wang}, {Wang}, {Wong}, {Yabe}, {Yamada}, {Yamanoi}, {Yasuda}, {Yeh}, {Yonehara}, \& {Yuma}}]{2018PASJ...70S...4A}
{Aihara}, H., {Arimoto}, N., {Armstrong}, R., {et~al.} 2018, \pasj, 70, S4

\bibitem[{{Amendola} {et~al.}(2018){Amendola}, {Appleby}, {Avgoustidis}, {Bacon}, {Baker}, {Baldi}, {Bartolo}, {Blanchard}, {Bonvin}, {Borgani}, {Branchini}, {Burrage}, \& {et al.}}]{Amendola18}
{Amendola}, L., {Appleby}, S., {Avgoustidis}, A., {et~al.} 2018, Living Reviews in Relativity, 21, 2

\bibitem[{Desai {et~al.}(2016)Desai, Mohr, Bertin, Kümmel, \& Wetzstein}]{DESAI201667}
Desai, S., Mohr, J., Bertin, E., Kümmel, M., \& Wetzstein, M. 2016, Astronomy and Computing, 16, 67

\bibitem[{{Euclid Collaboration: Aussel} {et~al.}(2026{\natexlab{a}}){Euclid Collaboration: Aussel}, {Tereno}, {Schirmer}, {et~al.}}]{Q1-TP001}
{Euclid Collaboration: Aussel}, H., {Tereno}, I., {Schirmer}, M., {et~al.} 2026{\natexlab{a}}, \aap, 711, A1

\bibitem[{{Euclid Collaboration: Aussel} {et~al.}(2026{\natexlab{b}})}]{DR1-TP001}
{Euclid Collaboration: Aussel}, H. {et~al.} 2026{\natexlab{b}}, arXiv, submitted

\bibitem[{{Euclid Collaboration: Blanchard} {et~al.}(2020){Euclid Collaboration: Blanchard}, {Camera}, {Carbone}, {et~al.}}]{Blanchard-EP7}
{Euclid Collaboration: Blanchard}, A., {Camera}, S., {Carbone}, C., {et~al.} 2020, \aap, 642, A191

\bibitem[{{Euclid Collaboration: Corcho-Caballero} {et~al.}(2026){Euclid Collaboration: Corcho-Caballero}, {Balbinot}, {De Jong}, {et~al.}}]{DR1-TP006}
{Euclid Collaboration: Corcho-Caballero}, P., {Balbinot}, E., {De Jong}, J., {et~al.} 2026, A\&A, submitted

\bibitem[{{Euclid Collaboration: Cropper} {et~al.}(2025){Euclid Collaboration: Cropper}, {Al-Bahlawan}, {Amiaux}, {et~al.}}]{EuclidSkyVIS}
{Euclid Collaboration: Cropper}, M., {Al-Bahlawan}, A., {Amiaux}, J., {et~al.} 2025, A\&A, 697, A2

\bibitem[{{Euclid Collaboration: Hartley} {et~al.}(2026){Euclid Collaboration: Hartley}, {Paltani}, {Dubath}, {et~al.}}]{DR1-TP011}
{Euclid Collaboration: Hartley}, W., {Paltani}, S., {Dubath}, F., {et~al.} 2026, A\&A, submitted

\bibitem[{{Euclid Collaboration: Jahnke} {et~al.}(2025){Euclid Collaboration: Jahnke}, {Gillard}, {Schirmer}, {et~al.}}]{EuclidSkyNISP}
{Euclid Collaboration: Jahnke}, K., {Gillard}, W., {Schirmer}, M., {et~al.} 2025, A\&A, 697, A3

\bibitem[{{Euclid Collaboration: Kubik} {et~al.}(2026){Euclid Collaboration: Kubik}, {Conseil}, {Ferriol}, {et~al.}}]{DR1-TP038}
{Euclid Collaboration: Kubik}, B., {Conseil}, S., {Ferriol}, S., {et~al.} 2026, A\&A, submitted

\bibitem[{{Euclid Collaboration: K\"ummel} {et~al.}(2026){Euclid Collaboration: K\"ummel}, {Caro}, {Dimauro}, {et~al.}}]{DR1-TP009}
{Euclid Collaboration: K\"ummel}, M., {Caro}, F., {Dimauro}, P., {et~al.} 2026, A\&A, submitted

\bibitem[{{Euclid Collaboration: Mellier} {et~al.}(2025){Euclid Collaboration: Mellier}, {Abdurro'uf}, {Acevedo~Barroso}, {et~al.}}]{EuclidSkyOverview}
{Euclid Collaboration: Mellier}, Y., {Abdurro'uf}, {Acevedo~Barroso}, J., {et~al.} 2025, A\&A, 697, A1

\bibitem[{{Euclid Collaboration: Radovich} {et~al.}(2026){Euclid Collaboration: Radovich}, {da Silva}, {Polenta}, {et~al.}}]{DR1-TP004}
{Euclid Collaboration: Radovich}, M., {da Silva}, R., {Polenta}, G., {et~al.} 2026, A\&A, submitted

\bibitem[{{Euclid Collaboration: Ruffenach} {et~al.}(2026){Euclid Collaboration: Ruffenach}, {Frailis}, {Zacchei}, {et~al.}}]{DR1-TP033}
{Euclid Collaboration: Ruffenach}, M., {Frailis}, M., {Zacchei}, A., {et~al.} 2026, A\&A, submitted

\bibitem[{{Euclid Collaboration: Scaramella} {et~al.}(2022){Euclid Collaboration: Scaramella}, {Amiaux}, {Mellier}, {et~al.}}]{Scaramella-EP1}
{Euclid Collaboration: Scaramella}, R., {Amiaux}, J., {Mellier}, Y., {et~al.} 2022, \aap, 662, A112

\bibitem[{{Euclid Collaboration: Tereno} {et~al.}(2026){Euclid Collaboration: Tereno}, {Dinis}, {Dupac}, {et~al.}}]{DR1-TP027}
{Euclid Collaboration: Tereno}, I., {Dinis}, J., {Dupac}, X., {et~al.} 2026, arXiv, submitted

\bibitem[{{Euclid Data Release 1}(2026)}]{DR1cite}
{Euclid Data Release 1}. 2026, \url{https://doi.org/10.57780/esa-d079c51}

\bibitem[{{Gaia Collaboration} {et~al.}(2016){Gaia Collaboration}, {Prusti}, {de Bruijne}, {Brown}, {Vallenari}, {Babusiaux}, {Bailer-Jones}, {Bastian}, {Biermann}, {Evans}, {Eyer}, {Jansen}, {Jordi}, {Klioner}, {Lammers}, {Lindegren}, {Luri}, {Mignard}, {Milligan}, {Panem}, {Poinsignon}, {Pourbaix}, {Randich}, {Sarri}, {Sartoretti}, {Siddiqui}, {Soubiran}, {Valette}, {van Leeuwen}, {Walton}, {Aerts}, {Arenou}, {Cropper}, {Drimmel}, {H{\o}g}, {Katz}, {Lattanzi}, {O'Mullane}, {Grebel}, {Holland}, {Huc}, {Passot}, {Bramante}, {Cacciari}, {Casta{\~n}eda}, {Chaoul}, {Cheek}, {De Angeli}, {Fabricius}, {Guerra}, {Hern{\'a}ndez}, {Jean-Antoine-Piccolo}, {Masana}, {Messineo}, {Mowlavi}, {Nienartowicz}, {Ord{\'o}{\~n}ez-Blanco}, {Panuzzo}, {Portell}, {Richards}, {Riello}, {Seabroke}, {Tanga}, {Th{\'e}venin}, {Torra}, {Els}, {Gracia-Abril}, {Comoretto}, {Garcia-Reinaldos}, {Lock}, {Mercier}, {Altmann}, {Andrae}, {Astraatmadja}, {Bellas-Velidis}, {Benson}, {Berthier}, {Blomme}, {Busso}, {Carry}, {Cellino}, {Clementini},
  {Cowell}, {Creevey}, {Cuypers}, {Davidson}, {De Ridder}, {de Torres}, {Delchambre}, {Dell'Oro}, {Ducourant}, {Fr{\'e}mat}, {Garc{\'\i}a-Torres}, {Gosset}, {Halbwachs}, {Hambly}, {Harrison}, {Hauser}, {Hestroffer}, {Hodgkin}, {Huckle}, {Hutton}, {Jasniewicz}, {Jordan}, {Kontizas}, {Korn}, {Lanzafame}, {Manteiga}, {Moitinho}, {Muinonen}, {Osinde}, {Pancino}, {Pauwels}, {Petit}, {Recio-Blanco}, {Robin}, {Sarro}, {Siopis}, {Smith}, {Smith}, {Sozzetti}, {Thuillot}, {van Reeven}, {Viala}, {Abbas}, {Abreu Aramburu}, {Accart}, {Aguado}, {Allan}, {Allasia}, {Altavilla}, {{\'A}lvarez}, {Alves}, {Anderson}, {Andrei}, {Anglada Varela}, {Antiche}, {Antoja}, {Ant{\'o}n}, {Arcay}, {Atzei}, {Ayache}, {Bach}, {Baker}, {Balaguer-N{\'u}{\~n}ez}, {Barache}, {Barata}, {Barbier}, {Barblan}, {Baroni}, {Barrado y Navascu{\'e}s}, {Barros}, {Barstow}, {Becciani}, {Bellazzini}, {Bellei}, {Bello Garc{\'\i}a}, {Belokurov}, {Bendjoya}, {Berihuete}, {Bianchi}, {Bienaym{\'e}}, {Billebaud}, {Blagorodnova}, {Blanco-Cuaresma}, {Boch},
  {Bombrun}, {Borrachero}, {Bouquillon}, {Bourda}, {Bouy}, {Bragaglia}, {Breddels}, {Brouillet}, {Br{\"u}semeister}, {Bucciarelli}, {Budnik}, {Burgess}, {Burgon}, {Burlacu}, {Busonero}, {Buzzi}, {Caffau}, {Cambras}, {Campbell}, {Cancelliere}, {Cantat-Gaudin}, {Carlucci}, {Carrasco}, {Castellani}, {Charlot}, {Charnas}, {Charvet}, {Chassat}, {Chiavassa}, {Clotet}, {Cocozza}, {Collins}, {Collins}, \& {Costigan}}]{2016A&A...595A...1G}
{Gaia Collaboration}, {Prusti}, T., {de Bruijne}, J.~H.~J., {et~al.} 2016, \aap, 595, A1

\bibitem[{{G{\'o}rski} {et~al.}(2005){G{\'o}rski}, {Hivon}, {Banday}, {Wandelt}, {Hansen}, {Reinecke}, \& {Bartelmann}}]{2005ApJ...622..759G}
{G{\'o}rski}, K.~M., {Hivon}, E., {Banday}, A.~J., {et~al.} 2005, \apj, 622, 759

\bibitem[{{Ibata} {et~al.}(2017){Ibata}, {McConnachie}, {Cuillandre}, {Fantin}, {Haywood}, {Martin}, {Bergeron}, {Beckmann}, {Bernard}, {Bonifacio}, {Caffau}, {Carlberg}, {C{\^o}t{\'e}}, {Cabanac}, {Chapman}, {Duc}, {Durret}, {Famaey}, {Fabbro}, {Gwyn}, {Hammer}, {Hill}, {Hudson}, {Lan{\c{c}}on}, {Lewis}, {Malhan}, {di Matteo}, {McCracken}, {Mei}, {Mellier}, {Navarro}, {Pires}, {Pritchet}, {Reyl{\'e}}, {Richer}, {Robin}, {S{\'a}nchez-Janssen}, {Sawicki}, {Scott}, {Scottez}, {Spekkens}, {Starkenburg}, {Thomas}, \& {Venn}}]{Ibata2017}
{Ibata}, R.~A., {McConnachie}, A., {Cuillandre}, J.-C., {et~al.} 2017, \apj, 848, 128

\bibitem[{{Magnier} {et~al.}(2020){Magnier}, {Schlafly}, {Finkbeiner}, {Tonry}, {Goldman}, {R{\"o}ser}, {Schilbach}, {Casertano}, {Chambers}, {Flewelling}, {Huber}, {Price}, {Sweeney}, {Waters}, {Denneau}, {Draper}, {Hodapp}, {Jedicke}, {Kaiser}, {Kudritzki}, {Metcalfe}, {Stubbs}, \& {Wainscoat}}]{Magnier2020}
{Magnier}, E.~A., {Schlafly}, E.~F., {Finkbeiner}, D.~P., {et~al.} 2020, \apjs, 251, 6

\bibitem[{{Weaverdyck} {et~al.}(2026){Weaverdyck}, {Rodr{\'\i}guez-Monroy}, {Elvin-Poole}, {Sevilla-Noarbe}, {Porredon}, {Avila}, {Lee}, {Riquelme}, {Tabbutt}, {Huterer}, {Prat}, {De Vicente}, {Mena-Fern{\'a}ndez}, {Crocce}, {S{\'a}nchez}, {Bernstein}, {Henning}, {Cawthon}, {Ross}, {Abbott}, {Aguena}, {Allam}, {Alves}, {Andrade-Oliveira}, {Bacon}, {Bechtol}, {Bertin}, {Blazek}, {Bocquet}, {Brooks}, {Camilleri}, {Carnero Rosell}, {Carretero}, {Castander}, {Choi}, {da Costa}, {da Silva Pereira}, {Davis}, {Diehl}, {Doux}, {Drlica-Wagner}, {Eifler}, {Everett}, {Evrard}, {Flaugher}, {Garc{\'\i}a-Bellido}, {Gatti}, {Gazta{\~n}aga}, {Giannini}, {Gruen}, {Gutierrez}, {Hinton}, {Hollowood}, {Honscheid}, {Jain}, {Kacprzak}, {Kuehn}, {Lahav}, {Marshall}, {Menanteau}, {Miquel}, {Mohr}, {Muir}, {Myles}, {Nichol}, {Ogando}, {Palmese}, {Paterno}, {Percival}, {Plazas Malag{\'o}n}, {Rosenfeld}, {Rykoff}, {Samuroff}, {Sanchez}, {Sanchez Cid}, {Sheldon}, {Sherman}, {Smith}, {Soares-Santos}, {Suchyta}, {Swanson}, {Gregory},
  {Thomas}, {To}, {Tucker}, {Vikram}, {Yamamoto}, \& {Yanny}}]{2026arXiv260114484W}
{Weaverdyck}, N., {Rodr{\'\i}guez-Monroy}, M., {Elvin-Poole}, J., {et~al.} 2026, arXiv e-prints, arXiv:2601.14484

\bibitem[{{Yan} {et~al.}(2025){Yan}, {Wright}, {Chisari}, {Georgiou}, {Joudaki}, {Loureiro}, {Reischke}, {Asgari}, {Bilicki}, {Dvornik}, {Heymans}, {Hildebrandt}, {Jalan}, {Joachimi}, {Lesci}, {Li}, {Linke}, {Mahony}, {Moscardini}, {Napolitano}, {St{\"o}lzner}, {Von Wietersheim-Kramsta}, \& {Yoon}}]{2025A&A...694A.259Y}
{Yan}, Z., {Wright}, A.~H., {Chisari}, N.~E., {et~al.} 2025, \aap, 694, A259

\end{thebibliography}

\clearpage

\begin{appendix}
\onecolumn

\section{Zoom-in on the \DRone EWS masks}
\label{sec:app_zoomin}
\begin{figure*}[htbp!]
    \begin{adjustwidth}{-1cm}{-1cm}
        \centering
        \includegraphics[angle=0,width=0.245\hsize]{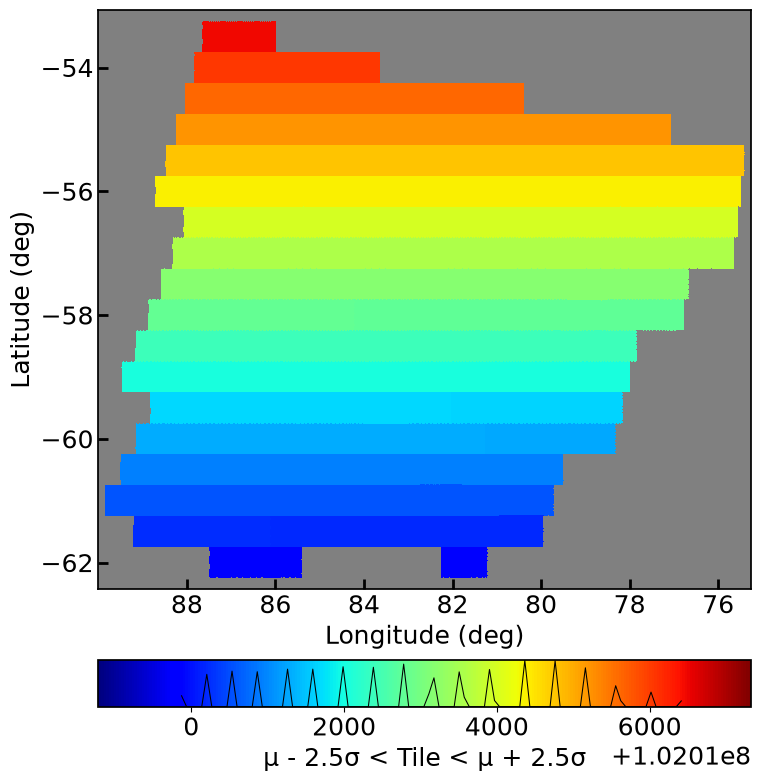}
        \includegraphics[angle=0,width=0.245\hsize]{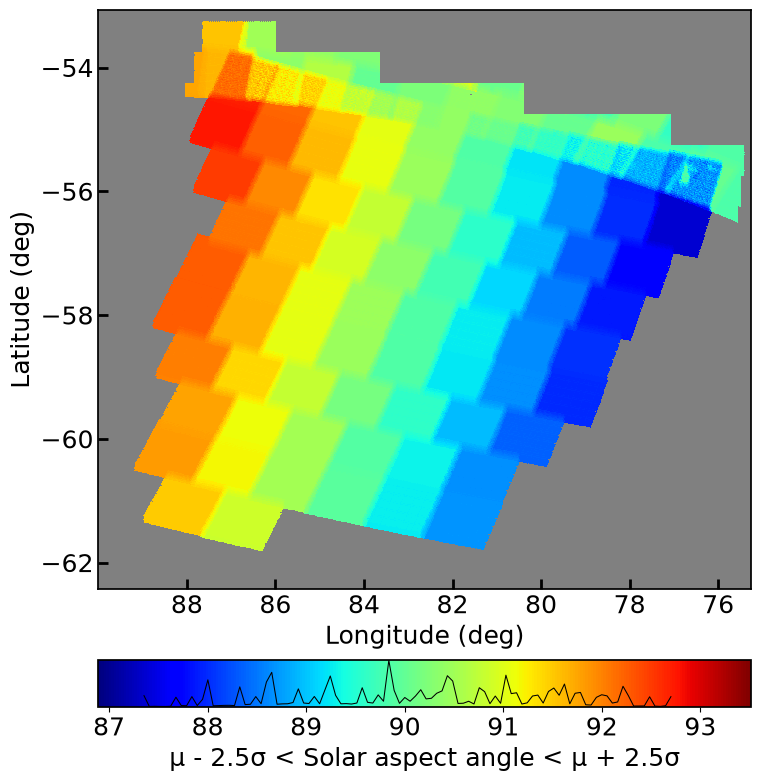}
        \includegraphics[angle=0,width=0.245\hsize]{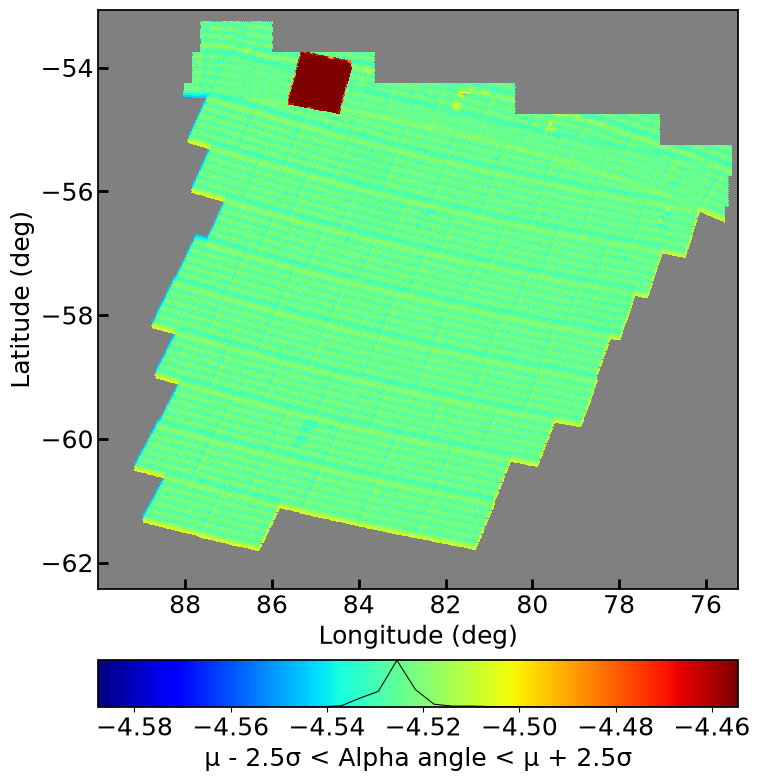}
        \includegraphics[angle=0,width=0.245\hsize]{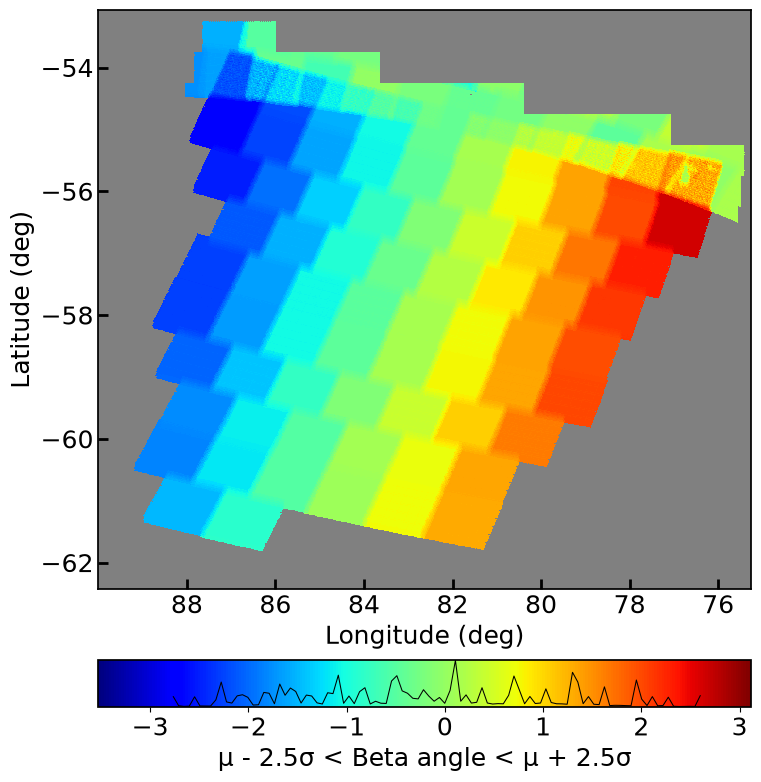}
    
        \includegraphics[angle=0,width=0.245\hsize]{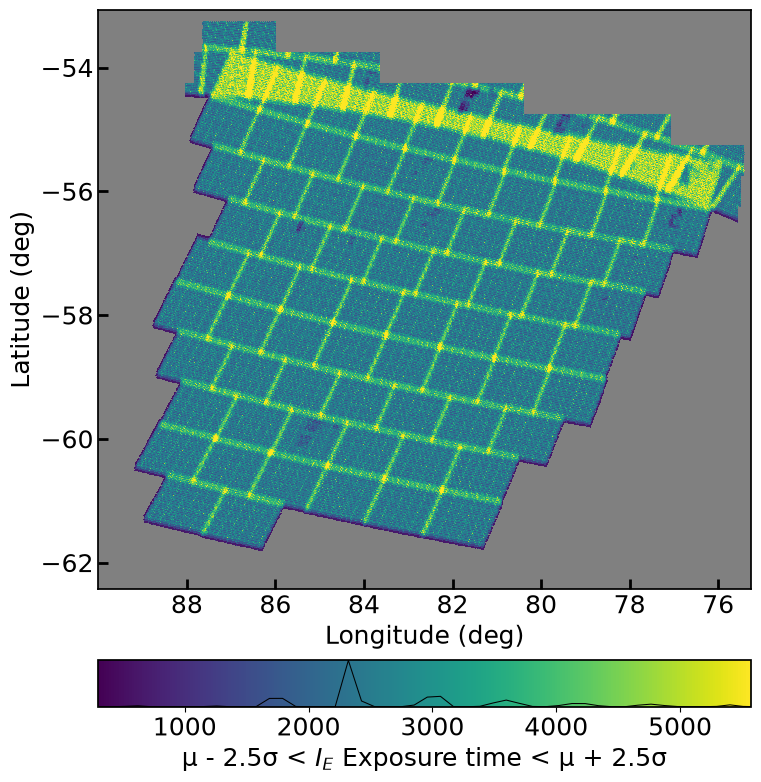}
        \includegraphics[angle=0,width=0.245\hsize]{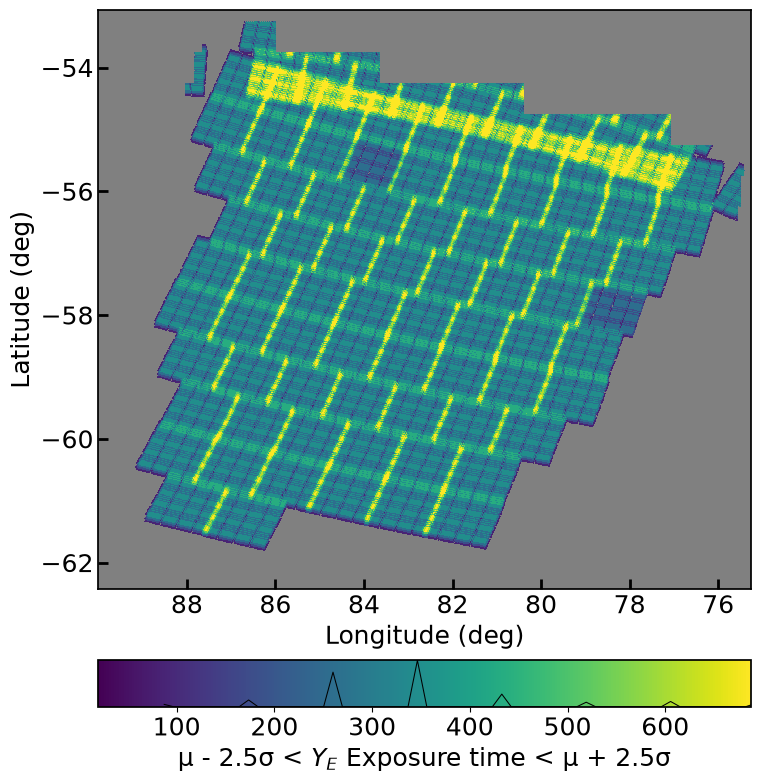}
        \includegraphics[angle=0,width=0.245\hsize]{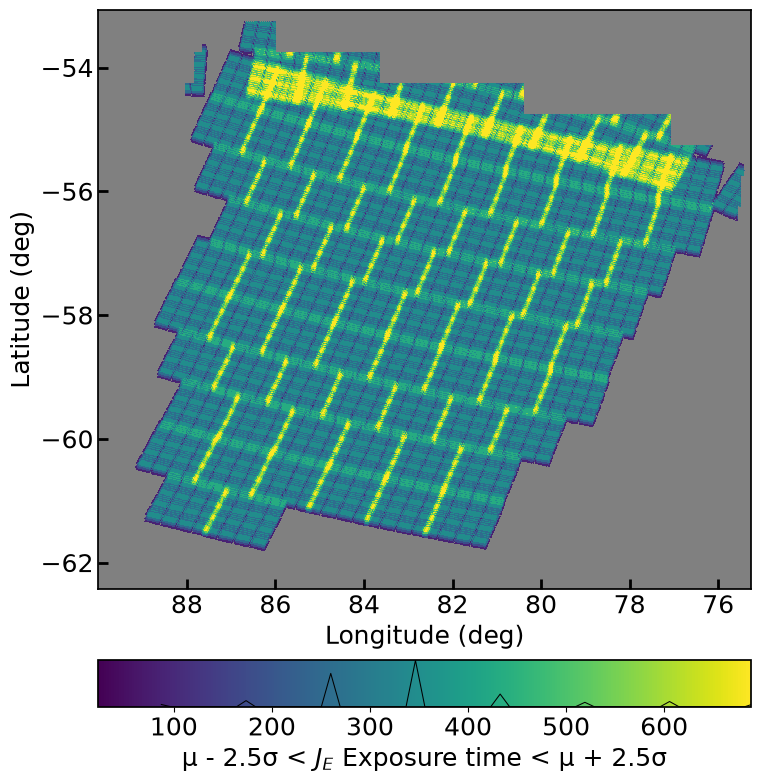}
        \includegraphics[angle=0,width=0.245\hsize]{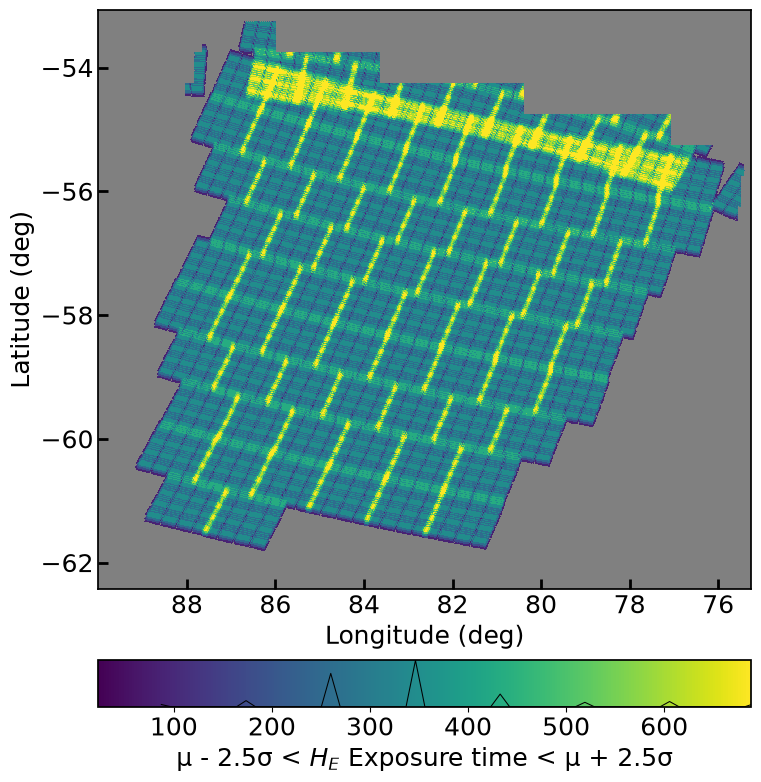} 

        \caption{Main survey masks for the DR1 EWS Patch ID 4. From left to right, the panels show the tile index, solar aspect angle, alpha angle, and beta angle (top row), followed by the integration time for the \IE, \YE, \JE, and \HE bands (bottom row). \label{fig:patch4_surveymasks}} 
    \end{adjustwidth}
\end{figure*}

\begin{figure*}[htbp!]
    \begin{adjustwidth}{-1cm}{-1cm}
        \centering
        \includegraphics[angle=0,width=0.245\hsize]{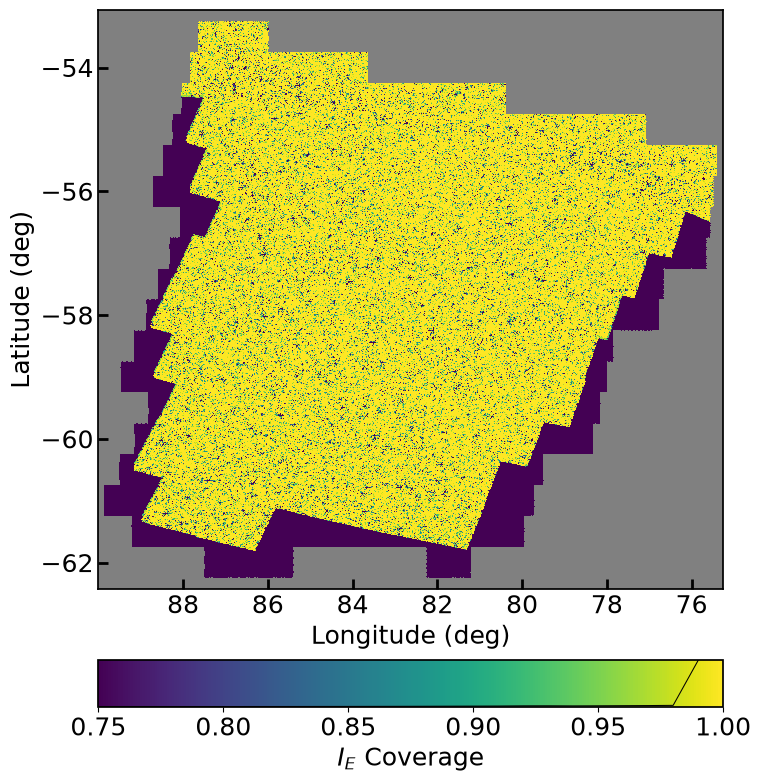}
        \includegraphics[angle=0,width=0.245\hsize]{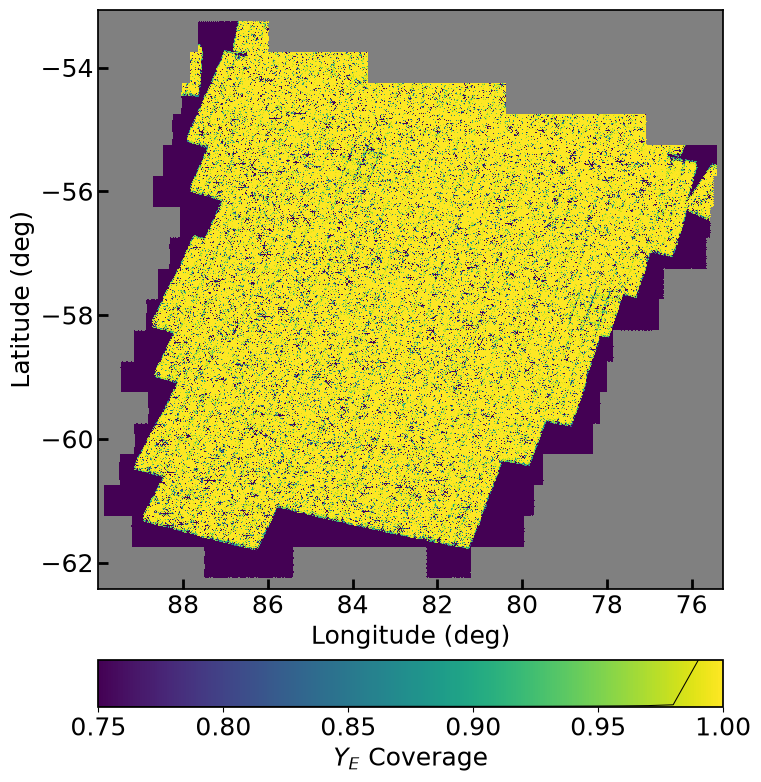}
        \includegraphics[angle=0,width=0.245\hsize]{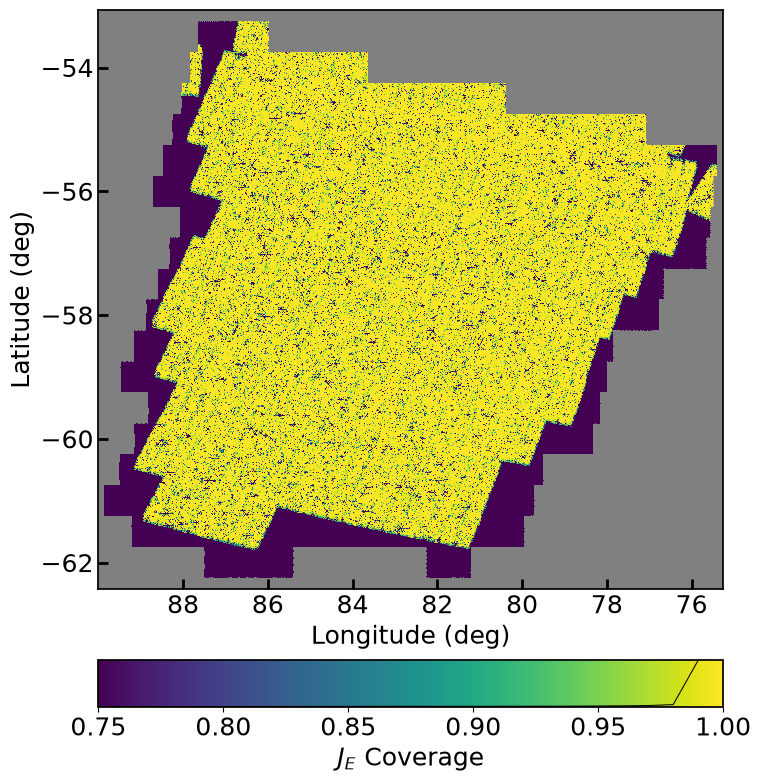}
        \includegraphics[angle=0,width=0.245\hsize]{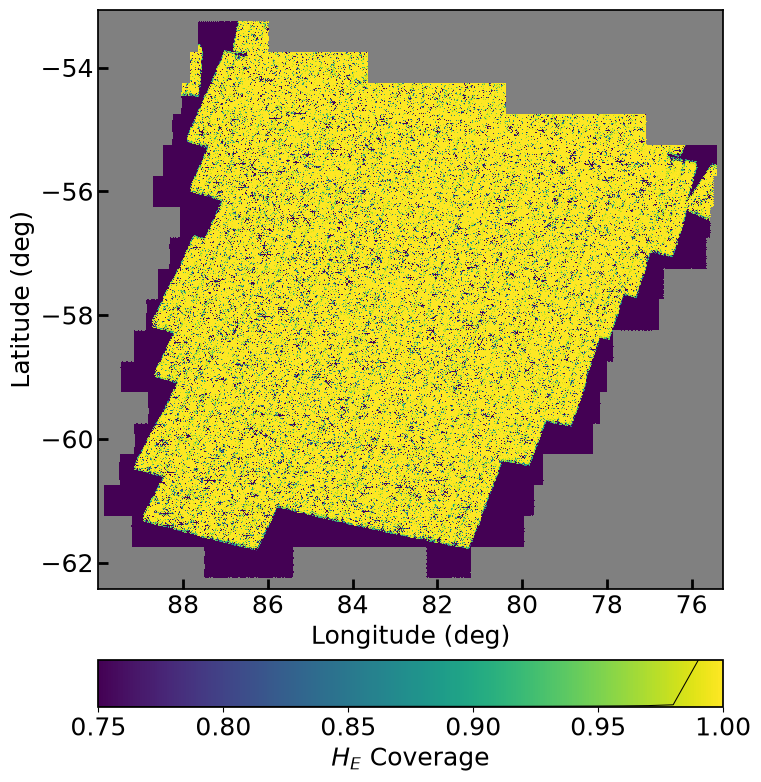}

        \includegraphics[angle=0,width=0.245\hsize]{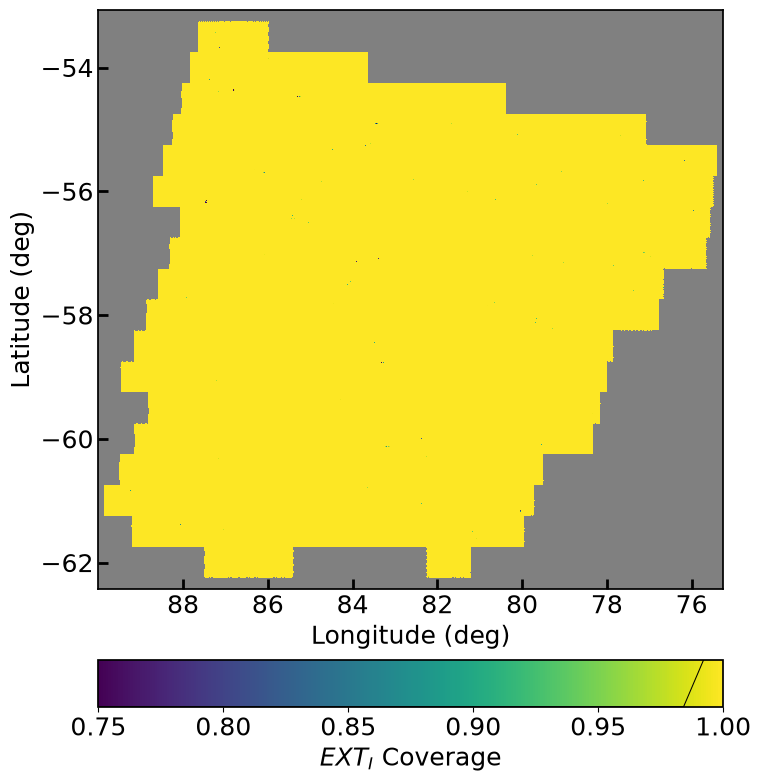}    
        \includegraphics[angle=0,width=0.245\hsize]{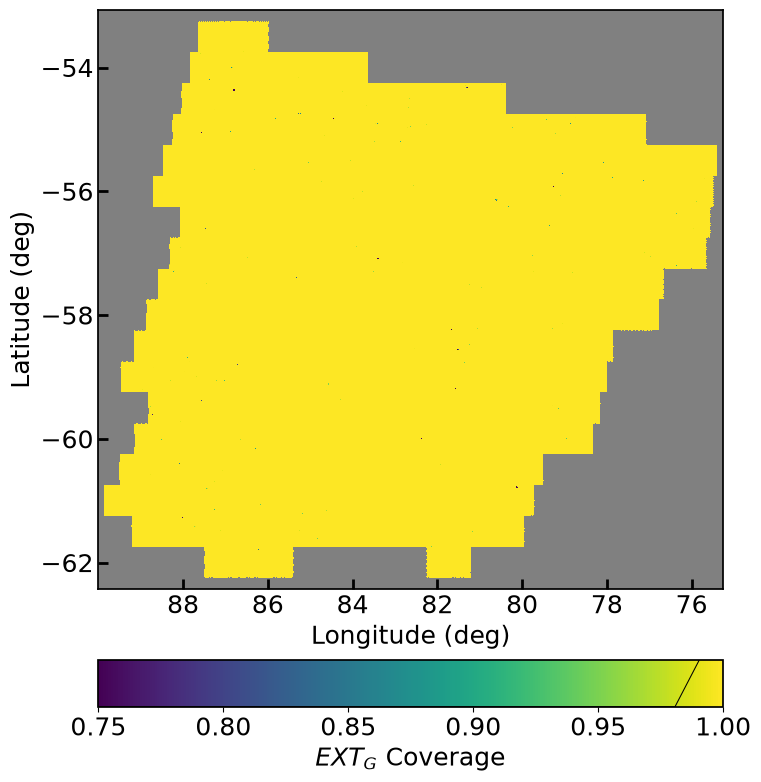}
        \includegraphics[angle=0,width=0.245\hsize]{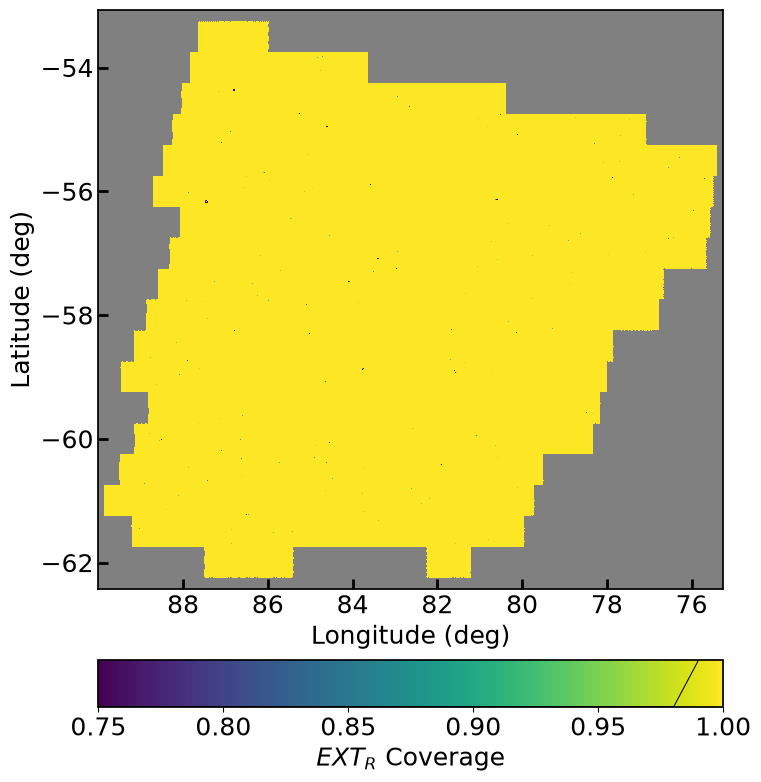}
        \includegraphics[angle=0,width=0.245\hsize]{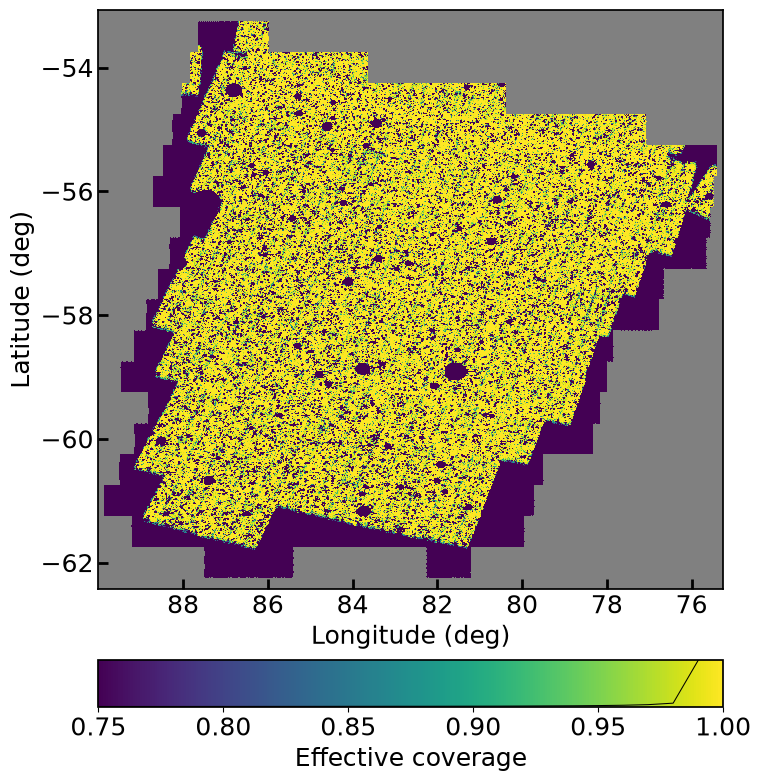}
    
        \caption{Examples of survey-definition masks for the \DRone EWS Patch ID 4. From left to right, the panels show the coverage for the \IE, \YE, \JE, and \HE bands (top row), followed by the coverage for the available \EXT\ bands and the combined effective coverage (bottom row). For \DRone, the \EXT\ coverage masks do not include the polygons used to mask regions affected by bright stars. This masking is currently applied only when constructing the combined effective coverage mask.}
        \label{fig:patch4_Coveragemasks}
   \end{adjustwidth} 
\end{figure*}

\begin{figure*}[htbp!]
    \begin{adjustwidth}{-1cm}{-1cm}
        \centering
        \includegraphics[angle=0,width=0.245\hsize]{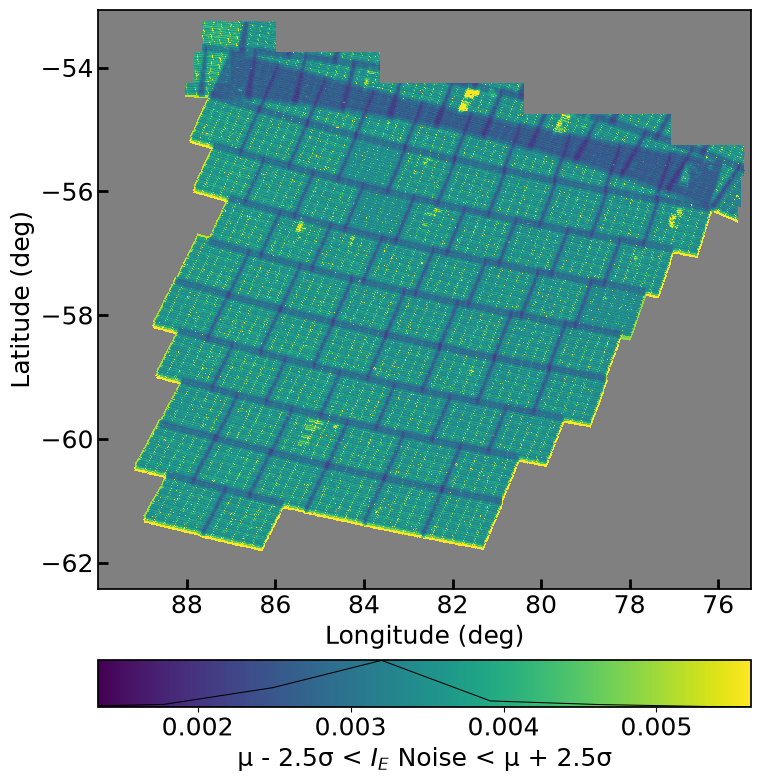}
        \includegraphics[angle=0,width=0.245\hsize]{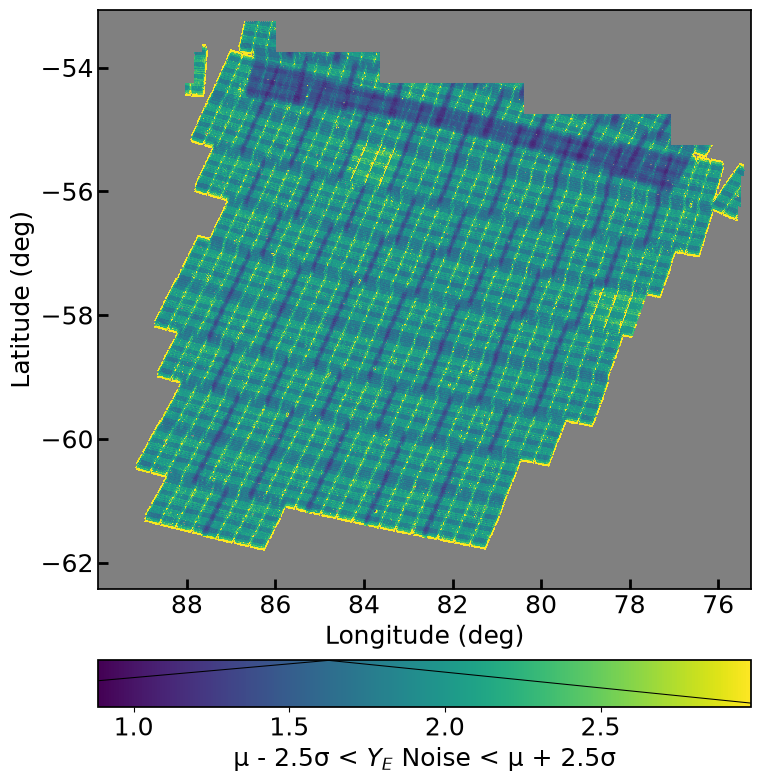}
        \includegraphics[angle=0,width=0.245\hsize]{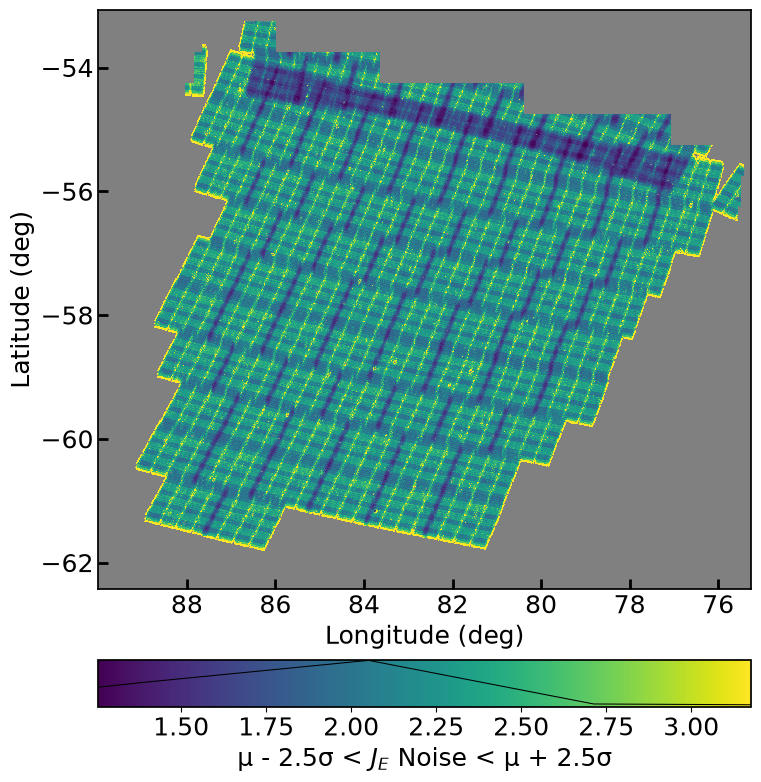}
        \includegraphics[angle=0,width=0.245\hsize]{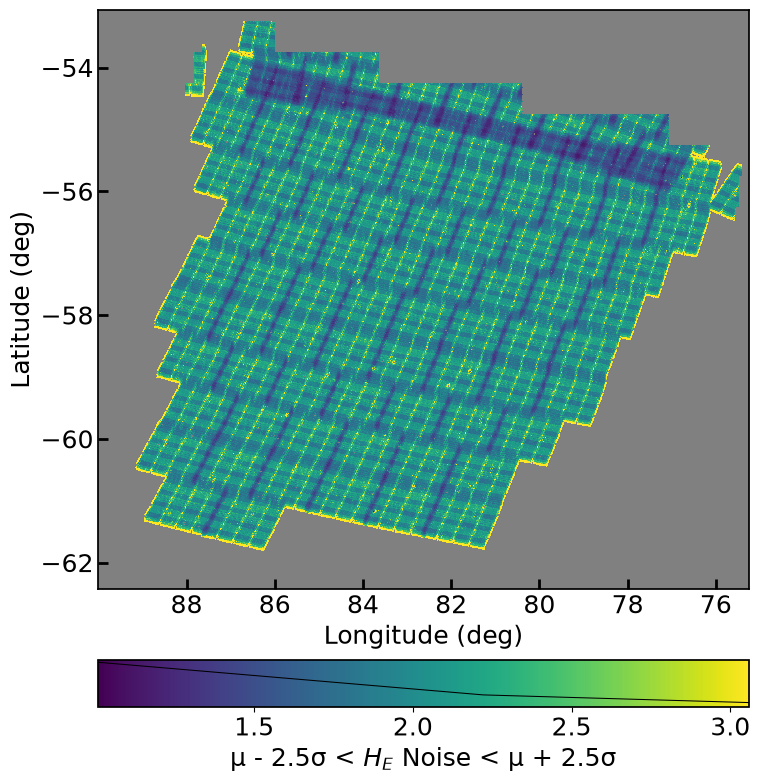} 
        
        \includegraphics[angle=0,width=0.245\hsize]{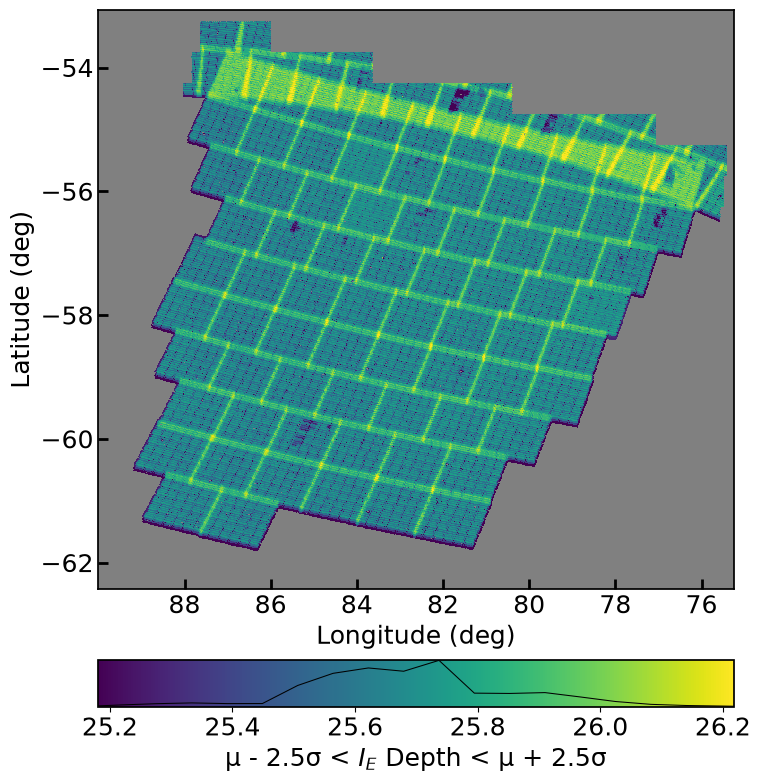}
        \includegraphics[angle=0,width=0.245\hsize]{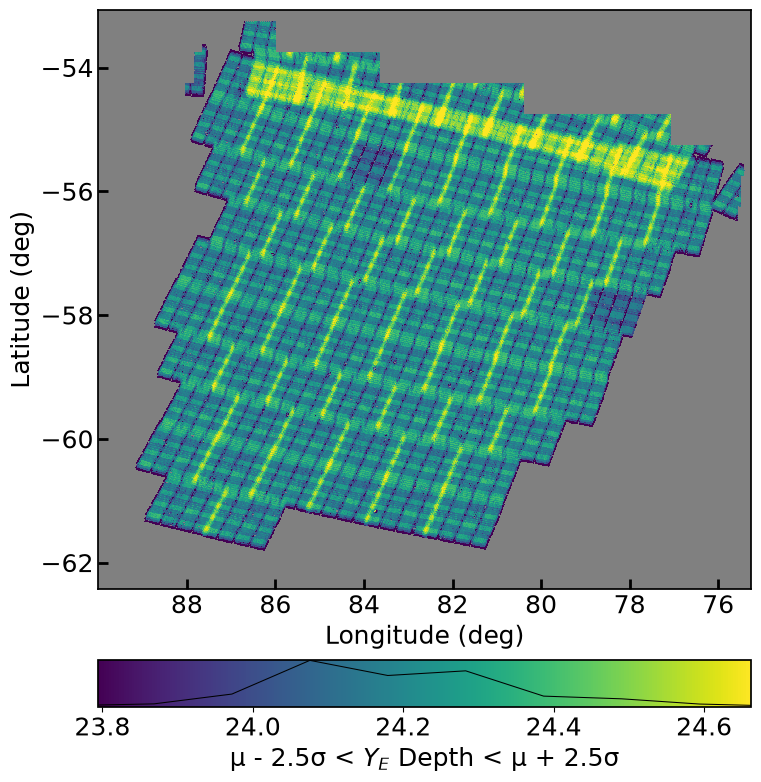}
        \includegraphics[angle=0,width=0.245\hsize]{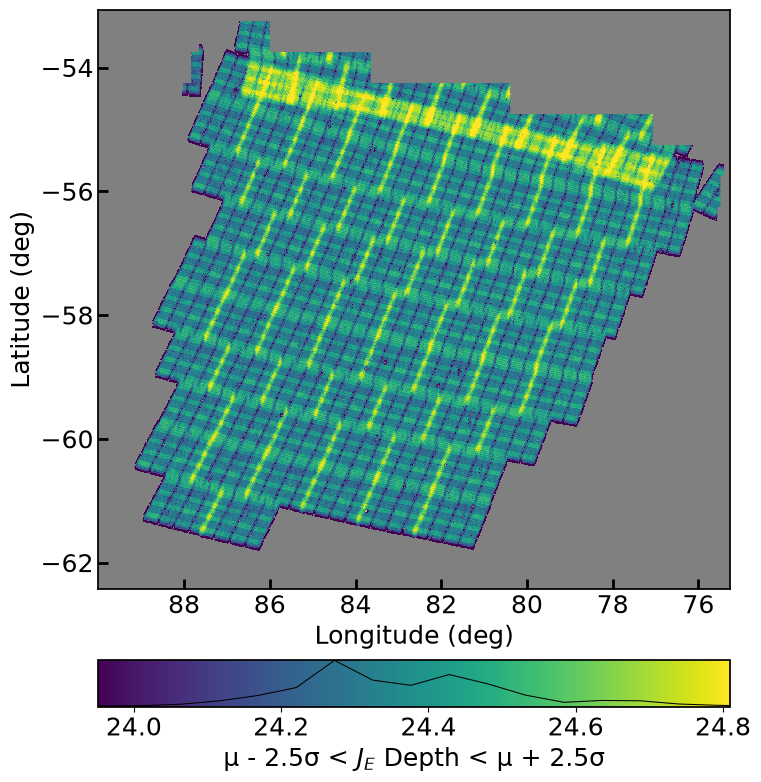}
        \includegraphics[angle=0,width=0.245\hsize]{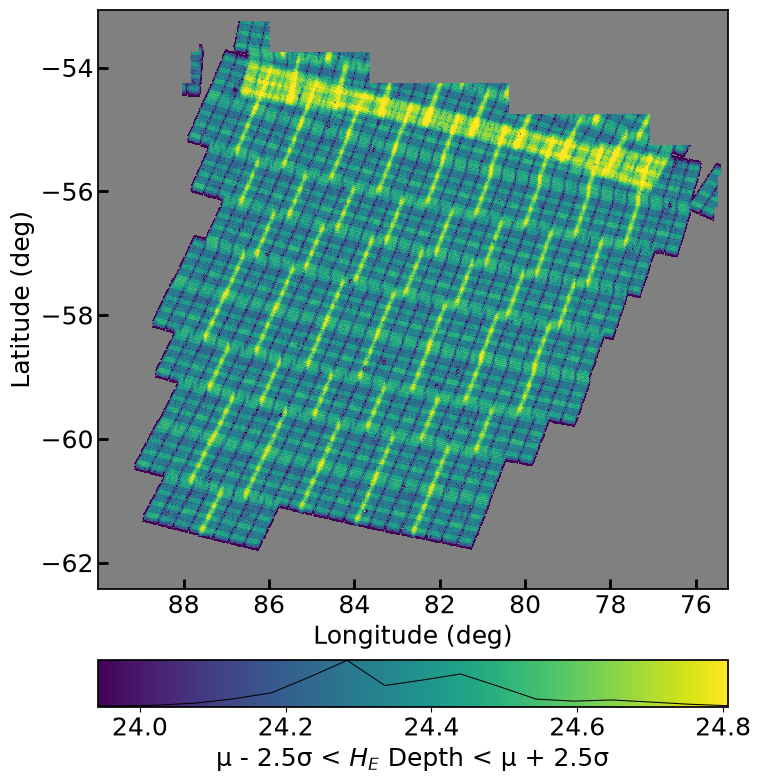}
        
        \includegraphics[angle=0,width=0.245\hsize]{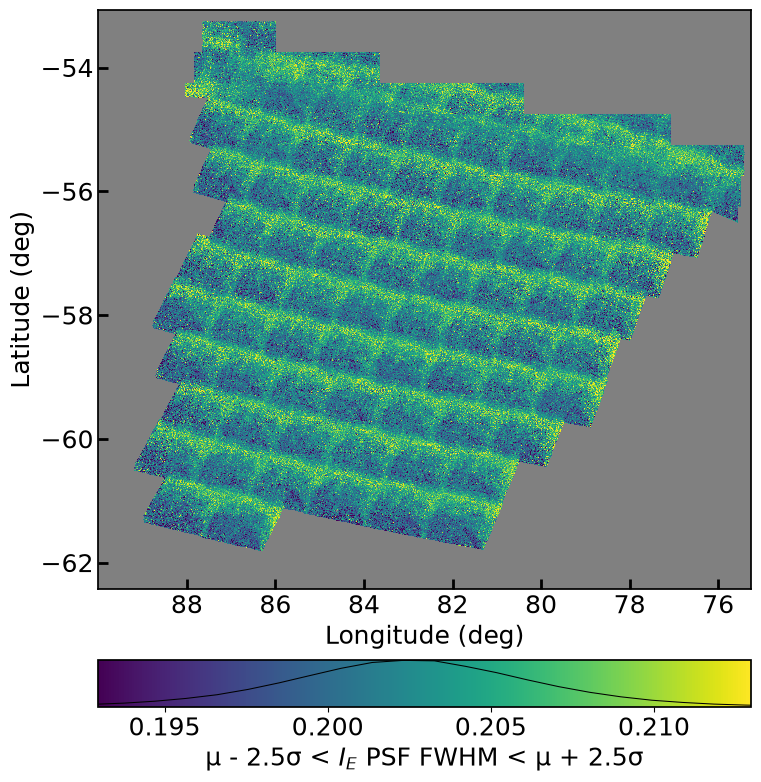}
        \includegraphics[angle=0,width=0.245\hsize]{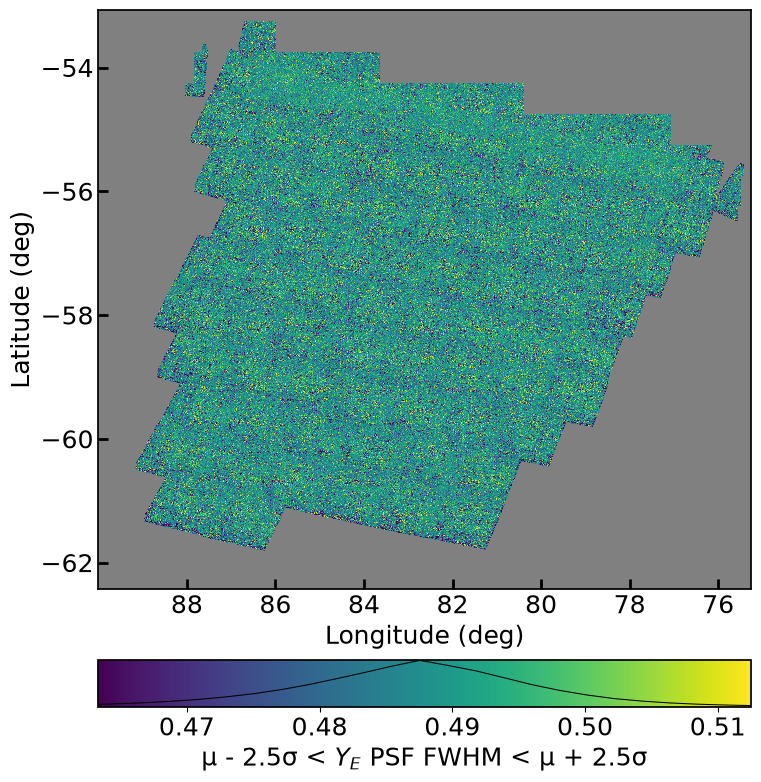}
        \includegraphics[angle=0,width=0.245\hsize]{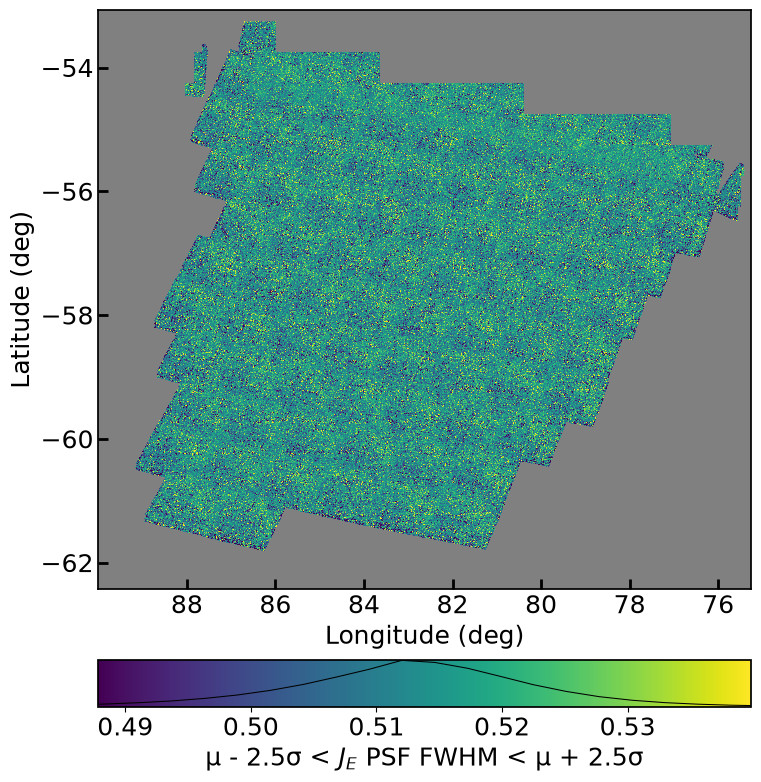}
        \includegraphics[angle=0,width=0.245\hsize]{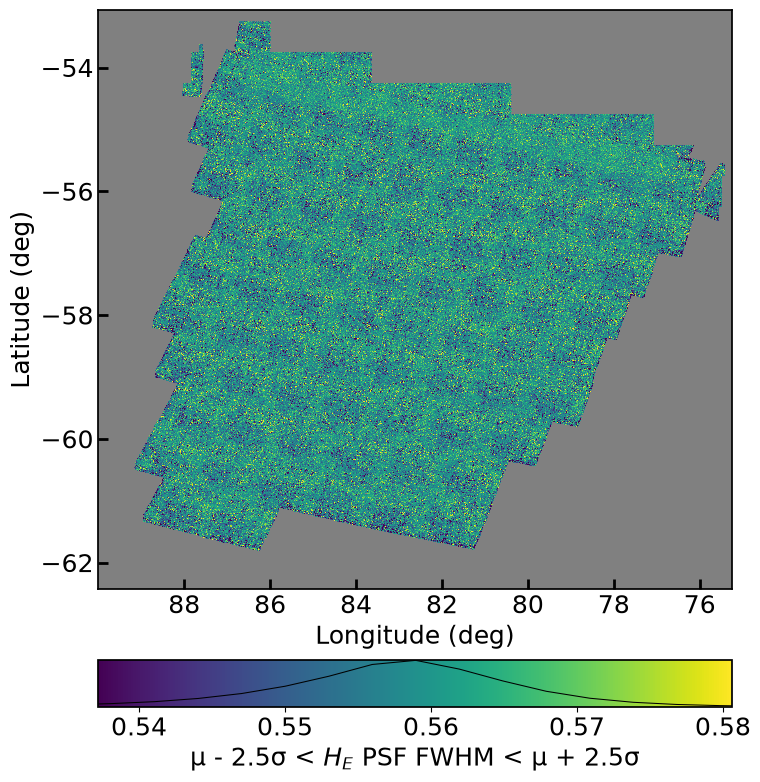}
        
        \includegraphics[angle=0,width=0.245\hsize]{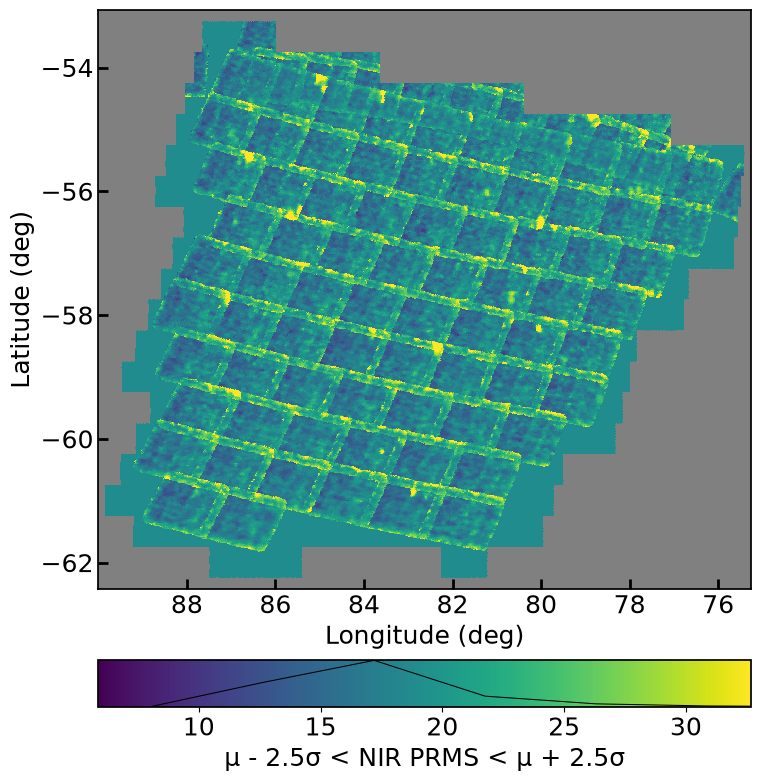}
        \includegraphics[angle=0,width=0.245\hsize]{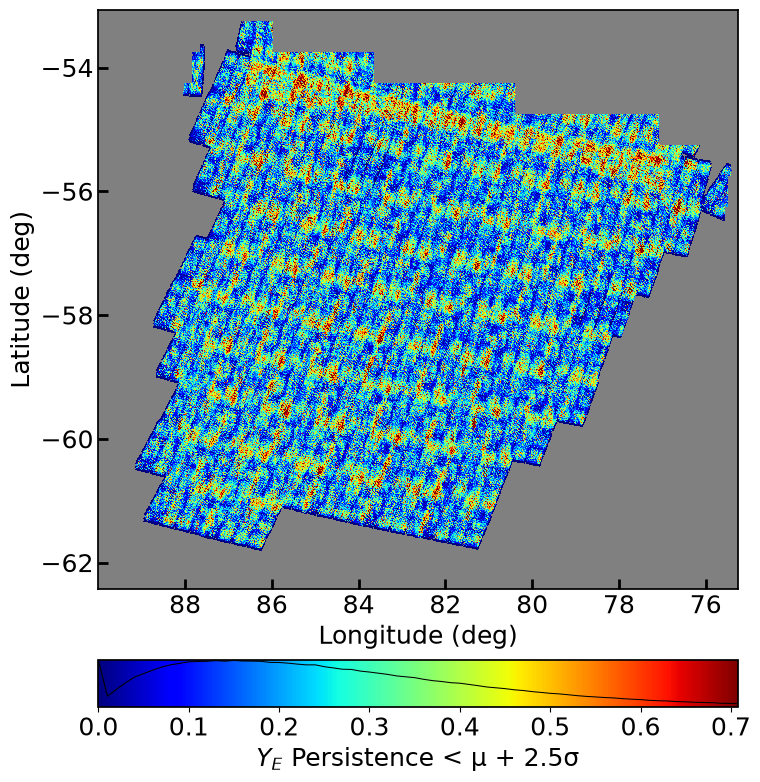}
        \includegraphics[angle=0,width=0.245\hsize]{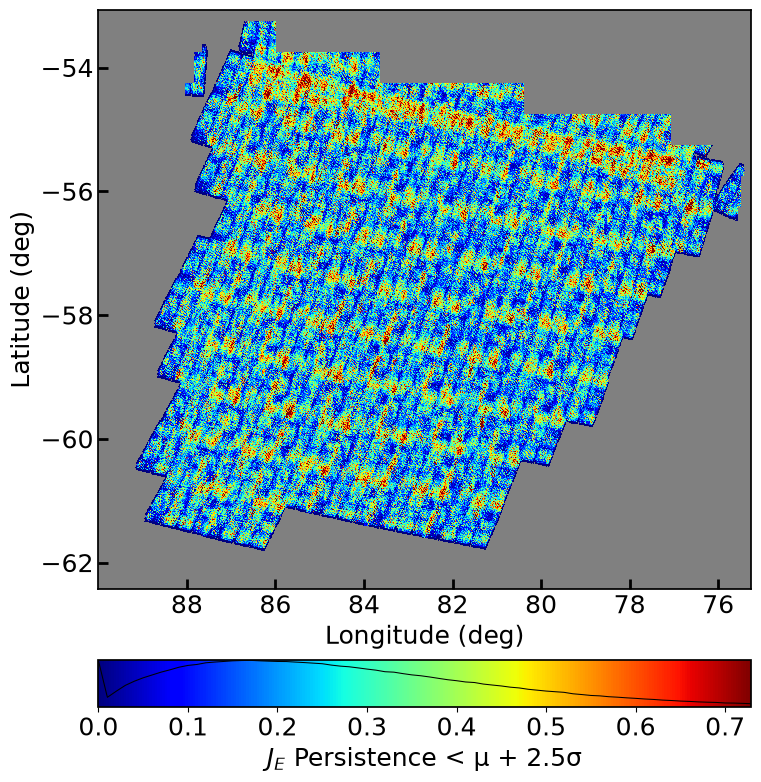}
        \includegraphics[angle=0,width=0.245\hsize]{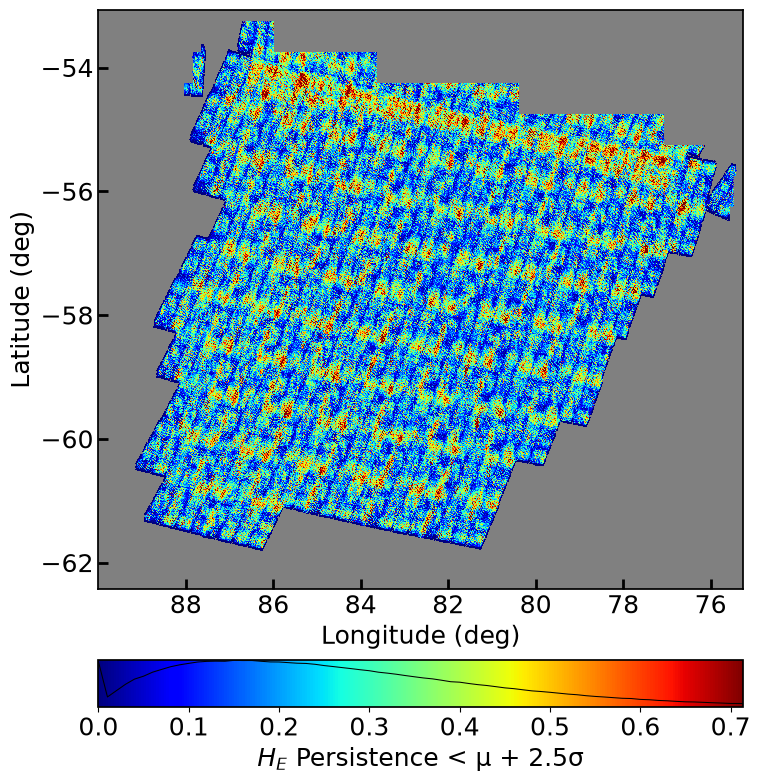}
        
        \caption{Examples of instrumental-property masks for the \DRone EWS Patch ID 4. From top to bottom, the panels show the RMS noise, depth, and PSF FWHM for the \IE, \YE, \JE, and \HE bands, followed by the NIR PRMS and persistence masks for the NIR bands.}
        \label{fig:patch4_instrumentmasks}
   \end{adjustwidth} 
\end{figure*}

\begin{figure}[htbp!]
    \begin{adjustwidth}{-1cm}{-1cm}
        \centering
        \includegraphics[angle=0,width=0.33\hsize]{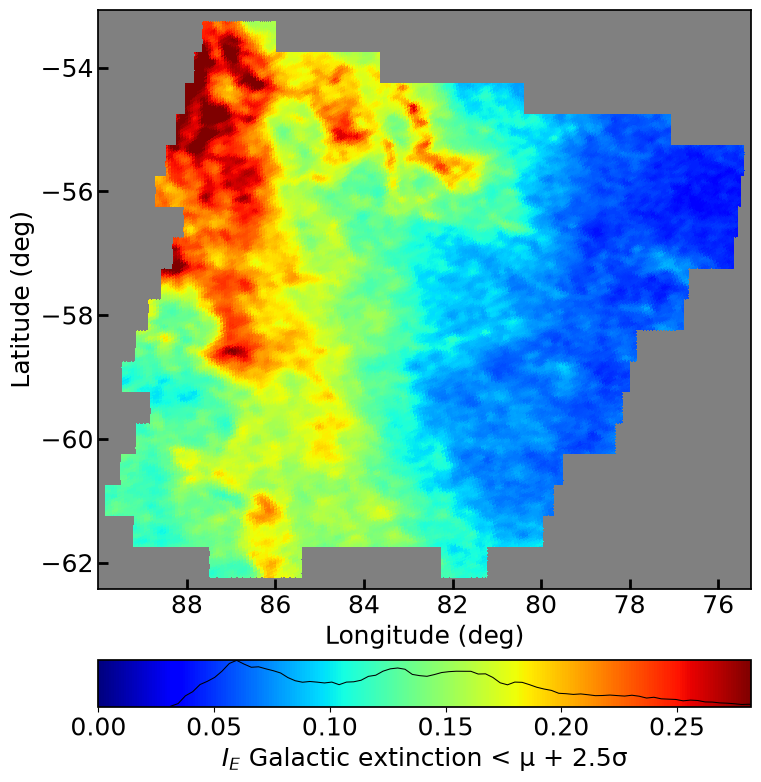}
        \includegraphics[angle=0,width=0.33\hsize]{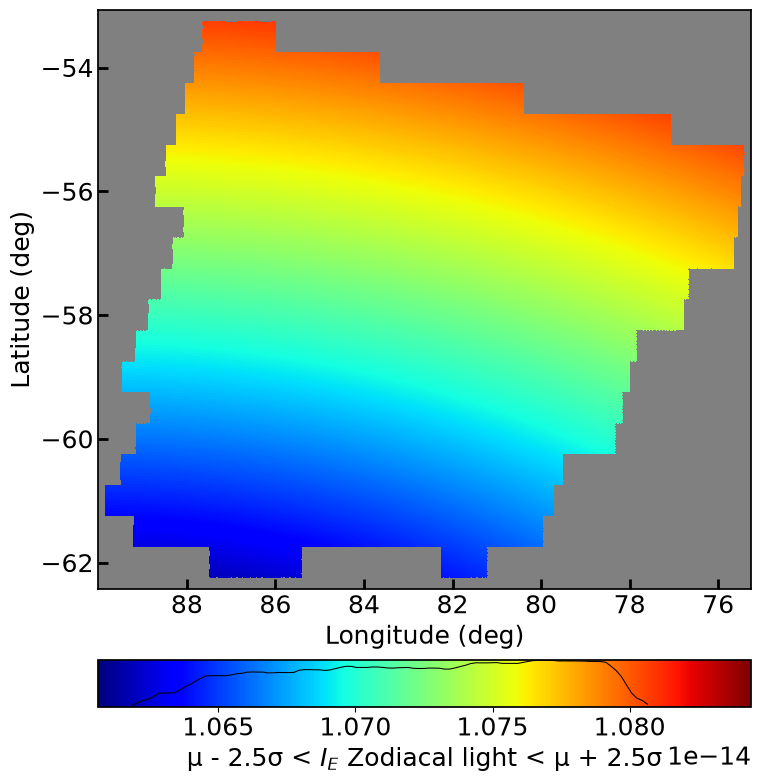}
        \includegraphics[angle=0,width=0.33\hsize]{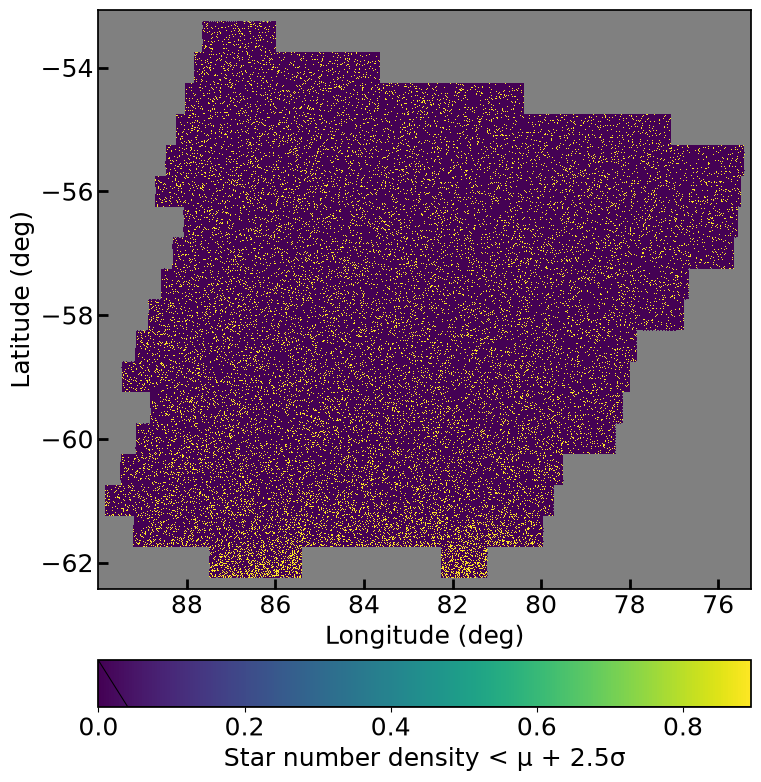}
        
        \includegraphics[angle=0,width=0.33\hsize]{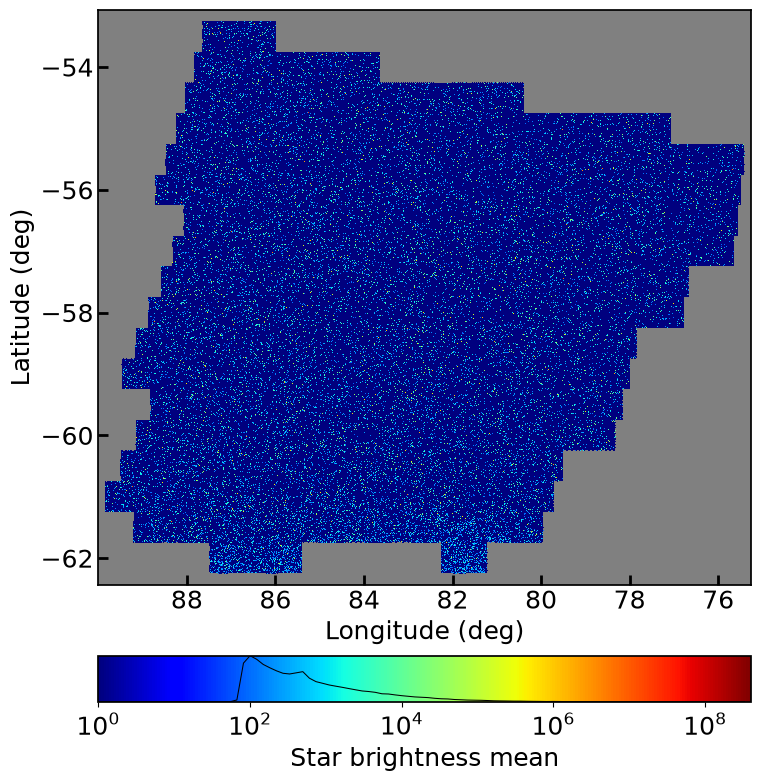}
        \includegraphics[angle=0,width=0.33\hsize]{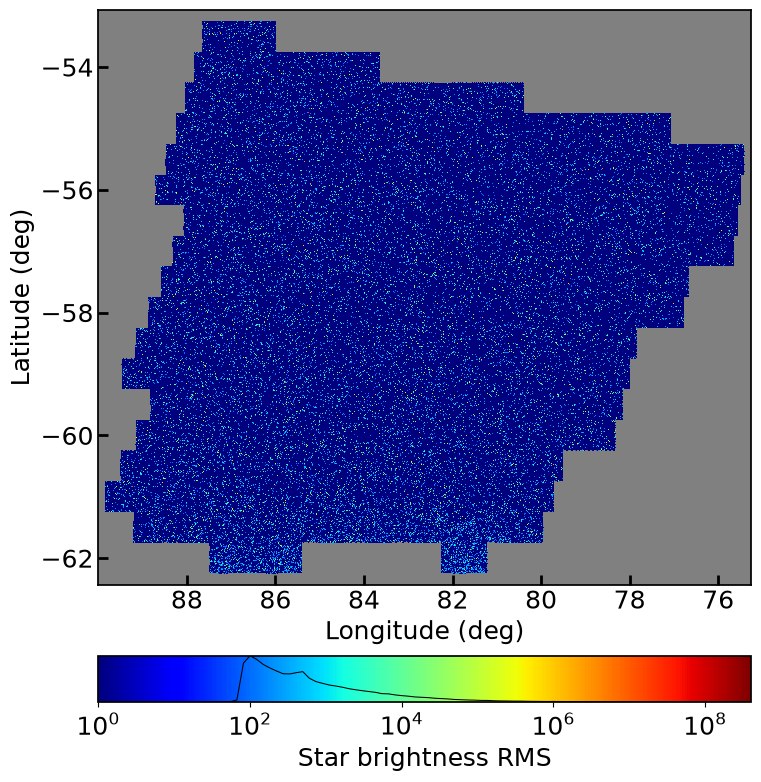}
        \includegraphics[angle=0,width=0.33\hsize]{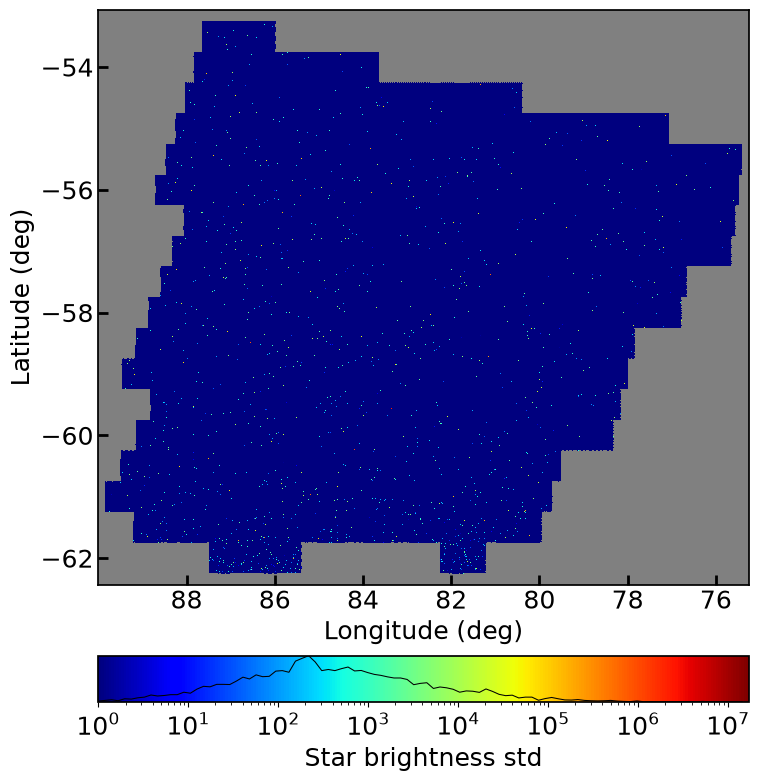}

        \caption{Examples of sky-emission and stellar-property masks for the \DRone EWS Patch ID 4. From left to right and from top to bottom, the panels show Galactic extinction, zodiacal light, star number density, mean star brightness, star brightness RMS, and star brightness standard deviation.\label{fig:patch4_skyemissionmasks}}    
    \end{adjustwidth}
\end{figure}

Examples of the \DRone EWS masks are presented here for observation Patch ID 4, in Figs.~\ref{fig:patch4_surveymasks}, \ref{fig:patch4_Coveragemasks}, \ref{fig:patch4_instrumentmasks}, and \ref{fig:patch4_skyemissionmasks}.

\FloatBarrier

\FloatBarrier

\section{Photometric and instrumental properties of the Euclid Deep Fields}
Tables~\ref{tab:instrumentprop_EDFF}, \ref{tab:instrumentprop_EDFN}, and \ref{tab:instrumentprop_EDFS} present statistical information for the EDFs.

\begin{table*}[htbp!]
    \caption{For \DRone EDF-F, the mean, median, standard deviation (Std), percentiles ($Q_{16}$, $Q_{25}$, $Q_{75}$, and $Q_{84}$), and IQR of the distributions of integration time, RMS Noise, depth, and PSF FWHM for the four \Euclid\ bands. \label{tab:instrumentprop_EDFF}}
    \centering
    \begin{small}
        \begin{tabular*}{\textwidth}{@{\extracolsep{\fill}}clrrrrrrrr}
            \hline\hline
            \noalign{\vskip 2pt}
            Band & Quantity & Mean & Median & Std & $Q_{16}$ & $Q_{25}$ & $Q_{75}$ & $Q_{84}$ & IQR \\
            \noalign{\vskip 1pt}
            \hline
            \noalign{\vskip 2pt}
            & Exposures [s] & 4642 & 4842 & 1070 & 4103 & 4192 & 5403 & 5492 & 1211 \\
            $ \IE $ & Noise [ADU] & 0.003197 & 0.003033 & 0.0008102 & 0.002836 & 0.002918 & 0.003204 & 0.003328 & 0.0002856 \\
            & Depth [AB mag] & 25.89 & 25.93 & 0.1783 & 25.83 & 25.87 & 25.97 & 26 & 0.1014 \\
            & PSF FWHM [arcsec] & 0.2035 & 0.2033 & 0.004154 & 0.1994 & 0.2005 & 0.2064 & 0.2077 & 0.005833 \\
            \hline
            \noalign{\vskip 2pt}
            & Exposures [s] & 582.5 & 610.7 & 134.7 & 523.5 & 523.5 & 698 & 698 & 174.5 \\
            $ \YE $ & Noise [ADU] & 1.74 & 1.675 & 0.3142 & 1.529 & 1.583 & 1.78 & 1.88 & 0.1968 \\
            & Depth [AB mag] & 24.34 & 24.37 & 0.159 & 24.25 & 24.31 & 24.43 & 24.47 & 0.1272 \\
            & PSF FWHM [arcsec] & 0.4896 & 0.4899 & 0.007884 & 0.4828 & 0.4853 & 0.4942 & 0.4965 & 0.008924 \\
            \hline
            \noalign{\vskip 2pt}
            & Exposures [s] & 579.4 & 610.7 & 135.3 & 523.5 & 523.5 & 698 & 698 & 174.5 \\
            $ \JE $ & Noise [ADU] & 2.017 & 1.939 & 0.3633 & 1.768 & 1.835 & 2.072 & 2.178 & 0.2366 \\
            & Depth [AB mag] & 24.48 & 24.51 & 0.1596 & 24.39 & 24.44 & 24.57 & 24.61 & 0.1317 \\
            & PSF FWHM [arcsec] & 0.5155 & 0.5158 & 0.008394 & 0.5083 & 0.511 & 0.5205 & 0.5229 & 0.009538 \\
            \hline
            \noalign{\vskip 2pt}
            & Exposures [s] & 570.3 & 610.7 & 144.1 & 436.2 & 523.5 & 698 & 698 & 174.5 \\
            $ \HE $ & Noise [ADU] & 1.864 & 1.776 & 0.3511 & 1.618 & 1.679 & 1.928 & 2.044 & 0.2489 \\
            & Depth [AB mag] & 24.47 & 24.51 & 0.1694 & 24.35 & 24.42 & 24.57 & 24.61 & 0.1501 \\
            & PSF FWHM [arcsec] & 0.5602 & 0.5606 & 0.00712 & 0.5543 & 0.5565 & 0.5644 & 0.5664 & 0.007904 \\
            \hline
        \end{tabular*}
    \end{small}
\end{table*}

\begin{table*}[htbp!]
    \caption{As Table~\ref{tab:instrumentprop_EDFF} for \DRone EDF-N. \label{tab:instrumentprop_EDFN}}
    \centering
    \begin{small}
        \begin{tabular*}{\textwidth}{@{\extracolsep{\fill}}llrrrrrrrr}
            \hline\hline
            \noalign{\vskip 2pt}
            Band & Quantity & Mean & Median & Std & $Q_{16}$ & $Q_{25}$ & $Q_{75}$ & $Q_{84}$ & IQR \\
            \noalign{\vskip 1pt}
            \hline
            \noalign{\vskip 2pt}
            & Exposures [s] & 21\,373 & 25\,332 & 8888 & 8092 & 17\,687 & 27\,282 & 28\,111 & 9595 \\
            $ \IE $ & Noise [ADU] & 0.001635 & 0.001221 & 0.001115 & 0.001162 & 0.001176 & 0.001454 & 0.002123 & 0.0002775 \\
            & Depth [AB mag] & 26.72 & 26.91 & 0.4366 & 26.31 & 26.72 & 26.95 & 26.97 & 0.23 \\
            &  PSF FWHM [arcsec] & 0.2033 & 0.2032 & 0.002332 & 0.2016 & 0.2021 & 0.2043 & 0.2049 & 0.002179 \\
            \hline
            \noalign{\vskip 2pt}
            \textcolor{white}{Band} & Exposures [s] & 2565 & 2966 & 1048 & 1047 & 2268 & 3228 & 3315 & 960 \\
            $ \YE $ & Noise [ADU] & 0.92 & 0.7123 & 0.8847 & 0.6729 & 0.684 & 0.8261 & 1.202 & 0.1421 \\
            & Depth [AB mag] & 25.12 & 25.3 & 0.402 & 24.73 & 25.14 & 25.34 & 25.36 & 0.205 \\
            &  PSF FWHM [arcsec] & 0.4918 & 0.4924 & 0.005346 & 0.4888 & 0.4902 & 0.4944 & 0.4954 & 0.004196 \\
            \hline
            \noalign{\vskip 2pt}
            & Exposures [s] & 2490 & 2966 & 1006 & 960 & 2181 & 3141 & 3228 & 960 \\
            $ \JE $ & Noise [ADU] & 1.055 & 0.8213 & 1.07 & 0.7784 & 0.7898 & 0.9519 & 1.372 & 0.1621 \\
            &  Depth [AB mag] & 25.27 & 25.44 & 0.3952 & 24.89 & 25.28 & 25.49 & 25.5 & 0.2026 \\
            & PSF FWHM [arcsec] & 0.5181 & 0.5186 & 0.005552 & 0.5149 & 0.5163 & 0.5207 & 0.5218 & 0.004441 \\
            \hline
            \noalign{\vskip 2pt}
            \textcolor{white}{Band} & Exposures [s] & 2554 & 2966 & 1044 & 1047 & 2181 & 3228 & 3315 & 1047 \\
            $ \HE $ & Noise [ADU] & 0.9624 & 0.7467 & 0.8437 & 0.705 & 0.7167 & 0.8694 & 1.257 & 0.1528 \\
            & Depth [AB mag] & 25.27 & 25.45 & 0.401 & 24.88 & 25.28 & 25.49 & 25.51 & 0.2098 \\
            & PSF FWHM [arcsec] & 0.5631 & 0.5637 & 0.004943 & 0.5605 & 0.5618 & 0.5655 & 0.5664 & 0.003773 \\
            \hline
        \end{tabular*}

    \end{small}

\end{table*}

\begin{table*}[htbp!]
    \caption{As Table~\ref{tab:instrumentprop_EDFF} for \DRone EDF-S. \label{tab:instrumentprop_EDFS}}
    \centering
    \begin{small}
        \begin{tabular*}{\textwidth}{@{\extracolsep{\fill}}clrrrrrrrr}
            \hline\hline
            \noalign{\vskip 2pt}
            Band & Quantity & Mean & Median & Std & $Q_{16}$ & $Q_{25}$ & $Q_{75}$ & $Q_{84}$ & IQR \\
            \noalign{\vskip 1pt}
            \hline
            \noalign{\vskip 2pt}
            & Exposures [s] & 20\,275 & 23\,561 & 7669 & 10\,245 & 18\,898 & 25\,040 & 25\,714 & 6142 \\
            $ \IE $ & Noise [ADU] & 0.00168 & 0.001296 & 0.001154 & 0.001246 & 0.00126 & 0.001451 & 0.001962 & 0.0001915 \\
            & Depth [AB mag] & 26.68 & 26.85 & 0.4079 & 26.4 & 26.73 & 26.88 & 26.89 & 0.1537 \\
            & PSF FWHM [arcsec] & 0.2035 & 0.2034 & 0.002228 & 0.2019 & 0.2024 & 0.2044 & 0.205 & 0.002045 \\
            \hline
            \noalign{\vskip 2pt}
            & Exposures [s] & 2505 & 2879 & 947 & 1221 & 2356 & 3141 & 3228 & 785 \\
            $ \YE $ & Noise [ADU] & 0.9221 & 0.7313 & 0.932 & 0.6972 & 0.7065 & 0.8173 & 1.099 & 0.1109 \\
            & Depth [AB mag] & 25.11 & 25.27 & 0.3784 & 24.83 & 25.15 & 25.31 & 25.32 & 0.1583 \\
            & PSF FWHM [arcsec] & 0.4916 & 0.4921 & 0.005175 & 0.4887 & 0.49 & 0.4941 & 0.4951 & 0.004157 \\
            \hline
            \noalign{\vskip 2pt}
            & Exposures [s] & 2514 & 2966 & 953 & 1221 & 2356 & 3141 & 3228 & 785 \\
            $ \JE $ & Noise [ADU] & 1.048 & 0.8318 & 0.9247 & 0.7931 & 0.8035 & 0.9321 & 1.253 & 0.1286 \\
            & Depth [AB mag] & 25.27 & 25.43 & 0.378 & 24.99 & 25.31 & 25.47 & 25.48 & 0.1612 \\
            & PSF FWHM [arcsec] & 0.5177 & 0.5182 & 0.005378 & 0.5146 & 0.5159 & 0.5203 & 0.5214 & 0.004376 \\
            \hline
            \noalign{\vskip 2pt}
            & Exposures [s] & 2511 & 2966 & 952.7 & 1221 & 2268 & 3141 & 3228 & 872 \\
            $ \HE $ & Noise [ADU] & 0.9673 & 0.7651 & 1.208 & 0.7291 & 0.7388 & 0.8596 & 1.155 & 0.1209 \\
            & Depth [AB mag] & 25.26 & 25.42 & 0.3795 & 24.97 & 25.3 & 25.46 & 25.47 & 0.1645 \\
            & PSF FWHM [arcsec] & 0.5629 & 0.5634 & 0.00469 & 0.5603 & 0.5615 & 0.5652 & 0.5661 & 0.003726 \\
            \hline
        \end{tabular*}
    \end{small}

\end{table*}

\FloatBarrier

\end{appendix}

\label{LastPage}
\end{document}